\documentclass[12pt]{article}

\usepackage[utf8]{inputenc}
\usepackage{textcomp}
\usepackage{chetnew}
\usepackage{tikz}
\usepackage{amsmath,amssymb,amsfonts,mathrsfs,dsfont,yfonts,bbm}\usetikzlibrary{calc}
\usepackage{xcolor}
\usepackage{braket}
\usepackage[display]{texpower}
\usepackage{graphicx}
\usepackage{relsize}
\usepackage{tabu}
\usepackage{multirow}
 \usepackage{makecell}
 \usepackage[T1]{fontenc}
 \usepackage{bm}
\usepackage{verbatim}
\usepackage{longtable}
\usepackage{colortbl}

\usepackage{cancel}

\usepackage{float}
\newcolumntype{L}{>{$}l<{$}}
 
\newcolumntype{C}{>{$}c<{$}}

\newcommand{\mk}{\cellcolor{gray!25} }

 \newcommand{\para}[1]{ \paragraph{$\bm { #1}$}     }

\newcommand{\be}{\begin{equation}}
\newcommand{\ee}{\end{equation}}
\newcommand{\bea}{\begin{eqnarray}}
\newcommand{\eea}{\end{eqnarray}}
\newcommand{\p}{\partial}

\newcommand{\CL}{\mathcal{L}}

\newcommand{\CT}{\mathcal{T}}

\newcommand{\CN}{\mathcal{N}}
\newcommand{\CS}{\mathcal{S}}

\newcommand{\beqn}{\begin{eqnarray}}
\newcommand{\eeqn}{\end{eqnarray}}

\newcommand*{\boxcoloro}{orange}
\makeatletter
\newcommand{\boxedo}[1]{\textcolor{\boxcoloro}{%
\tikz[baseline={([yshift=-1ex]current bounding box.center)}] \node [rectangle, minimum width=1ex,rounded corners,draw] {\normalcolor\m@th$\displaystyle#1$};}}
 \makeatother
\newcommand*{\boxcolorr}{red}
\makeatletter
\newcommand{\boxedr}[1]{\textcolor{\boxcolorr}{%
\tikz[baseline={([yshift=-1ex]current bounding box.center)}] \node [rectangle, minimum width=1ex,rounded corners,draw] {\normalcolor\m@th$\displaystyle#1$};}}
 \makeatother
\newcommand*{\boxcolorb}{blue}
\makeatletter
\newcommand{\boxedb}[1]{\textcolor{\boxcolorb}{%
\tikz[baseline={([yshift=-1ex]current bounding box.center)}] \node [rectangle, minimum width=1ex,rounded corners,draw] {\normalcolor\m@th$\displaystyle#1$};}}
 \makeatother
\newcommand*{\boxcolorg}{green}
\makeatletter
\newcommand{\boxedg}[1]{\textcolor{\boxcolorg}{%
\tikz[baseline={([yshift=-1ex]current bounding box.center)}] \node [rectangle, minimum width=1ex,rounded corners,draw] {\normalcolor\m@th$\displaystyle#1$};}}
 \makeatother
 \newcommand*{\boxcolorp}{purple}
\makeatletter
\newcommand{\boxedp}[1]{\textcolor{\boxcolorp}{%
\tikz[baseline={([yshift=-1ex]current bounding box.center)}] \node [rectangle, minimum width=1ex,rounded corners,draw] {\normalcolor\m@th$\displaystyle#1$};}}
 \makeatother
  \newcommand*{\boxcolorc}{cyan}
\makeatletter
\newcommand{\boxedc}[1]{\textcolor{\boxcolorc}{%
\tikz[baseline={([yshift=-1ex]current bounding box.center)}] \node [rectangle, minimum width=1ex,rounded corners,draw] {\normalcolor\m@th$\displaystyle#1$};}}
 \makeatother
  \newcommand*{\boxcolory}{yellow}
\makeatletter
\newcommand{\boxedy}[1]{\textcolor{\boxcolory}{%
\tikz[baseline={([yshift=-1ex]current bounding box.center)}] \node [rectangle, minimum width=1ex,rounded corners,draw] {\normalcolor\m@th$\displaystyle#1$};}}
 \makeatother

\usepackage{amsmath}	
\begin{document}
\preprint{QMUL-PH-21-27}

\title{1-Form Symmetry, Isolated $\CN=2$ SCFTs,\\[2mm] and Calabi-Yau Threefolds}

\author{Matthew Buican$^{\diamondsuit}$ and Hongliang Jiang$^{\clubsuit}$}

\affiliation{\smallskip CRST and School of Physics and Astronomy\\
Queen Mary University of London, London E1 4NS, UK
\emails{$^{\diamondsuit}$m.buican@qmul.ac.uk, $^{\clubsuit}$h.jiang@qmul.ac.uk}}

\abstract{We systematically study 4D $\CN=2$ superconformal field theories (SCFTs) that can be constructed via type IIB string theory on isolated hypersurface singularities (IHSs) embedded in $\mathbb{C}^4$. We show that if a theory in this class has no $\CN=2$-preserving exactly marginal deformation (i.e., the theory is isolated as an $\CN=2$ SCFT), then it has no 1-form symmetry. This situation is somewhat reminiscent of 1-form symmetry and decomposition  in 2D quantum field theory. Moreover, our result suggests that, for theories arising from IHSs, 1-form symmetries originate from gauge groups (with vanishing beta functions). One corollary of our discussion is that there is no 1-form symmetry in IHS theories that have all Coulomb branch chiral ring generators of scaling dimension less than two. In terms of the $a$ and $c$ central charges, this condition implies that IHS theories satisfying $a<{1\over24}(15r+2f)$ and $c<{1\over6}(3r+f)$ (where $r$ is the complex dimension of the Coulomb branch, and $f$ is the rank of the continuous 0-form flavor symmetry) have no 1-form symmetry.  After reviewing the 1-form symmetries of other classes of theories, we are motivated to conjecture that general interacting 4D $\CN=2$ SCFTs with all Coulomb branch chiral ring generators of dimension less than two have no 1-form symmetry.

}
\date{June 2021}

\setcounter{tocdepth}{2}

\maketitle
\toc

\newsec{Introduction}
While a complete classification of 4D quantum field theories (QFTs) seems well out of reach, one gets the sense that 4D $\CN=2$ superconformal field theories (SCFTs) may be more amenable to classification. This feeling arises due to a confluence of factors: 4D $\CN=2$ SCFTs are closely related to special 2D QFTs (e.g., see \cite{Cecotti:2010fi,Beem:2013sza,Nekrasov:2015wsu}), integrable structures naturally arise in the study of these theories (e.g., see \cite{Donagi:1995cf,Gaiotto:2012xa}), 4D $\CN=2$ conformal manifolds (when they exist) are highly constrained (e.g., see \cite{Papadodimas:2009eu,Gerchkovitz:2014gta}), and $\CN=2$ moduli spaces feature various rigid structures (e.g., see \cite{Seiberg:1994rs,Seiberg:1994aj,Argyres:1996eh,Argyres:2015ffa,Argyres:2015gha,Argyres:2016xmc,Bourget:2019aer}).

The intersection of these last two points forms a starting point of our story. In particular, many 4D $\CN=2$ conformal manifolds can be constructed by gauging a $G$ flavor symmetry of one or more isolated 4D $\CN=2$ SCFTs\footnote{In this context, by \lq\lq isolated'' we mean $\CN=2$ theories with no $\CN=2$-preserving exactly marginal deformation.} (combined with unitarity, the existence of such a symmetry implies the existence of non-nilpotent moment map operators\footnote{By non-nilpotent moment map operators, we have in mind holomorphic $\mu$ such that $\mu^N\ne0$ in the chiral ring for any $N>0$ (we do not contract indices on $\mu$).} \cite{Buican:2016arp} and, by standard lore, a Higgs branch of moduli space in each of the isolated theories we gauge). The resulting $G$ gauge group, with vanishing beta functions, is a natural source for line operators (Wilson and 't Hooft lines) charged under 1-form symmetries \cite{Gaiotto:2014kfa}.

Therefore, in order to gain an additional handle on the space of 4D $\CN=2$ SCFTs, it is useful to understand whether these conformal gauge groups are the only sources of 1-form symmetries. To answer this question, we should study whether (1) isolated $\CN=2$ SCFTs can have 1-form symmetries and (2) whether all exactly marginal couplings in 4D $\CN=2$ are indeed gauge couplings.

Our focus in this paper will mostly be on the first point.\footnote{Regarding the second point in the previous paragraph, we are not aware of gauge coupling interpretations for most of the conformal manifolds we will come across, but we are also not aware of arguments forbidding such interpretations (similar comments apply to the vast majority of the theories discussed in \cite{Buican:2018ddk,Buican:2019kba}). Indeed, there are conjectures that an $\CN=2$-preserving exactly marginal deformation always corresponds to a gauge coupling \cite{Beem:2014zpa,Perlmutter:2020buo}.} Moreover, we will mainly restrict our attention to the case of 4D $\CN=2$ SCFTs arising from the low-energy limit of type IIB string theory on three-complex-dimensional isolated hypersurface singularities (IHSs) embedded in $\mathbb{C}^4$ and characterized by quasi-homogeneous polynomials satisfying\cite{Shapere:1999xr}
 \be\label{IHS}
W(x_i):=W(x_1,x_2,x_3,x_4)=0~,\ \  W(\lambda^{q_i} x_i) =\lambda W(x_i)~, \ \ x_i \in \mathbb C~,\ \ \lambda \in \mathbb C^{*}~,\ \ q_i\in\mathbb{Q}~,
 \ee
 with
 \begin{equation}
 dW(x_i)=W(x_i)=0\ \Leftrightarrow\ x_{1,2,3,4}=0~,
 \end{equation}
 and
 \begin{equation}
q_i>0~,\ \ \ \sum_iq_i>1~. 
 \end{equation}
 In terms of these variables, the holomorphic 3-form is given by
 \begin{equation}\label{3formC4}
\Omega={dx_1\wedge dx_2\wedge dx_3\wedge dx_4\over dW}~. 
 \end{equation}
 
This vast class of SCFTs includes, among many others, the well-known $(\mathfrak{g},\mathfrak{g}')$ theories of \cite{Cecotti:2010fi}. The early Argyres-Douglas (AD) SCFTs \cite{Argyres:1995jj,Argyres:1995xn} are prominent members of this family of theories: the $(A_1, A_2)$, $(A_1, A_3)$, and $(A_1, D_4)$ theories correspond to SCFTs on the Coulomb branches of $SU(2)$ with $N_f=0,1,2$ respectively (or alternatively, in the first case, to pure $SU(3)$ $\CN=2$ SYM).

For general IHS theories of this type, our main claim in this paper is the following:

\medskip
\noindent
{\bf Claim 1 (main claim):} \emph{Any isolated 4D $\CN=2$ SCFT (i.e., a theory without $\CN=2$-preserving exactly marginal deformations\footnote{We allow for exactly marginal deformations that only preserve $\CN=1$.}) arising from type IIB string theory on an IHS embedded in $\mathbb{C}^4$ has trivial 1-form symmetry.\footnote{Note that here we have in mind the SCFTs specified by the singularities themselves. In particular, we do not consider decoupled $U(1)$ factors that may arise in RG flows emanating from these SCFTs as IHS theories in their own right.}} (Unless otherwise noted, when we refer to IHS theories below, we mean theories of this type).
\noindent
\medskip

\noindent
We believe this result extends to theories arising on IHSs embedded in other spaces like $\mathbb{C}^{*}\times\mathbb{C}^3$, and we will discuss some of the resulting $\CN=2$ SCFTs explicitly in section \ref{gencomments}. We will also comment on other $\CN=2$ SCFTs which are not  realized via IHSs.

A formalism for finding the 1-form symmetries in the context of the theories covered by our main claim was developed in \cite{Closset:2020scj,DelZotto:2020esg,Closset:2020afy}. Using these tools, along with mathematical results in \cite{yau2005classification} and their physical interpretation in \cite{Xie:2015rpa}, we give a proof of the main claim for the subset of IHS theories corresponding to $(\mathfrak{g},\mathfrak{g}')$ SCFTs in section \ref{IHSproof}. We relegate a complete proof of our claim for all the IHS theories to the appendix.

However, our results do not fully rely on \cite{yau2005classification}. In particular, motivated by \cite{Davenport:2016ggc,Closset:2020scj}, we allow for potentially more general IHS-defining polynomials with at least five terms. As we describe in the appendix, we are able to bypass the precise nature of these additional monomials, because we are only interested in questions regarding 1-form symmetries in isolated theories.

One intriguing aspect of the theories we study is that they are related to 2D $\CN=(2,2)$ Landau-Ginzburg (LG) models via the correspondence in \cite{Cecotti:2010fi} (indeed, the analysis of \cite{Davenport:2016ggc} mentioned in the previous paragraph proceeds from this connection). Therefore, one may wonder if there is a relation between the 1-form symmetries we study and symmetries in the corresponding LG models.

While we do not currently have a precise answer to this question, we note an intriguing parallel between our story for 4D $\CN=2$ SCFTs and the idea of decomposition in 2D QFT (e.g., see \cite{Hellerman:2006zs} and the recent workshop \cite{d2021}). In essence, we are arguing that, on their own, isolated 4D $\CN=2$ SCFTs arising from IHSs have no 1-form symmetry. Instead, we need to combine such SCFTs into larger theories via gauging in order to have such symmetry.\footnote{Note that this larger theory may or may not have an IHS description.} Equivalently, if we start from a gauged collection of such isolated SCFTs and decompose the theory into its isolated SCFT constituents at zero coupling, each isolated constituent on its own will lack 1-form symmetry. Somewhat similarly, in the context of 2D QFT, the idea of decomposition states that a 2D QFT with 1-form symmetry can be rewritten as a disjoint union of QFTs without 1-form symmetry. 

Given this discussion, several immediate questions arise:
\begin{itemize}
\item{Is the converse of our main claim true? In other words, do all conformal manifolds arising from IHSs have one-form symmetries? No. For example, we can have local operators in the isolated matter sectors transforming non-trivially under the centers of the exactly marginal gauge groups (thereby breaking the 1-form symmetry). More concretely, consider the realization of $(A_5, A_7)$ given in \cite{Giacomelli:2020ryy} as a diagonal $SU(3)$ gauging of the $D_7(SU(3))$ SCFT and the $D_7(SU(4))$ SCFT.\footnote{The $D_p(G)$ theories were discussed in \cite{Cecotti:2012jx,Cecotti:2013lda}.} The $su(4)$ moment map of $D_7(SU(4))$ decomposes into the following representations of $su(3)$
\begin{equation}
\mu^A:\ \ \ {\bf15}\to{\bf8}\oplus{\bf3}\oplus{\bf\bar3}\oplus{\bf1}~.
\end{equation}
As a result, the moment map has components that transform under the $\mathbb{Z}_3$ center of $SU(3)$, and the theory does not have 1-form symmetry (e.g., Wilson lines in the fundamental can end on $D_7(SU(4))$ moment map operators). It is possible to argue that $D_7(SU(3))$ and $D_7(SU(4))$ have trivial 1-form symmetry,\footnote{For one argument, see \cite{Hosseini:2021ged}. We will discuss these theories further in section \ref{gencomments}.} and so $(A_5, A_7)$ has trivial one-form symmetry even though it has a conformal manifold and a gauge group with vanishing beta function (the result for $(A_5, A_7)$ also follows from the methods in \cite{Closset:2020scj,DelZotto:2020esg,Closset:2020afy}).
}

\item{Does our main claim hold for more general $\CN=2$ theories? As we will see in section \ref{gencomments}, the arguments in \cite{Tachikawa:2013hya,Gukov:2020btk,Bhardwaj:2021pfz} suggest that many isolated (as $\CN=2$ SCFTs) class $\CS$ theories (here we mean isolated theories coming from the  twisted compactification of the 6D $(2,0)$ theory on surfaces that do not have irregular punctures) also have no 1-form symmetry. 
On the other hand, recent arguments suggest that the $\CN=3$ theory related to the $G(3,3,3)$ complex reflection group \cite{Aharony:2016kai} might have $\mathbb{Z}_3$ 1-form symmetry \cite{Zafrir:2020epd}. While, on general grounds \cite{Aharony:2015oyb}, this theory is isolated as an $\CN\ge2$ theory, it may potentially be part of an $\CN=1$ conformal manifold with a 1-form symmetry arising from an $\CN=1$ gauging of the $E_6$ Minahan-Nemeschansky (MN) theory coupled to some additional matter fields. If it is true that the $G(3,3,3)$ $\CN=3$ theory has 1-form symmetry, then perhaps one can show that, for $\CN\ge2$ theories that are isolated as $\CN=1$ theories and have no \lq\lq conformal gauge theory origin'' (i.e., do not have a formulation in terms of an $\CN=1$ gauge group with vanishing beta function), there is no 1-form symmetry. At present, we do not have a proof of this statement.\footnote{One should be careful with such a conjecture. Indeed, if there are $\CN=3$ theories of this type with 1-form symmetry, then the universal mass deformation $\delta W=\lambda\mu$, with $\mu$ the moment map of the $U(1)$ $\CN=2$ flavor symmetry all $\CN=3$ SCFTs possess, will generate flows to theories that have (in the absence of SUSY enhancement) $\CN=2$ SUSY. These theories may inherit the 1-form symmetry of the UV $\CN=3$ SCFT and might be isolated (even as $\CN=1$ theories). In this case, a more general conjecture would be that isolated theories have 1-form symmetry only if they have some (possibly UV) gauge theory origin. However, it is not clear to us how meaningful this statement is, since it may be that all 4D QFTs can be obtained from RG flows emanating from gauge theories.}}
\item{Does our result hold more generally for 4D $\CN=2$ SCFTs related to 2D LG theories via the correspondence in \cite{Cecotti:2010fi}? For example, there are theories of this type involving more than four variables that don't typically have a presentation in terms of type IIB on an IHS. While such a result seems plausible, we do not currently have a conclusive argument one way or the other.}
\item{Can we deduce any universal constraints from our main claim? As we will discuss in section \ref{conjectures}, there are some tantalizing hints the answer may be yes. Indeed, one corollary that follows from our claim is that, in our class of SCFTs, any theory with all Coulomb branch generators of dimension less than two must have trivial 1-form symmetry. After reviewing other classes of theories in section \ref{gencomments}, we are motivated to conjecture that this statement is more generally true for interacting 4D $\CN=2$ SCFTs. Using results in \cite{Shapere:2008zf} (see also the application in \cite{Xie:2015xva}) we can then argue this condition implies that for IHS theories
\begin{equation}\label{acbound}
c<{1\over6}(3r+f)~,\ \ \ a<{15r+2f\over24}~,
\end{equation}
where $r$ is the complex dimension of the Coulomb branch (i.e., the \lq\lq rank'' of the theory), and $f$ is the rank of the continuous 0-form flavor symmetry.  Note that this discussion can be phrased abstractly for any 4D $\CN=2$ SCFT,\footnote{The flavor symmetry rank is an unambiguous non-perturbative quantity. The rank of the Coulomb branch may be replaced by the number of generators of the $\CN=2$ chiral ring modulo relations (here we define the chiral ring operators to be annihilated by the full anti-chiral set of $\CN=2$ supercharges).} and it would be interesting to understand if these bounds imply lack of 1-form symmetry more generally.
}
\end{itemize}

The plan of this paper is as follows. First we briefly review the tools constructed in \cite{Closset:2020scj,DelZotto:2020esg,Closset:2020afy} for 1-form symmetry detection. We then give a proof of our main claim for isolated $\CN=2$ SCFTs arising from the subset of $(\mathfrak{g},\mathfrak{g}')$ SCFTs and explain how these results fit into our broader proof. After that, we derive some useful general results for our IHS theories including a lemma on the absence of 1-form symmetries in IHS theories that admit certain bilinears. Then, we give a very rough sketch of the 1-form symmetry content of more general 4D SCFTs with $\CN\ge2$ SUSY (including those arising from IHSs in $\mathbb{C}^{*}\times\mathbb{C}^3$). Before concluding and mentioning open problems, we discuss the corollary described around \eqref{acbound} and its potential implications for the space of 4D $\CN=2$ SCFTs. We relegate a full proof of our main claim for all theories arising from IHSs in $\mathbb{C}^4$ to appendix \ref{SCFTclassify}.

\newsec{Trivial 1-form symmetries in isolated SCFTs arising from IHSs}\label{revProof}
Given a realization of a 4D $\CN=2$ SCFT, $\CT_{X_6}$, as the low-energy limit of type IIB string theory on an IHS 3-fold, $X_6$, the authors of \cite{Closset:2020scj,DelZotto:2020esg,Closset:2020afy} found a prescription for computing the corresponding 1-form symmetry. The basic idea is to compute the component of the defect group, $\mathbb{D}^{(1)}$, that arises from D3 branes wrapping certain 3-cycles in $X_6$ (we refer the interested reader to the original papers for more details). Here we merely state the result:
\begin{equation}\label{D1def}
\mathbb{D}^{(1)}=\mathbb{Z}^{\kappa}\oplus\bigoplus_{i=1}^n\left(\mathbb{Z}_{r_i}^{2g_i}\right)~,
\end{equation} 
where the 4D 1-form symmetry group is determined by choosing maximal isotropic subgroups of the finite groups in the summand. This procedure ensures mutual locality of the spectrum (in analogy with the procedure in \cite{Aharony:2013hda}).

It turns out that these finite groups can be fixed in terms of the singularity data \eqref{IHS}. In particular, taking
\be\label{qVU}
 q_i= \frac{V_i}{U_i}~,\ \ \  \gcd(U_i,V_i)=1~,
\ee 
yields \cite{Closset:2020scj,DelZotto:2020esg,Closset:2020afy}
 \be\label{r1}
 r_i =\gcd(w_1, \cdots, \hat w_i, \cdots w_4)~,\ \ \ w_i:=D q_i~, \ \ \ D:={\rm lcm}(U_1,U_2,U_3,U_4)~,
 \ee
 and
 \be\label{g1}
 2g_i= -1 +\frac{D^2 r_i}{w_1 \cdots  \hat w_i  \cdots w_4 }+\sum_{j\neq i} \frac{ \gcd(D,w_j)}{w_j}
-\sum_{j<k, j,k\neq i} \frac{D\; \gcd(w_j,w_k)}{w_jw_k}~.
 \ee
Note that in \eqref{r1} and \eqref{g1}, \lq\lq$\hat x$'' refers to dropping the variable $x$ from the corresponding expressions.

For our purposes below, it is slightly more useful to rewrite the previous two equations in a way that does not refer directly to $D$. In particular, as discussed in \cite{boyer2007sasaki}, one has
\be\label{Ri}
 r_i =\frac{U_i }{\gcd (U_1,U_2,U_3,U_4)}  \frac{ \prod_{k<l,k,l\neq i}\gcd(U_i,U_k,U_l)}{ \prod_{j\neq i }    \gcd(U_i,U_j)}~,
\ee
and 
 \be\label{Gi}
 2g_i=-1+\sum_{j\neq i} \frac{1}{V_j} - \sum_{k<l,k,l\neq i} \frac{\gcd(U_k,U_l)}{V_k V_l}+\frac{ \prod_{k<l,k,l\neq i}\gcd(  U_k,U_l)}{ \prod_{j<k<l, j,k,l\neq i }    V_jV_kV_l\gcd(U_j,U_k,U_l)}~.
 \ee
 To argue that a theory has no 1-form symmetry, it suffices to show that there is no $i$ such that $r_i>1$ and $g_i>0$.

To check this absence of 1-form symmetry holds for isolated theories, we need to show that the above conditions hold for any theory that has no dimension-two Coulomb branch operator (the level-four superconformal descendant arising from the integration over all of chiral superspace becomes the dimension-four exactly marginal deformation).\footnote{By the general discussion in \cite{Buican:2013ica,Buican:2014qla}, such operators are uncharged under $\CN=2$ flavor symmetry. Therefore, using the results of \cite{Green:2010da} (see \cite{buicanKITP} for a discussion), such deformations in the $\CN=2$ prepotential are, in addition to being marginal, exactly marginal (these operators cannot pair up with other operators to become long multiplets).} In terms of the singularity data, this condition amounts to checking that there is no deformation 
\begin{equation}\label{EM}
\delta W(x_i)=\lambda\prod_{i=1}^{4}x_i^{m_i}\in\mathcal R~,\ \ \ Q=\sum_i q_im_i=1~,
\end{equation}
in the Milnor ring of the singularity
\begin{equation}
\mathcal R:= \mathbb C[x_1,x_2,x_3,x_4]/dW~.
\end{equation}
Note that the number of independent deformations in $\mathcal{R}$ is given by the Milnor number
 \be\label{Milnum}
 \mu=\prod_i (1/q_i-1)~,
 \ee
and that these deformations are encoded in the Poincar\'e polynomial
 \be\label{PoincarePoly}
 P(t) =\sum_Q \dim H_Q\; t^Q =\prod_{i=1}^4 \frac{1-t^{1-q_i}}{1-t^{q_i}}~,
 \ee
where each $t^Q$ represents a non-trivial deformation of the theory with weight $Q$.\footnote{Since the $q_i$ are rational, \eqref{PoincarePoly} is not technically a polynomial. On the other hand, to get a polynomial we can simply replace $t\to t^D$. Given this simple relation, we will abuse terminology and refer to \eqref{PoincarePoly} as the Poincar\'e polynomial.} In particular, the marginal deformations correspond to the term $\dim H_1t\subset P(t)$. Therefore, the condition for the SCFT to be isolated is simply $\dim H_1=0$. 
 
Equivalently, in the language of the 2D $\CN=(2,2)$ SCFT that the LG model of \cite{Cecotti:2010fi} flows to, there should be no exactly marginal deformation.\footnote{It would be interesting to understand if theories with conformal 4D gauge couplings translate into special 2D conformal manifolds.} Indeed, we have that the scaling dimension of the coupling $\lambda$ vanishes
\begin{equation}
[\lambda]={2(1-Q )\over2-\hat c}=0~,
\end{equation}
where the 2D central charge, $\hat c=\sum_{i=1}^4(1-2q_i)$. Therefore, $\lambda$ can be thought of as corresponding to a coordinate on the conformal manifold.

Before making further comments, let us briefly recapitulate. Our goal is to show that when a deformation of the form \eqref{EM} does not exist (i.e., when the theory is isolated as an $\CN=2$ SCFT), we have no $i$ such that $r_i>1$ in \eqref{Ri} and $g_i>0$ in \eqref{Gi}.

Note that the 1-form symmetry of an IHS theory is completely fixed in terms of the weights of the coordinates (i.e., the $q_i$). These weights, in turn, specify the Poincar\'e polynomial of the singularity and, therefore, the allowed deformations. Since these geometrical deformations correspond to $\CN=2$-preserving deformations of the SCFT and since (conformal) gauge groups often give rise to 1-form symmetry, it is natural to expect some connection between 1-form symmetry and exactly marginal deformations. On the other hand, the 1-form symmetry itself only depends on a very rough number-theoretical characterization of the weights. Therefore, very different theories (some isolated and some not) can have the same 1-form symmetry content. Moreover, given a set of weight vectors, it is not apriori obvious that there is a well-defined IHS SCFT realizing them.

Therefore, our basic strategy will be to apply the above formulas, in combination with the classification of IHS theories in \cite{yau2005classification,Xie:2015rpa} (allowing for potential generalizations as in \cite{Davenport:2016ggc,Closset:2020afy} alluded to in the introduction), to prove our main claim. In the next section, we apply this strategy to the $(\mathfrak{g},\mathfrak{g}')$ subset of IHS theories and leave an exhaustive proof for all IHS theories to the appendix.

\subsec{$(\mathfrak{g},\mathfrak{g'})$ SCFTs and comments on type I theories}\label{IHSproof}
In this section we will explicitly focus on the case of the $(\mathfrak{g},\mathfrak{g}')$ SCFTs \cite{Cecotti:2010fi}. Our main reasons for doing so are that these theories are well-studied (as described in the introduction, they include the original AD theories; in addition, \cite{Closset:2020scj,DelZotto:2020esg,Closset:2020afy} explicitly studied 1-form symmetries in these SCFTs), they illustrate some of the main techniques of this paper, and they also provide a neat entry to the classification of \cite{yau2005classification,Xie:2015rpa,Davenport:2016ggc}.

To that end, consider type IIB string theory on a hypersurface singularity defined via 
 \be
 W_{(\mathfrak{g},\mathfrak{g}')}(x,y,u,v)=W_{\mathfrak{g}}(x,y)+W_{\mathfrak{g}'}(u,v)~,
 \ee
where $W_{\mathfrak{g}}$ and $W_{\mathfrak{g}'}$ are given in table~\ref{tableADEdata}.
\begin{table}
 \begin{center}
\begin{tabular} {|c|c|c|c|cc}   \hline
 $\mathfrak{g}$ & $W_{\mathfrak{g}}(x,y)$ & Milnor ring monomial element & $Q_{\mathfrak{g}}$   \\ \hline
 $ A_{p-1}$ & $x^2+y^{p }$  & $y^k  \;(k=0, \cdots p-2)$ & $k/p$  \\ \hline
  $ D_{p+1}$ & $x^2 y+y^{p }$  & $x~,\;y^k  \; (k=0, \cdots p-1)$ & $(p-1)/(2p)~, \; k/p$  \\ \hline
 $ E_6 $&  $x^3 +y^4$  & $x^ky^l  \; ( k=0, 1, \; l=0,1,2)$ & $k/3+ l/4$\\ \hline
 $ E_7 $&  $x^3 +xy^3$  & $x^k y^l  \; ( k,l=0, 1, 2, k+l\le 2 \text{ or } k=2,l=1)$ & $k/3+2 l/9$\\ \hline
 $ E_8 $&  $x^3 +y^5$  & $x^ky^l  \; ( k=0, 1, \; l=0,1,2,3)$ & $k/3+l/5$\\ \hline
 \end{tabular} 
 \caption{Useful data for the $(\mathfrak{g},\mathfrak{g}')$ SCFTs.}
\label{tableADEdata}
\end{center}
\end{table}
Here $W_{\mathfrak{g}}$ is chosen so that $w^2+W_{\mathfrak{g}}$ is the $\mathfrak{g}$-type Du Val singularity. 

Let us go through these theories in turn. First we impose absence of exactly marginal couplings as in the discussion around \eqref{EM}. Then, we check that the 1-form symmetry is indeed trivial.

Let us first consider the $(A_{p-1}, A_{q-1})$ SCFTs. Although the 1-form symmetry here is known to be trivial \cite{Closset:2020scj,DelZotto:2020esg,Closset:2020afy}, these simple theories form a good starting point. A generic deformation in such a theory comes from taking a product of entries in the first row of table \ref{tableADEdata} (one for $A_{p-1}$ and another for $A_{q-1}$). The corresponding weight for this term takes the form
\be
Q=\frac{k}{p}+\frac{l}{q}, \qquad  k=0, \cdots, p-2, \quad   l=0, \cdots, q-2~.
\ee
To find a solution of $Q=1$, and therefore an exactly marginal deformation, we need to solve
\be\label{QAA}
1=\frac{k}{p}+\frac{l}{q}, \qquad  k=0, \cdots, p-2~, \quad   l=0, \cdots, q-2~.
\ee
Clearly, $k,l$ cannot both be zero.  Moreover, when $\gcd(p,q)=1$, this equation has no solution (as can be seen by multiplying both sides by a factor of $p$ or $q$ and using the fact that $p/q, q/p$ are never integer if  $p,q$ are co-prime\footnote{More precisely, $pl/q, qk/p\not\in\mathbb{Z}$. For these quantities to be integers, we need to have $q|pl $, which is equivalent to $\gcd(q,pl)=q$. However, $\gcd(pl,q)=\gcd(l,q)<q$ as $l<q$ and $p,q$ are coprime.}).

Therefore, theories with $\gcd(p,q)=1$ are isolated $\CN=2$ SCFTs. Are they the only isolated $(A_{p-1},A_{q-1})$ SCFTs? To see this is indeed true, we can proceed by considering $s=\gcd(p,q)>1$ and defining $p=sp', q=sq'$.  Then we have
\be
s=\frac{k}{p'}+\frac{l}{q'}~, \qquad  k=0, \cdots p-2~, \quad   l=0, \cdots q-2~.
\ee
Since $\gcd(p',q')=1$, the only possible solution is given by 
\be\label{spq}
k=ap' =ap/s~, \ \ \  l =b q' =bq/s~, \ \ \ \text{s.t.}\ \ \   \frac{a}{s} +\frac{b}{s}=\frac{k}{p}+\frac{l}{q}=1~.
\ee
However, we also need to take into account that $k,l$ are in the restricted set of integers $k=0, \cdots,\ p-2$ and $l=0, \cdots,\ q-2$. As a result, when $p=2$ or $q=2$ or $(p,q)=(3,3)$, there is no solution to \eqref{spq}. Otherwise, one can always find a solution to \eqref{spq}. 

To summarize, the isolated SCFTs of type  $(A_{p-1},A_{q-1})$ are given by 
\be\label{ApAqisolated}
\gcd(p,q)=1~,\  \text{ or }\ p= 2~,  \ \text{ or }\ q= 2~,\   \text{ or }  \ (p,q)= (3,3)~.
\ee
By plugging these results into \eqref{Ri} and \eqref{Gi} it is easy to check that either $r_i=1$ or $g_i=0$ for all $i$ and so there is no 1-form symmetry. Alternatively one can, say, consult table 1 in \cite{DelZotto:2020esg}. Therefore, our main claim is true in these theories (as discussed above, this statement is somewhat trivial given the fact that all $(A_{p-1}, A_{q-1})$ SCFTs have no 1-form symmetry, even if they are not isolated). 

Next let us consider the $(A_{p-1},D_{q+1})$ SCFTs. This time, from table \ref{tableADEdata}, we see that the $Q$'s are given by 
\beqn
Q&=&\frac k p+\frac l q ~, \qquad\nonumber
 \qquad k=0, \cdots, p-2, \quad l=0, \cdots, q-1~,
\\  
Q&=&\frac k p+\frac {q-1}{2 q} ~,
 \qquad k=0, \cdots, p-2   \label{AD2ndQ}
\eeqn
Let us consider the first case (which arises from terms independent of $u$). The equation we want to study is 
\be
1=\frac k p+\frac l q~, \qquad
 \qquad k=0, \cdots~, p-2~, \quad l=0~, \cdots, q-1~.  
\ee
This equation is almost identical to \eqref{QAA}, except for the range of $l$. Therefore, it has no solution if and only if (note we consider $q\ge 3$ here) 
\be
\gcd(p,q)=1~,  \text{ or }\ p= 2~.
\ee
This is a necessary condition for having an isolated SCFT. To find a sufficient condition, we must also consider the second equation in \eqref{AD2ndQ} with $Q=1$, namely 
\be
\frac {2k}{ p}=\frac {q+1}{  q}~, \qquad
  k=0, \cdots, p-2~,
\ee
which arises from $u$-dependent deformations. For $\gcd(p,q)=1$ this equation obviously has no solution.  Moreover, for $p=2$, it also has no solution because $(q+1)/q$ is not an integer for $q\ge 3$.

Therefore we conclude that the isolated SCFTs of type  $(A_{p-1},D_{q+1})$ are given by (see also the recent results in \cite{Carta:2021whq})
\be
\gcd(p,q)=1,  \text{ or }\qquad p= 2~.  
\ee
Using \eqref{Ri} and \eqref{Gi} it is again easy to check that these isolated theories have no 1-form symmetry. Alternatively, one can again consult table 1 of \cite{DelZotto:2020esg} to verify this statement. This result is somewhat more non-trivial, since there are non-isolated $(A_{p-1},D_{q+1})$ SCFTs with 1-form symmetry (see \cite{Closset:2020scj,DelZotto:2020esg,Closset:2020afy}). Our discussion suggests this 1-form symmetry might arise from a conformal gauge group.

Let us now move on to the $(D_{p+1},D_{q+1})$ SCFTs. From table \ref{tableADEdata}, the set of $Q$'s are given by 
\beqn\label{QDD}
Q&=&\frac k p+\frac l q~, \qquad
 \qquad k=0, \cdots, p-1, \quad l=0, \cdots, q-1  \nonumber
\\  
Q&=&\frac k p+\frac {q-1}{2 q}~, \qquad \nonumber
 \qquad k=0, \cdots, p-1   
 \\  
Q&=&\frac l q+\frac {p-1}{2p}~, \qquad \nonumber
 \qquad l=0, \cdots, q-1   
 \\  
Q&=&\frac {p-1}{2p} +\frac {q-1}{2 q}~.\qquad 
 \eeqn
 Much as before, absence of a solution to setting the first set of $Q$'s to one is equivalent to requiring
 \be
 \gcd(p,q)=1~.
 \ee
 This is a necessary condition for having an isolated SCFT. To see it is also sufficient, note that it immediately implies that none of the remaining three sets of $Q$'s in \eqref{QDD} can equal one.
 
Therefore, we conclude that the isolated SCFTs of type  $(D_{p+1},D_{q+1})$ are given by 
\be
\gcd(p,q)=1~.
\ee
It is again easy to check that \eqref{Ri} and \eqref{Gi} imply that the 1-form symmetry is trivial in these cases. As before, one may also verify this statement by consulting table 1 of \cite{DelZotto:2020esg}. Note that non-isolated $(D_{p+1},D_{q+1})$ theories can, in general, have non-trivial 1-form symmetry. 

Let us move on to the exceptional cases. In particular, consider $(A_{p-1},E_6)$. In this case, table \ref{tableADEdata} implies that
\be
Q=\frac kp +\frac l 3 +\frac h 4~, \qquad k=0~, \cdots~, p-2~, \qquad l=0, 1~, \qquad h=0,1,2~.
\ee
 Setting $Q=1$ is equivalent to
\be
0=1-k/p~, \qquad 2/3-k/p~, \qquad 3/4-k/p~, \qquad 1/2-k/p~, \qquad  5/12-k/p~, \qquad 1/6-k/p~.
\ee
Obviously, when $2$ and $3$ are coprime to $p$, the above equation has no solution. But we also need to take into account the range of $k$. It is not difficult to find that a sufficient and necessary condition for isolated $(A_{p-1},E_6)$ SCFTs is
\be
\gcd(2,p)=\gcd(3,p)=1~,\    \text { or } p=2~,\    \text { or } p=3~.
\ee
Equivalently, this condition can be written as 
 \be
 p=2~,  \text { or }   p=3~,   \text { or }  p=6k\pm 1~, \quad k=1,2,3, \cdots~.
 \ee
 Either by consulting table 1 of \cite{DelZotto:2020esg} or explicitly checking \eqref{Ri} and \eqref{Gi}, it is easy to see that these isolated SCFTs have no 1-form symmetry. Again, non-isolated theories in this class can have 1-form symmetry.

Next we check $(A_{p-1}, E_7)$. From table \ref{tableADEdata} we read off the monomial weights
\be
Q=\frac kp +\frac l 3 + \frac{2h}{9}~, \qquad k=0~, \cdots p-2~, \qquad  h,l=0,1,2~, \quad l+h\le 2 \text{ or } l=2,h=1~.
\ee
 An analysis similar to the one in the previous case leads to the following necessary and sufficient condition for isolated SCFTs:
\be
\gcd(3,p)=1~.
\ee
 It is again easy to check that \eqref{Ri} and \eqref{Gi} imply that the 1-form symmetry is trivial in these cases. As before, one may also verify this statement by consulting table 1 of \cite{DelZotto:2020esg}. Note that non-isolated $(A_{p-1},E_7)$ theories can, in general, have non-trivial 1-form symmetry. 
 
Let us move on to $(A_{p-1}, E_8)$. From table \ref{tableADEdata} we read off the monomial weights
\be
Q=\frac kp +\frac l 3 +\frac h 5~, \qquad k=0~, \cdots p-2~, \qquad l=0, 1~, \qquad h=0,1,2,3~.
\ee
Analogously to the $(A_{p-1},E_6)$ case, the isolated SCFTs in this class are given by 
\be
\gcd(3,p)=\gcd(5,p)=1~,\ \text { or }\ p=3~.
\ee
Yet again we can use \eqref{Ri} and \eqref{Gi} to check that these theories have no 1-form symmetry. Equivalently, we may verify this statement by consulting table 1 of \cite{DelZotto:2020esg}. Note that non-isolated $(A_{p-1},E_8)$ theories can, in general, have non-trivial 1-form symmetry.  

Next we discuss the $(D_{p+1},E_6)$ SCFTs. From table \ref{tableADEdata} we need to study solutions of
 \be
1= Q =  \frac{k}{p} +\frac{l}{3}+\frac{h}{4}~, \qquad k=0~, \cdots p-1~, \frac{p-1}{2}~, \quad l=0,1~, \quad h=0,1,2~,
\ee
where $k=(p-1)/2$ is for the case in which $x$ appears in the deformation. To be isolated, we   have the necessary and sufficient  condition
 \be
 \gcd(2,p)=1~, \qquad \gcd(3,p)=1~,
\ee
where the $p=2$ or $p=3$ option in the $(A_{p-1},E_6)$ case no longer exists here due to the larger range of $k$. Again, we an check using either \eqref{Ri} and \eqref{Gi} or table 1 of \cite{DelZotto:2020esg} that such theories have no 1-form symmetries. This is again in contrast to the general exactly marginal case where such 1-form symmetries are allowed.

Now we move on to the $(D_{p+1},E_7)$ SCFTs. From table \ref{tableADEdata} we need to study solutions of
   \be
1=Q=\frac kp +\frac l 3 + \frac{2h}{9}~, \ \ k=0, \cdots, p-1~, \frac{p-1}{2}~, \ \ l,h=0,1,2~, \ \ l+h\le 2 \text{ or } l=2~,h=1~.
 \ee
Similarly to the $(A_{p-1},E_7)$ case, the condition of being isolated is given by
 \be
  \qquad \gcd(3,p)=1~.
 \ee
As in the previous examples, using either \eqref{Ri} and \eqref{Gi} or table 1 of \cite{DelZotto:2020esg} shows that such theories have no 1-form symmetries. This is again in contrast to the general exactly marginal case where such 1-form symmetries are allowed.

Let us now discuss the $(D_{p+1},E_8)$ SCFTs. From table \ref{tableADEdata}, we need to study solutions of
   \be
1= Q =  \frac{k}{p}
 +\frac{l}{3}+\frac{h}{5}~, \qquad k=0, \cdots p-1, \frac{p-1}{2}~, \quad l=0,1~, \quad h=0,1,2,3~.
 \ee
 To be isolated, a sufficient and necessary condition is
 \be
\gcd(3,p)=\gcd(5,p)=1~.
 \ee
The $p=3$ option in the $(A_{p-1}, E_8)$ case is absent here due to the larger range of $k$. As in the previous examples, using either \eqref{Ri} and \eqref{Gi} or table 1 of \cite{DelZotto:2020esg} shows that such theories have no 1-form symmetries. This is again in contrast to the general exactly marginal case where such 1-form symmetries are allowed.

Finally, we come to the $(E_n,E_m)$ cases.  For these 6 types, we note that 
   \be
\left\{\frac13,  \frac12\right\} \subset Q_{E_6}, \qquad \left\{\frac13,  \frac23\right\} \subset Q_{E_7},  \qquad \left\{\frac13,  \frac25, {3\over5}\right\} \subset Q_{E_8}, 
   \ee
 Since $Q_{(E_n,E_m)} =Q_{E_n}+Q_{E_m} $, obviously $(E_7, E_{6,7,8}), (E_6,E_6)$, and $(E_8,E_8)$ are not isolated. On the other hand, one can verify that the isolated set is given by 
   \be
   (n,m)=(6,8)\simeq(8,6)~. 
   \ee
From \eqref{Ri} and \eqref{Gi} or table 1 of \cite{DelZotto:2020esg}, we see that such theories have no 1-form symmetries. In fact, since the $(E_n, E_m)$ theories more generally cannot have 1-form symmetry, this result is somewhat trivial.

Therefore, we see that all isolated (as $\CN=2$ SCFTs) $(\mathfrak{g},\mathfrak{g}')$ theories have no 1-form symmetry. This establishes our main claim in this subset of the theories. In the next section, we explain how this statement fits into a similar result for type I IHS theories in the nomenclature of \cite{Xie:2015rpa} (or  the related $S_{1,1}^{\times 4}$ LG theories in \cite{Davenport:2016ggc}). We also discuss certain facts about 1-form symmetries in a subclass of type I IHS theories with exactly marginal deformations and relate these results to ones discussed in \cite{DelZotto:2015rca,Buican:2020moo,Closset:2020scj,Closset:2020afy}. These conformal manifolds all have known gauge coupling interpretations, which gives us an opportunity to discuss certain global properties of the corresponding matter sectors (this type of analysis will be relevant when discussing a particular $\CN=3$ theory in section \ref{gencomments}). In the appendix we consider the remaining IHS theories discussed in \cite{Xie:2015rpa,Davenport:2016ggc,Closset:2020afy} and complete the general proof of our main claim.

\subsubsec{More general type I theories}
In this section, we briefly discuss how some of the $(\mathfrak{g},\mathfrak{g}')$ theories fit into the classification scheme of \cite{yau2005classification,Xie:2015rpa,Davenport:2016ggc,Closset:2020afy} and explain phenomena associated with some of the non-isolated theories that will be useful later. To that end, note that the $(A_{p-1}, A_{q-1})$, $(D_4, A_{k-1})$, $(E_6, A_{k-1})$, and $(E_8, A_{k-1})$ SCFTs can all be written in the following form
\be\label{t1}
W(x,y,u,v)=x^a + y^b + u^c + v^d~,
\ee
with $a$, $b$, $c$, and $d$ as in table \ref{Iinf0}.\footnote{For the case of $(D_4, A_{k-1})$, we use a change of variables to bring it to this form.}
\begin{table}
\begin{center}
\begin{tabular}{|C|c| }
\hline
 {(a,b,c,d)} & other name \\ \hline
(2,2,p,q)&$(A_{p-1}, A_{q-1})$ \\ \hline
(2,3,3,k)&$(D_4, A_{k-1})$ \\ \hline
(2,3,4,k)&$(E_6, A_{k-1})$ \\ \hline
(2,3,5,k)& $(E_8,A_{k-1})$ \\ \hline
(3,3,3,k)& $(3,k)$ \\ \hline
(2,4,4,k)& $(4,k )$ \\ \hline
(2,3,6,k)& $(6,k)$ \\
\hline
\end{tabular}
\caption{The various infinite sequences of type I singularities. Here the entries of the four-component vector in the left-hand column represent, respectively, the different powers of $x$, $y$, $u$, and $v$ in \eqref{t1}. Each set of solutions leads to an isolated singularity at a finite distance in moduli space \cite{yau2005classification,Xie:2015rpa}. The first four entries are of the $(\mathfrak{g},\mathfrak{g}')$ type indicated in the second column. The $(3,k)$, $(4,k)$, and $(6,k)$ theories are studied in \cite{DelZotto:2015rca,Buican:2020moo,Closset:2020scj,Closset:2020afy} (we use the notation of \cite{Buican:2020moo}). There is some overlap between the different theories (e.g., $(D_4, A_2)\simeq(3,2)$). There are also a finite set of additional sporadic type I singularities and corresponding theories. This table is adapted from  \cite{Xie:2015rpa}.}
\label{Iinf0}
\end{center}
\end{table}

More generally, \cite{Xie:2015rpa} classified so-called type I theories that come from singularities of the form in \eqref{t1}. Requiring that the singularity is at finite distance in moduli space yields the seven infinite sequences of singularities in table \ref{Iinf0} in addition to a finite collection of sporadic solutions. We have checked that the sporadic solutions satisfy our main claim.

The three remaining classes of solutions are studied in \cite{DelZotto:2015rca,Buican:2020moo,Closset:2020scj,Closset:2020afy} and are called the $(3,k)$, $(4,k)$ and $(6,k)$ theories (here we use the nomenclature of \cite{Buican:2020moo}). These theories all have exactly marginal gauge couplings and so our main claim does not apply to them. Still it is interesting to investigate some of these theories further for reasons that will become apparent in section \ref{gencomments}.

In particular, the $(3,k)$ theories consist of an exactly marginal diagonal $su(k)$ gauging of a collection of three $D_3(SU(k))$ theories (these latter theories are defined in \cite{Cecotti:2012jx,Cecotti:2013lda}). For ${\rm gcd}(3,k)=1$, the $D_3(SU(k))$ matter sectors are isolated $\CN=2$ theories. The resulting IHSs take the form
\begin{equation}
W(x,y,u,v)=x^3+y^3+u^3+v^k~.
\end{equation}
From these equations, it is easy to check that the finite part of \eqref{D1def} is $\mathbb{Z}_k^2\subset\mathbb{D}^{(1)}$ in the case of ${\rm gcd}(3,k)=1$. In particular, the 1-form symmetry can be fixed by picking a maximal isotropic sub-group.

That this result is exactly the same as for $su(k)$ $\CN=4$ super Yang-Mills (SYM) does not seem to be a coincidence. For example, let us take $k=2$. The $(3,2)$ theory has been studied in detail in \cite{Buican:2020moo}. There it was shown that a subset of the local operator content (i.e., the Schur sector) is in one-to-one correspondence (via a special map) with the Schur sector of $su(2)$ $\CN=4$ SYM.

Moreover, just as the adjoint hypermultiplet matter of $SU(2)$ $\CN=4$ SYM ensures that the $\mathbb{Z}_2$ one-form symmetry is unbroken, we can argue the same is true for the $D_3(SU(2))$ matter in the $(3,2)$ SCFT with gauge group $SU(2)$. In particular, we claim that none of the local operators (even unprotected operators in long multiplets) of the $D_3(SU(2))$ theory transform under the center of $SU(2)$. Therefore, the flavor symmetry of this theory is    $SO(3)$ (see \cite{Bhardwaj:2021ojs,Apruzzi:2021vcu} for related recent discussions in other classes of theories from a different perspective).\footnote{This statement is compatible with the 2D chiral algebra related to the theory via the general correspondence in \cite{Beem:2013sza}: it is generated by the $su(2)$ moment map \cite{Buican:2015ina,Buican:2015tda}. This statement is also consistent with the dimensional reduction: the 3D theory is $U(1)$ with $N_f=2$ and has a manifest $SO(3)$ global symmetry \cite{Xie:2012hs,Buican:2015hsa}.}

To understand this statement, let us define $D_3(SU(2))$ in the way it was first constructed in \cite{Argyres:1995xn}: as the maximally singular point on the Coulomb branch of $SU(2)$ $\CN=2$ SQCD with $N_f=2$. In this case, the UV matter hypermultiplets can be written as $Q_i^a\oplus\tilde Q^j_b$, where $i,j=1,2$ are color indices and $a,b=1,2$ are $su(2)$ flavor indices. The remaining SQCD degrees of freedom (in the $\CN=2$ vector multiplet) do not transform under the flavor symmetry. In terms of these variables, the IR flavor symmetry of the $D_3(SU(2))$ theory is already manifest in the UV. Now, it is easy to show that there are no UV operators transforming in odd-spin representations of the flavor $su(2)$. Indeed, all gauge-invariant operators transforming under flavor must have an even number of hypermultiplet fields. Since this flavor symmetry is not anomalous, we see that the IR $D_3(SU(2))$ theory has no local operators tranforming in the center of $SU(2)$ (i.e., the flavor symmetry group is $SO(3)$). In particular, this logic implies that if we choose an $SU(2)$ gauge group for the $(3,2)$ theory, the $\mathbb{Z}_2$ 1-form symmetry corresponding to the center is preserved.

We will return to a similar discussion in a different non-Lagrangian theory in section \ref{gencomments}. In any case, we see that (modulo explicit checks we have done in the case of sporadic theories and have not discussed here), our claim 1 in the introduction is true for type I IHS theories. In the appendix, we deal with the remaining cases and complete the proof of our main claim. In the next section we introduce some broader results that apply to certain theories we have just discussed and are also helpful in the proof of our main claim for more general theories appearing in the appendix. 

\newsec{Useful results beyond $(\mathfrak{g},\mathfrak{g}')$ and type I}\label{LGmass}
Recall from the introduction that our 4D $\CN=2$ SCFTs are related to 2D LG models via the correspondence in \cite{Cecotti:2010fi}. In particular, these latter theories are described by 2D chiral superfields $x$, $y$, $u$, and $v$ with superpotential given by the $W$ describing our IHS. 

From this perspective, it is intuitively clear that if $W\supset X_iX_j$, where $X_{i,j}\in\left\{x,y,u,v\right\}$ are distinct and satisfy $q_i+q_j=1$, the theory and symmetry content should simplify (e.g., we expect a simpler Seiberg-Witten description). Indeed, from the 2D perspective we can then integrate out two of the four fields describing our theory without changing any of the scaling dimensions of operators (similar comments apply if $W$ admits a deformation $\delta W=\lambda X_iX_j$ preserving the $U(1)_R$ symmetry).

It turns out this intuition is correct, in the sense that the 1-form symmetry content of such a theory is trivial:

\bigskip
\noindent
{\bf Lemma: \label{lemma1}} \emph{If the sum of any two weights equals one, then the 1-form symmetry is trivial.}
 
\medskip
\noindent
\emph{Proof:} Without loss of generality, we can assume that
 \be
 \vec q=\left( \frac{V_1}{U_1}, \;  \frac{V_2}{U_2 },\frac{V_3}{U_3 }, \; \frac{U_3- V_3}{U_3} \right)~.
 \ee
 This weight vector is irreducible as long as $\gcd(V_i, U_i)=1$. In particular, the last fractional number is irreducible.

 Applying the formula \eqref{Ri}, \eqref{Gi}  directly, one finds 
 \be
 r_3=r_4=1, \quad 2g_1=-1+\frac{1}{V_2}, \quad 2g_2=-1+\frac{1}{V_1}~.
 \ee
 Obviously $g_1,g_2\le 0$. This statement implies that the 1-form symmetry is always trivial. $\square$

\bigskip
This claim is independent of any IHS SCFT classification. Of course, not all IHS theories satisfy the conditions of the lemma.\footnote{Examples of theories satisfying this lemma include the $(A_{p-1},A_{q-1})$ SCFTs. In particular, the lemma gives a simple proof that these theories have trivial 1-form symmetry (even when there is an exactly marginal deformation).} However, we will see that in large classes of seemingly complicated theories considered in the appendix, SCFTs that do not satisfy this lemma also have an exactly marginal deformation. This fact will lead to many simplifications.

More generally, any IHS theory either satisfies the conditions of the lemma or else admits a relevant mass deformation of the form $\delta W = mX_iX_j$. Indeed, suppose $q_{X_i}+q_{X_j}>1$. Without loss of generality, we can assume $q_{X_i}>1/2$. Then, from the argument of lemma 2.5 of \cite{Davenport:2016ggc}, there must be a field $X_k$ with $q_{X_i}+q_{X_k}=1$ (otherwise $W$ will not describe an isolated singularity). On the other hand, if $q_{X_i}+q_{X_j}<1$, the deformation $\delta W=mX_iX_j$ is relevant and initiates an RG flow. The end point of this RG flow will be a $W$ corresponding to an SCFT with trivial 1-form symmetry (although there will typically be other sources of 1-form symmetry in the IR; for example, we will generically have various decoupled $U(1)$ factors).

Another simple consequence of the above lemma is the following claim that will be useful later and is entirely independent of any classification of IHS theories:

\bigskip
\noindent
{\bf Claim 2:} \emph{Any IHS theory with at most two different weights has trivial 1-form symmetry or else admits a conformal manifold.}

\medskip
\noindent
\emph{Proof:} See appendix \ref{twoq}. $\square$

\newsec{Comments on 1-form symmetries in more general isolated $\CN\ge2$ SCFTs}\label{gencomments}
In this section, we briefly discuss 1-form symmetries in more general isolated 4D $\CN\ge2$ SCFTs. We leave a more detailed accounting for later work. 

Let us start with the maximal amount of SUSY: $\CN=4$ SCFTs. These theories may or may not have 1-form symmetry (e.g., $g_2$ $\CN=4$ SYM does not have 1-form symmetry, but $su(N)$ does). However, if we assume locality, then $\CN=4$ SCFTs are never isolated, since the energy-momentum tensor is in the same multiplet with an exactly marginal deformation. As a result, our claim has nothing to say about local $\CN=4$ theories.\footnote{We expect non-local $\CN=4$ theories to be quite constrained in their structure. For example, it seems unlikely that we can engineer them as boundaries of 5D QFTs, since Nahm's classification \cite{Nahm:1977tg} forbids bulk SCFTs with sufficiently many supercharges. Similar comments apply if we try to engineer non-local $\CN=4$ theories on defects in $D\ge5$ dimensional QFTs.} 

Next let us discuss $\CN=3$ theories. On general grounds, these theories are isolated as $\CN\ge2$ SCFTs \cite{Aharony:2015oyb}. However, they may be non-isolated as $\CN=1$ SCFTs.

In fact, from the $\CN=1$ Lagrangian conformal manifold construction in \cite{Zafrir:2020epd}, one can argue that a $c=5/4$ rank-one $\CN=3$ theory (related to the $G(3,1,1)$ complex reflection group in the language of \cite{Aharony:2016kai}) has trivial one-form symmetry.

On the other hand, the proposed $\CN=1$ conformal manifold for the $\CN=3$ rank-three theory related to the $G(3,3,3)$ complex reflection group has $\mathbb{Z}_3$ one-form symmetry \cite{Zafrir:2020epd}. This statement follows from the fact that the $\CN=1$ theory involves gauging a diagonal $SU(3)^3\subset E_6$ flavor subgroup of the $E_6$ MN theory along with some additional chiral superfields that are manifestly invariant under the diagonal $\mathbb{Z}_3$ center. To check whether the strongly interacting $E_6$ MN sector is invariant under this diagonal center is not too difficult. Indeed, since the fundamental of $E_6$ decomposes as ${\bf 27}=({\bf3},{\bf\bar 3},{\bf1})\oplus({\bf\bar 3},{\bf1},{\bf3})\oplus({\bf1},{\bf3},{\bf\bar 3})$, no local operator in the $E_6$ MN theory is charged under the diagonal $\mathbb{Z}_3$ center (this agrees also with discussions in \cite{Tachikawa:2013hya,Bhardwaj:2021ojs}).\footnote{We thank G.~Zafrir for this argument.} This conclusion on the local operator spectrum is reminiscent of our discussion for the $D_3(SU(2))$ matter sector of the $(3,2)$ SCFT in the previous section (although the argument here is purely group theoretical while our argument in the previous section involved invoking an RG flow).

As a result, if the $G(3,3,3)$ SCFT indeed lies on this $\CN=1$ conformal manifold, it would likely be an example of an $\CN=3$ theory with a non-trivial one-form symmetry. Clearly, it is worth answering this question definitively (perhaps by a computation in the associated chiral algebra).

Finally, it might be that some of the $\CN=3$ theories that can be engineered by gauging certain discrete $su(N)$ $\CN=4$ SYM 0-form symmetries may potentially have non-trivial 1-form symmetry (whether such 1-form symmetry is inherited or not would clearly be interesting to check). If these theories possess 1-form symmetry, then one might be able to argue that isolated $\CN\ge2$ theories have 1-form symmetry only if they have a gauge theory origin (i.e., via flowing from a gauge theory and/or discrete gauging of global symmetries of a gauge theory). Of course, this seems like a very expansive class of theories.

Next let us focus on other constructions of $\CN=2$ theories. In particular, let us briefly discuss how some of the class $\CS$ results of \cite{Tachikawa:2013hya,Bhardwaj:2021pfz} fit in with our discussion. Of the theories we have checked in these references, all have trivial one-form symmetry when they are isolated. For example, consider (4.6) of \cite{Bhardwaj:2021pfz}, which describes the abelian group of line operators, $\CL$, (including mutually non-local lines) of the $(2,0)$ theory compactified on a genus $g$ surface with $n$ regular twisted punctures
\begin{equation}
\CL\simeq\mathbb{Z}_2^g\times\mathbb{Z}_2^g~.
\end{equation}
Since $\CL$ is trivial for $g=0$, we require $g>0$ in order to have 1-form symmetry. Having non-trivial genus leads in turn to a theory with an exactly marginal gauge coupling.

Somewhat more generally, consider (4.21) in \cite{Bhardwaj:2021pfz}
\begin{equation}
\CL\simeq\prod_{i=1}^{k-1}\CL_{i,i+1}^A\times\prod_{i=1}^{k-1}\CL_i^B\times\widehat{Z}_A^g\times\widehat{Z}_B^g~.
\end{equation}
This expression corresponds to $\CL$ for the case of the $(2,0)$ theory compactified on a Riemann surface of genus $g$ with $n$ regular twisted punctures and $2k$ $\mathbb{Z}_2$-twisted regular punctures. In order for $\CL$ to be non-trivial, we require that $g>1$ or $k>1$. If $g>1$, then the theory is not an isolated $\CN=2$ SCFT. On the other hand, if $g=0$ and $k>1$, then the theory again has an exactly marginal gauge coupling. 
Therefore, these theories seem to be compatible with a generalization of our claim on isolated IHS $\CN=2$ SCFTs.

Finally, we come to a set of theories that flow to the $(A_{k-1},G)$ theories we have studied in section \ref{IHSproof}. In particular, there are well-known RG flows of the following type \cite{Maruyoshi:2016tqk,Maruyoshi:2016aim,Agarwal:2016pjo,Giacomelli:2017ckh}\footnote{Here we drop decoupled matter fields.} 
\begin{equation}\label{RG1}
D_k^{h^{\vee}}(G)\simeq D_k(G)\to (A_{k-1},G)\simeq G^{h^{\vee}}(k)~,
\end{equation} 
where $D_k(G)$ are 4D $\CN=2$ SCFTs with (at least) $G\in\left\{A_n,D_n,E_n\right\}$ global symmetry defined in \cite{Cecotti:2012jx,Cecotti:2013lda}, $h^{\vee}$ is the dual Coxeter number, and $D_k^{h^{\vee}}(G)$, $G^{h^{\vee}}(k)$ are alternate names for the UV and IR theories that will connect with further generalizations we will discuss.

As emphasized in \cite{Wang:2015mra,Giacomelli:2017ckh}, the UV $D_k(G)$ theory has a type IIB string theory realization in terms of the same IHS singularity as the IR $(A_{k-1},G)$ SCFT provided we embed the former in $\mathbb{C}^{*}\times\mathbb{C}^3$
\begin{eqnarray}
W_{G^{h^{\vee}}(k)}(x,y,u,v)&:=&W_{A_{k-1}}(x,y)+W_{G}(u,v)~, \ \ \ G\in\left\{A_{n},D_n,E_n\right\}~,\cr W_{D_k^{h^{\vee}}(G)}(x,t,u,v)&:=&W_{A_{k-1}}(x,t)+W_{G}(u,v)~, \ \ \ t=e^{y}\in\mathbb{C}^{*}~,
\end{eqnarray}
where $t\ne0$ and $x,y,u,v\in\mathbb C$. In terms of these variables, we have
\begin{equation}\label{Omega1}
\Omega_{G^{h^{\vee}}(k)}={dx\wedge dy\wedge du\wedge dv\over dW}~,\ \ \ \Omega_{D_k^{h^{\vee}}(G)}={dx\wedge dt\wedge du\wedge dv\over tdW}~.
\end{equation}

It is natural to ask if the UV theories in \eqref{RG1} obey our main claim as well. We will shortly see that the answer is yes, provided we can identify the UV and IR 1-form symmetries.\footnote{Note that under the RG flow \eqref{RG1}, the UV and IR ranks are the same, so there is no possibility of decoupled free $U(1)$'s and accidental continuous IR 1-form symmetry. This statement means that (up to caveats we omit) one can reasonably guess that the UV and IR 1-form symmetries match. This guess is confirmed by the BPS quiver computations in \cite{Hosseini:2021ged} for all the many cases that were checked. \label{noAcc}} The reason is that the UV and IR Milnor rings are closely related:
\begin{eqnarray}
\mathcal{R}_{G^{h^{\vee}}(k)}&:=&\mathbb{C}[x,y,u,v]/\langle{\partial_xW,\partial_yW,\partial_uW,\partial_vW}\rangle~,\cr\mathcal{R}_{D_k^{h^{\vee}}(G)}&:=&\mathbb{C}[x,t,u,v]/\langle{\partial_xW,t\partial_tW,\partial_uW,\partial_vW}\rangle~.
\end{eqnarray}
In particular, we see that the UV Milnor ring is larger than the IR Milnor ring since, in essence, we take $\partial_yW\to t\partial_tW$. Said differently, monomials involving $t^{k-1}$ are non-trivial in the UV, while monomials involving $y^{k-1}$ are trivial in the IR. Hence, we can canonically identify the IR ring with the UV subring involving monomials with $t^{\beta}$ and $\beta<k-1$.\footnote{As a concrete example, consider the Maruyoshi-Song flow (see also \cite{Benvenuti:2017lle})  $D_2^4(SU(4))\to A_3^4(2)$ (where the UV is $\CN=2$ $SU(2)$ SQCD with $N_f=4$, and the IR is the $(A_1, A_3)$ SCFT). The UV theory admits six deformations (four related to mass parameters, one related to the dimension two Coulomb branch vev, and one related to the gauge coupling) while the IR theory admits three deformations (one related to the $SU(2)$ mass parameter, one related to the dimension $4/3$ vev, and one related to the corresponding dimension $2/3$ coupling). The UV deformations $\lambda_0$, $\lambda_1v$, and $\lambda_2v^2$ can be identified with the three IR deformations (though they are related to UV mass parameters!), while $\tilde\lambda_0t$, $\tilde\lambda_1vt$, and $\tilde\lambda_2v^2t$ are set to zero in the IR Milnor ring (after substituting $t\to y$).\label{example}}

While the scaling dimensions of the UV coordinates differ from those of the IR (this fact follows from demanding that the different $\Omega$ in \eqref{Omega1} have scaling dimension one), the weight vectors, $\vec{q}$, are the same. Therefore, we see that if the UV theory lacks a marginal deformation (i.e., a parameter, $\lambda$, with weight zero), then so too does the IR theory.\footnote{Note that the converse does not hold. Indeed, see the example in footnote \ref{example}. The issue is that UV deformations may trivialize in the IR.} Assuming we can equate the UV and IR 1-form symmetry (see footnote \ref{noAcc} for evidence in favor of this hypothesis for general $D_k(G)$ and conclusive proof in certain cases), we see that our UV theories can have 1-form symmetry only if they have conformal manifolds. Therefore, we strongly suspect that the $D_k(G)$ theories satisfy our main claim as well, even though they are related to singularities embedded in a different ambient space.\footnote{Note that some $D_p(G)$ theories can be realized via (different) IHSs embedded in either ambient space. For example, $D_3(SU(2))$ also has a realization as $(A_1, A_3)\simeq A_3^{4}(2)$. This subset of $D_p(G)$ theories is, of course, directly subject to our main claim.}

More generally, we have flows of the form \cite{Giacomelli:2017ckh}
\begin{equation}\label{RG2}
D_k^{b}(G)\to G^{b}(k)~,
\end{equation} 
where we can have $b\ne h^{\vee}$, and $G^b(k)$ is in our class of SCFTs. We may use similar logic to that used for $b=h^{\vee}$ to conclude that, if we can equate the UV and IR 1-form symmetries (as assumed in \cite{Hosseini:2021ged} and as hinted at by the agreement of the UV and IR ranks), the $D_k^b(G)$ theories have 1-form symmetry only if they are part of a conformal manifold.

Furthermore, if one can identify 1-form symmetry in certain more general flows between theories related to IHSs embedded in $\mathbb{C}^{*}\times\mathbb{C}^3$ and those related to IHSs embedded in $\mathbb{C}^4$, then we expect the above comments to generalize.

\newsec{Some rigorous bounds and a (less rigorous) conjecture}\label{conjectures}
In this section, we first establish a simple but useful claim showing that theories with $\sum_iq_i>3/2$ are isolated SCFTs. From this result and claim 1, it follows that theories with $\sum_iq_i>3/2$ have no 1-form symmetry. Using these results we suggest a few conjectures for more general 4D $\CN=2$ SCFTs.

\bigskip
\noindent
{\bf Claim 3:} \emph{IHS theories with $\sum_iq_i>3/2$ are isolated 4D $\CN=2$ SCFTs.}\footnote{Again, we allow for the possibility that they are special points on an $\CN=1$ conformal manifold.}

\medskip
\noindent
\emph{Proof:} We give two proofs of this statement. The first follows from unitarity bounds in the (possibly trivial!) $(2,2)$ 2D SCFT that the LG model with superpotential $W$ flows to.\footnote{We thank Z.~Komargodski for bringing these bounds to our attention in an unrelated context.} The second proof follows from an analysis of the Poincar\'e polynomial.

Let us consider the proof by unitarity bounds first. To that end, take an IHS 4D $\CN=2$ SCFT described by $W(x,y,u,v)=0$. The corresponding LG model with chiral superfields $x,y,u,v$ and superpotential $W$ will flow to some (possibly trivial) $(2,2)$ SCFT in the IR.

In this  IR SCFT, unitarity bounds require that $(2,2)$ chiral operators have left scaling dimension (e.g., see \cite{Chen:2017fvl})\footnote{If the IR SCFT is trivial, $c_{2d}=0$, and all chiral operators have $h=0$. This is consistent with the statement that the 2D theory is trivial.}
\begin{equation}
h\le {c_{2d}\over6}~,
\end{equation}
where $c_{2d}$ is the 2D central charge. Using the fact that $c_{2d}=3\hat c=6(2-\sum_iq_i)$ via the correspondence in \cite{Cecotti:2010fi} and the fact that the superconformal $R$-charge for a chiral operator satisfies $Q=2h$, we see that
\begin{equation}
Q\le2\left(2-\sum_iq_i\right)~.
\end{equation}
In particular, if $\sum_iq_i>3/2$, we are done since all chiral operators have $Q<1$ and therefore cannot give rise to exactly marginal deformations.

\smallskip
\noindent
{\it Alternate proof:} We can arrive at the same result by considering the Poincar\'e polynomial introduced in \eqref{PoincarePoly}. In particular, the weights appearing in \eqref{PoincarePoly} are bounded  \cite{Xie:2015rpa}
\be
0 \le  Q\le 4-2\sum_i q_i~.
\ee
Therefore, if $\sum_iq_i>3/2$, we see that $Q<1$, and the SCFT is isolated. $\square$

\bigskip
\noindent
Combining this result with the proof of our main claim (see the appendix), we have that

\bigskip
\noindent
{\bf Corollary 1:} \emph{IHS theories with $\sum_iq_i>3/2$ have no 1-form symmetry. Equivalently, IHS theories with all $\CN=2$ chiral ring generators (i.e., Coulomb branch chiral ring generators) having scaling dimension less than two have trivial 1-form symmetry.}

This corollary follows from our main claim, but we can also prove it directly:

\medskip
\noindent\emph{Direct proof:}
First, we should require all $q_i\le1/2$ in order to have a hope of finding non-trivial 1-form symmetry.\footnote{If there is a $j$ such that $q_j >1/2$, then lemma 2.5 of \cite{Davenport:2016ggc} implies there is a $k$ such that $q_k=1-q_j$. The 1-form symmetry is trivial by the lemma in section \ref{LGmass}.} Then, if there are two or more weights equal to $1/2$, the 1-form symmetry is still trivial because $1/2+1/2=1$.

Without loss of generality, consider the case that $q_4 =1/2$ with all other $q_i<1/2$. Clearly, we should have $q_1+q_2+q_3>1$ in order for $\sum _i q_i>3/2$. As a result, at least one of the weights, say $q_1$, should be bigger than $1/3$ (i.e., $1/2>q_1>1/3$). Lemma 2.8 of \cite{Davenport:2016ggc} then implies that, say, $q_2=1-2q_1$. As a result, $1/2>q_3>1/3$.\footnote{More explicitly, substituting $q_2=1-2q_1$ into $q_1+q_2+q_3>1$ shows that $q_3>q_1$. Therefore, $1/2>q_3>1/3$.} The same argument again implies that there must be an $i$ such that $q_i=1-2q_3<1/3$. Only $q_2<1/3$, but $q_3>q_1$, so this is impossible. Therefore, we need to have all $q_i<1/2$.

If no weight is bigger than $1/3$, namely all $q_i\le1/3$, then $\sum_i q_i\le4/3<3/2$, violating the assumption. 

If there are at least two weights $1/2>q_1,q_2>1/3$, then, without loss of generality, we should have $q_3 =1-2q_1<1/3$ and $q_4=1-2q_2<1/3$ according to  lemma 2.8   of \cite{Davenport:2016ggc}.  Thus we have $\sum_i q_i=2-q_1-q_2<2-2/3<3/2$, again violating the assumption.

Thus we only need to consider the case where there is just one weight   bigger than $1/3$. Let us assume     that $1/2>q_1>1/3$ and $q_{2,3,4}<1/3 $. Then we must have $q_2=1-2q_1$. Thus $q_1+q_2=1-q_1<2/3$. To have $\sum q_i>3/2$, we have, without loss of generality, $q_3+q_4>5/6$. This is in contradiction with the assumption $q_3,q_4<1/3$.  $\square$

\bigskip
\noindent
Assuming the discussion in \cite{Shapere:2008zf} applies to our theories of interest we can rephrase the above corollary in terms of bounds on $a$ and $c$. To understand this statement, we substitute $\sum_iq_i>3/2$ into (4.5) and (4.6) of \cite{Xie:2015xva} to obtain
\begin{equation}\label{acbound2}
c<{1\over6}(3r+f)~,\ \ \ a<{11\over24}r+{4r+2f\over24}={15r+2f\over24}~,
\end{equation}
where $r$ is the complex Coulomb branch dimension of our SCFT (i.e., the rank of the theory), and $f$ is the rank of its continuous 0-form flavor symmetry. 

One may wonder how robust \eqref{acbound2} is. For example, recall that for theories arising from gauging a discrete 0-form symmetry in a parent SCFT, the formalism of \cite{Shapere:2008zf} need not apply \cite{Aharony:2016kai,Argyres:2016yzz}. However, we do not expect discrete gauging to increase $r$ or $f$. Moreover, the central charges are unaffected. Now, if we start from a parent theory that has 1-form symmetry and therefore does not obey \eqref{acbound2}, the daughter theory will not obey \eqref{acbound2} either. As a result, we believe that for theories satisfying \eqref{acbound2}, the one-form symmetry is indeed trivial.\footnote{Since our theories are strongly coupled, it is natural to give a more universal criterion for when the analysis in \cite{Shapere:2008zf} applies. We are agnostic on this point, but one interesting possibility, following the discussion in \cite{Aharony:2016kai}, is that the correct criterion is the absence of 2-form symmetry (and that gauging this 2-form symmetry takes us back to the parent theory where \cite{Shapere:2008zf} applies).}

How general can we make the above statements? One reasonable conjecture is as follows:

\bigskip
\noindent
{\bf Conjecture:} \emph{Interacting 4D $\CN=2$ SCFTs with all $\CN=2$ chiral ring generators (i.e., Coulomb branch chiral ring generators) of scaling dimension less than two have trivial 1-form symmetry.}

\bigskip
\noindent
We have assembled some evidence for this conjecture. Indeed, it is satisfied by theories related to IHSs embedded in $\mathbb{C}^4$ (we have also seen evidence that this statement extends to more general ambient spaces).\footnote{More explicitly, here we have  $[\mathcal O]=2(1-Q)/(2-\hat c)<2$ for Coulomb branch operator $\mathcal  O$ as  $\hat c=4-2\sum_i q_i<1$ and $0<Q\le\hat c<1$.} Moreover, typical class $\CS$ theories (with regular punctures) involve at least some $\CN=2$ chiral ring generators with scaling dimensions $\Delta\ge3$ and therefore satisfy the conjecture trivially. Finally, $\CN=3$ SCFTs (and, in particular, the potentially troublesome $G(3,3,3)$ theory discussed in section \ref{gencomments}) satisfy this conjecture trivially as well. Indeed, all Coulomb branch operators in these theories have $\Delta\ge3$.\footnote{This statement follows from the fact that Coulomb branch operators in $\CN=3$ theories have integer dimension \cite{Nishinaka:2016hbw}. Since these operators cannot have dimension one (this would correspond to a free decoupled vector multiplet) or two ($\CN=3$ theories are isolated as $\CN=2$ SCFTs), we have $\Delta\ge3$.} The fact that $\CN=3$ theories---and particularly those related to $\CN=4$ SCFTs via discrete gauging---satisfy this constraint is important. Indeed, this statement implies that the fate of the conjecture does not depend on what happens to any $\CN=4$ 1-form symmetries in the discrete 0-form symmetry gauging process (i.e., on whether there are mixed 0-form / 1-form 't Hooft anomalies).

In the case of IHS theories (and any discrete gaugings thereof), we have seen that the central charge bounds \eqref{acbound2} also imply trivial 1-form symmetry. It would be interesting to understand if this statement is more universally true in the space of interacting 4D $\CN=2$ SCFTs.

Finally, let us conclude with a discussion of other possible bounds and their implications for 1-form symmetry. For example, it may well be true that, for sufficiently small $a$ or $c$ (independently of the rank and flavor symmetry of the theory), interacting 4D $\CN=2$ SCFTs do not have 1-form symmetry. Indeed, we know from \cite{Liendo:2015ofa} that for interacting theories $c\ge11/30$ and that this value of the central charge is saturated by the $(A_1, A_2)$ SCFT. We know this AD theory has trivial 1-form symmetry. Therefore, if one can show that $(A_1, A_2)$ is the unique lowest $c$ theory, we have the somewhat trivial result that for sufficiently small $c$ (i.e., $c=11/30$), interacting SCFTs have no 1-form symmetry.

On the other hand, we do not expect an absolute $a$ or $c$ bound on 1-form symmetry to be much larger than $11/30$. Indeed, $su(2)$ $\CN=4$ SYM has a $\mathbb{Z}_2$ 1-form symmetry and has $a=c=3/4$. Still, it would be interesting to use a version of the bootstrap including extended operators and prove from first principles that interacting theories with $a,c<3/4$ have no 1-form symmetry (or else to find a counterexample).

\newsec{Conclusions}
We have argued that 4D $\CN=2$ SCFTs arising from type IIB string theory on IHSs have 1-form symmetry only if they also have an exactly marginal deformation (see the appendix for the bulk of the proof). We saw some evidence that this behavior extends to certain other classes of $\CN\ge2$ theories, but there are some potential obstructions for the $G(3,3,3)$ $\CN=3$ theory. It would be interesting to understand whether this $\CN=3$ theory has 1-form symmetry or not. If one could indeed prove that it has 1-form symmetry, then one possibility might be that our main claim applies to  $\CN=2$ theories subject to the 4D/2D correspondence in \cite{Cecotti:2010fi} and that the $\CN=3$ theory in question does not fall into this class (one consequence of our results here is that this theory would not have a realization as an IHS theory).\footnote{Perhaps ideas related to those in \cite{Hellerman:2006zs} will be relevant for demonstrating this claim. In addition, perhaps one can make contact with the discussion in \cite{Tanizaki:2019rbk}.}

In section \ref{conjectures} we saw that by focusing on theories with all $\CN=2$ chiral ring generators of dimension less than two we could formulate a conjecture for interacting SCFTs having no 1-form symmetry that is compatible with all the data we are aware of (including being trivially satisfied by theories for which we are not able to explicitly compute the 1-form symmetry). Moreover, we argued that for IHS theories, this statement could be reformulated in terms of bounds on $a$ and $c$ (it would be interesting to understand if this is true more generally).

In addition to trying to prove (or disprove) the conjecture, there is much work to be done. For example, it would be interesting to explore what happens with less (or no) supersymmetry or to understand if constraints discussed in \cite{Perlmutter:2020buo} and \cite{Heidenreich:2021tna} are relevant here.

\ack{We are grateful to P.~Argyres, S.~Giacomelli, M.~Martone, T.~Nishinaka, D.~Xie, and G.~Zafrir for discussions. We also thank T.~Nishinaka for collaboration on related projects. Our work was funded by the Royal Society under the grant, “Relations, Transformations, and Emergence in Quantum Field Theory” and the STFC under the grant, “String Theory, Gauge Theory and Duality.”}

\newpage
\begin{appendix}
\section{Theories with at most two different weights}\label{twoq}
In this section we consider SCFTs with at most two distinct $q_i$ and prove claim 2 in section \ref{LGmass}. While these theories are also covered by the proof we give in section \ref{IHSproof} and appendix \ref{SCFTclassify}, our result here is entirely independent of any IHS classification. Therefore, we consider it worthwhile to exhibit this proof independently. Moreover, we will see that this result has its uses in appendix \ref{SCFTclassify}.

To that end, without loss of generality, we may take $q_x\ge q_y\ge q_u\ge q_v$. Let us first discuss the case where all $q_i=q$. Then all $U_i=U$ and so, from \eqref{Ri}, we have that $r_i=1$ for all $i$ and the 1-form symmetry is trivial. 

Next let us consider the situation in which there are two different weights. This scenario breaks up into various cases:

\bigskip
\noindent
{\bf Case 1 ($q_x>q_y=q_u=q_v=q$):} We have $q_x=Nq$ for $N>1$. Let us define $q:=V/U$ and $q_x:= V_x/U_x$ (where all fractions are reduced). 

First, suppose that $N\in\mathbb{Z}$. It then follows that $U=pU_x$ for some integer $p\ge1$ and so $\gcd(U,U_x)=U_x$. As a result, applying \eqref{Ri} yields
\begin{eqnarray}
r_y&=&r_u=r_v={U\over\gcd(U,U,U,U_x)}\cdot{\gcd(U,U_x,U)\gcd(U,U,U_x)\gcd(U,U,U)\over\gcd(U,U_x)\gcd(U,U)\gcd(U,U)}=1~,\cr r_x&=&{U_x\over\gcd(U,U,U,U_x)}\cdot{\gcd(U_x,U,U)\gcd(U_x,U,U)\gcd(U_x,U,U)\over\gcd(U_x,U)\gcd(U_x,U)\gcd(U_x,U)}=1~,
\end{eqnarray}
and so the 1-form symmetry is trivial.

Let us now consider $N\not\in\mathbb{Z}$.\footnote{In this case, we can have $\gcd(U,U_x)=1$. Then we would find $r_y=r_u=r_v=1$ and $r_x=U_x>1$.} An upper bound on the number of independent constraints on marginal (i.e., $Q=1$) terms arises from considering\footnote{In principle, some of these constraints could be redundant.}
\begin{equation}
0=\left\{\partial_yW,\partial_uW,\partial_vW\right\}\times\left\{y,u,v\right\}~,\ \ \ 0=x\partial_xW~.
\end{equation}
This amounts to at most ten constraints. How many marginal terms can we generate? Let us first recall that any variable appearing in $W$ must be a root or a pointer \cite{Davenport:2016ggc,Kreuzer:1992np}.\footnote{Recall that a variable is a root if it appears as $W\supset X^{N_x}$ and is a pointer if $W\supset X^{N_x'}Y$. In this latter case, we say that $X$ points to $Y$.}

Suppose first that $x$ is a root so that $x^{N_x}\subset W$. Let us also suppose that $x^{N_x'}y^a$ is marginal for some value of the exponents. Then
\begin{equation}
N_x'q_x+aq=1\ \Rightarrow\ q={1\over a}(N_xq_x-N_x'q_x)={1\over a}(N_x-N_x')q_x<q_x~.
\end{equation}
As a result, $a>1$ since otherwise we contradict the statement that $q<q_x$. In fact $a\ne2$ since otherwise we would have $N_x-N_x'=1$ which would imply that $N=2$, which contradicts $N\not\in\mathbb{Z}$. Therefore, $x^{N_x'}y^eu^fv^g$ are all marginal for $e+f+g=a\ge3$. In all this amounts to at least eleven marginal monomials and at most ten constraints. So, the theory has at least one exactly marginal deformation.

Let us now suppose that there is no $a$ such that the above holds. In this case, we still need marginal terms built out of $y$, $u$, and $v$. Since $N_x\ge2$ (otherwise the singularity is not isolated; from the 2D perspective, $x$ would be a supersymmetry-breaking Polonyi field), we must have terms of the form $y^eu^fv^g$ with $e+f+g\ge3$ (since $q_x>q$). This again amounts to at least eleven marginal monomials and at most ten constraints. Therefore, we have a conformal manifold.

Finally, let us consider the case that $x$ is a pointer. Without loss of generality, we may assume $x^{N_x'}y\subset W$. Then clearly $x^{N_x'}u$ and $x^{N_x'}v$ are also marginal. Therefore, we need at least one more marginal term and it cannot depend on $x$ (otherwise, setting $x=0$ implies that $\partial_uW=\partial_vW=\partial_yW=0$, and the singularity will not be isolated). As a result, we need an $x$-independent term. Since $q_x>q$, we need again at least ten marginal terms of the form $y^eu^fv^g$ with $e+f+g\ge3$. This gives at least thirteen marginal terms and at most ten constraints. Therefore the theory has a conformal manifold.

\bigskip
\noindent
{\bf Case 2 ($q'=q_x=q_y>q_u=q_v=q$):} We have $q'=Nq$ for $N>1$. Let us define $q':=V'/U'$ and $q:= V/U$ (where all fractions are reduced). As a result, applying \eqref{Ri} yields
\begin{eqnarray}
r_u&=&r_v={U\over\gcd(U,U,U',U')}\cdot{\gcd(U,U',U)\gcd(U,U',U)\gcd(U,U',U')\over\gcd(U,U)\gcd(U,U')\gcd(U,U')}=1~,\cr r_x&=&r_y={U'\over\gcd(U,U,U',U')}\cdot{\gcd(U',U',U)\gcd(U',U',U)\gcd(U',U,U)\over\gcd(U',U)\gcd(U',U)\gcd(U',U')}=1~,
\end{eqnarray}
and so the 1-form symmetry is trivial.

\bigskip
\noindent
{\bf Case 3 ($q'=q_x=q_y=q_u>q_v$):} We have $q'=Nq_v$ for $N>1$. Let us define $q':=V'/U'$ and $q_v:= V_v/U_v$ (where all fractions are reduced). 

First, suppose that $N\in\mathbb{Z}$. It then follows that $U_v=pU'$ for some integer $p\ge1$ and so $\gcd(U',U_v)=U_v$. As a result, applying \eqref{Ri} yields
\begin{eqnarray}
r_x&=&r_y=r_u={U'\over\gcd(U',U',U',U_v)}\cdot{\gcd(U',U_v,U')\gcd(U',U',U_v)\gcd(U',U',U')\over\gcd(U',U_v)\gcd(U',U')\gcd(U',U')}=1~,\cr r_v&=&{U_v\over\gcd(U',U',U',U_v)}\cdot{\gcd(U_v,U',U')\gcd(U_v,U',U')\gcd(U_v,U',U')\over\gcd(U_v,U')\gcd(U_v,U')\gcd(U_v,U')}=1~,
\end{eqnarray}
and so the 1-form symmetry is trivial (if $N\not\in\mathbb{Z}$, this statement need not hold).

Let us now consider $N\not\in\mathbb{Z}$. An upper bound on the number of independent constraints on marginal (i.e., $Q=1$) terms arises from considering
\begin{equation}
0=\left\{\partial_xW,\partial_yW,\partial_uW\right\}\times\left\{x,y,u\right\}~,\ \ \ 0=v\partial_vW~.
\end{equation}
This amounts to at most ten constraints. How many marginal terms can we generate? As in case 1, $x$ must be a root or a pointer.

First suppose that $x$ is a root so that $x^{N_x}\subset W$ (with $N_x\ge2$). If $N_x>2$, we are done: we get at least ten marginal terms of the form $x^ey^fu^g$ with $e+f+g\ge3$. In addition, we require at least one more term involving $v$ (otherwise the singularity is not isolated). On the other hand, if $N_x=2$ we are also done: $q'=1/2$ and lemma 1 guarantees the 1-form symmetry is trivial. 

Next, suppose that $x$ is a pointer. If we have $x^{N_x'}y\subset W$ we are again done if $N_x>2$ by the same logic as above. Similarly, if $N_x'=1$, we are done by lemma 1 (again $q'=1/2$), since the 1-form symmetry is trivial. The same argument applies if $x$ points to $u$ instead.

The final case to check is that $x^{N_x'}v\subset W$, where $N_x'>1$ (otherwise lemma 1 again guarantees trivial 1-form symmetry). We need a $v$-independent term to add to $W$ since otherwise the singularity is not isolated (setting $v=0$ solves $\partial_xW=\partial_yW=\partial_uW=0$ without any further constraints). Such a term cannot be consistent with $q_v<q'$ and marginality. $\square$

\section{Completing the proof of the main claim} \label{SCFTclassify}
The authors of \cite{Xie:2015rpa} argued for a classification of 4D $\CN=2$ SCFTs realized via type IIB string theory based on the classification of singularities in  \cite{yau2005classification}. In parts \ref{fourM} and \ref{moreM} of the appendix, we follow this classification and prove our main claim in this set of theories.

In appendix \ref{beyondClass}, we give arguments that do not depend on the classification of \cite{yau2005classification}. In particular, we allow for potentially more general terms in the IHS polynomials than those considered in \ref{moreM} (compatible with discussion in \cite{Davenport:2016ggc,Closset:2020afy}).

Up to these subtleties which we will address in \ref{beyondClass}, there are 19 types of theories we need to consider (see table~\ref{Full19type}). To define an SCFT, one needs to impose the constraint $\sum_i q_i>1$. For each type of singularity, the solution falls into two classes: some infinite sequences and a finite number of sporadic cases. To check the conjecture, one can consider the sporadic cases one by one since there are only a finite number of them. Indeed, we have checked the conjecture is true for these sporadic cases. Therefore, in the following, we will only explicitly discuss the infinite sequences.\footnote{More precisely, our treatment in \ref{moreM} implicitly covers all sporadic cases for this subset of theories. For the theories in \ref{fourM} and \ref{beyondClass}, we omit a discussion of sporadic cases (which we have checked separately).}

The 19 types of singularities can be divided into two broad classes. In the first class, there are only four monomials in the defining polynomials (e.g., as in the case of type I discussed in \eqref{t1}). We discuss these theories in section \ref{fourM}. In the second class, there are five or more monomials in the defining polynomials (e.g., as in type VIII). We will discuss these theories in section \ref{moreM}. We consider generalizations of these latter theories in section \ref{beyondClass}.

Throughout this appendix, we will make use of the notion of {\it reduction} or of reducing one theory to an other. By this we mean that the weights of the singularities corresponding to the theories in question coincide. As a result, the Poicar\'e polynomials and deformation spectra coincide as well.

  \bgroup
\def\arraystretch{1.0}%
\begin{table}   
\begin{center}
\begin{tabular}{|l|c|c| }
\hline
Type & $W(x,y,u,v)$ & $q_x+q_y+q_u+q_v$ \\ \hline
I & $x^a+y^b+u^c+v^d$& ${1\over a}+ {1\over b}+{1\over c}+{1\over d}$\\ \hline
II & $x^a+y^b+u^c+uv^d$& ${1\over a}+ {1\over b}+{1\over c}+{c-1\over cd}$\\ \hline
III & $x^a+y^b+u^cv+uv^d$& ${1\over a}+ {1\over b}+{d-1\over cd-1}+{c-1\over cd-1}$\\ \hline
IV & $x^a+xy^b+u^c+uv^d$& ${1\over a}+ {a-1\over a b}+{1\over c}+{c-1\over cd}$\\ \hline
V & $x^ay+xy^b+u^c+uv^d$& ${b-1\over ab-1}+ {a-1\over a b-1}+{1\over c}+{c-1\over cd}$\\ \hline
VI & $x^ay+xy^b+u^cv+uv^d$& ${b-1\over ab-1}+ {a-1\over a b-1}+{d-1\over c}+{cd-1\over cd}$\\ \hline
VII & $x^a+y^b+yu^c+uv^d$& ${1\over a}+ {1\over b}+{b-1\over b c}+{b(c-1)+1\over bcd}$\\ \hline
  VIII & $x^a+y^b+yu^c+yv^d+u^pv^q$, & ${1\over a}+ {1\over b}+{b-1\over b c}+{b-1\over bd}$\\
~ & ${p(b-1)\over bc}+{q(b-1)\over bd}=1$& ~\\ \hline
IX & $x^a+y^bv+u^cv+yv^d+y^pu^q$, & ${1\over a}+ {d-1\over bd-1}+{b(d-1)\over c(bd-1)}+{b-1\over bd-1}$\\
~ & ${p(d-1)\over bd-1}+{qb(d-1)\over c(bd-1)}=1$& ~\\ \hline
X & $x^a+y^bu+u^cv+yv^d$& ${1\over a}+ {d(c-1)+1\over bcd+1}+{b(d-1)+1\over bcd+1}+{c(b-1)+1\over bcd+1}$\\ \hline
XI & $x^a+xy^b+yu^c+uv^d$& ${1\over a}+ {a-1\over ab}+{a(b-1)+1\over abc}+{ab(c-1)+(a-1)\over abcd}$\\ \hline

XII & $x^a+xy^b+xu^c+yv^d+y^pu^q$& ${1\over a}+ {a-1\over ab}+{a-1\over ac}+{a(b-1)+1\over abd}$\\
~&${p(a-1)\over ab}+{q(a-1)\over ac}=1$&~ \\ \hline
XIII & $x^a+xy^b+yu^c+yv^d+u^pv^q$& ${1\over a}+ {a-1\over ab}+{a(b-1)+1\over abc}+{a(b-1)+1\over abd}$\\
~&${p(a(b-1)+1)\over abc}+{q(a(b-1)+1)\over abd}=1$&~ \\ \hline
XIV & $x^a+xy^b+xu^c+xv^d+y^pu^q+u^rv^s$& ${1\over a}+ {a-1\over ab}+{a-1\over ac}+{a-1\over ad}$\\
~&${p(a-1)\over ab}+{q(a-1)\over ac}=1={r(a-1)\over ac}+{s(a-1)\over ad}$&~ \\ \hline
XV&$x^ay+xy^b+xu^c+uv^d+y^pu^q$& ${b-1\over ab-1}+{a-1\over ab-1}+{b(a-1)\over c(ab-1)}+{c(ab-1)-b(a-1)\over cd(ab-1)}$ \\
~&${p(a-1)\over ab-1}+{qb(a-1)\over c(ab-1)}=1$&~ \\ \hline
XVI&$x^ay+xy^b+xu^c+xv^d+y^pu^q+u^rv^s$& ${b-1\over ab-1}+{a-1\over ab-1}+{b(a-1)\over c(ab-1)}+{b(a-1)\over d(ab-1)}$ \\
~&${p(a-1)\over ab-1}+{qb(a-1)\over c(ab-1)}=1={r(a-1)\over ac}+{s(a-1)\over ad}$&~ \\ \hline
XVII&$x^ay+xy^b+yu^c+xv^d+y^pv^q+x^ru^s$& ${b-1\over ab-1}+{a-1\over ab-1}+{a(b-1)\over c(ab-1)}+{b(a-1)\over d(ab-1)}$ \\
~&${p(a-1)\over ab-1}+{qb(a-1)\over d(ab-1)}=1={r(b-1)\over ab-1}+{sa(b-1)\over c(ab-1)}$&~ \\ \hline
XVIII&$x^au+xy^b+yu^c+yv^d+u^pv^q$&${b(c-1)+1\over abc+1}+{c(a-1)+1\over abc+1}+{a(b-1)+1\over c(abc+1)}+{c(a(b-1)+1)\over d(abc+1)}$ \\
~&${p(a(b-1)+1)\over abc+1}+{qc(a(b-1)+1)\over d(abc+1)}=1$&~ \\ \hline
XIX &$x^a u+xy^b+vu^c  +yv^d$ &${ b(d(c-1)+1)-1\over abcd-1}+{ d(c(a-1)+1)-1\over abcd-1}$\\
~&~&$+{ a(b(d-1)+1)-1\over abcd-1}+{ c(a(b-1)+1)-1\over abcd-1}$ \\
\hline
\end{tabular}
\caption{The canonical form for IHSs with a good $\mathbb{C}^*$ action. This table is adapted from  \cite{Xie:2015rpa}. Note that for those types of singularities with five or more monomials, the specific monomials are chosen to make the singularity isolated following  \cite{Xie:2015rpa}. But there are also other choices of these additional monomials, as mentioned in \cite{Davenport:2016ggc,Closset:2020afy}. Our proof  in \ref{beyondClass} does not rely on the structure of these additional terms. 
}
\label{Full19type}
\end{center}
\end{table}
\egroup

\subsection{Singularities with four monomials}\label{fourM}
In this section we prove our main claim for IHS theories with at most four monomials in $W$. In the body of the paper we already showed that theories of type I satisfy our claim. Therefore, after briefly recapitulating this case, we move onto the remaining types of theories with four monomials.

 Before discussing each singularity one by one, we observe that the type  I--VI singularities in table~\ref{Full19type} can be regarded as the composition of two subpolynomials in terms of $x,y$ and $u,v$. More precisely,  we can think of the singularities of types  I, II, $\cdots$, VI as types $\mathsf{XX,XY,XZ,YY, ZY,ZZ}$ where the $\mathsf{X,Y,Z}$ data is given in table~\ref{XYZtype}.
 
With this decomposition, one has $W(x,y,u,v)=W'(x,y)+W''(u,v)$ where $W',W'' \in W_{\mathsf{X,Y,Z}}$, and $Q=Q'+Q''$ where   $Q',Q'' \in Q_{\mathsf{X,Y,Z}}$ (recall that $Q$ is the scaling weight of a monomial deformation built from sub-monomials of scaling weights $Q'$ and $Q''$ in $x,y$ and $u,v$ respectively). To find the isolated set of SCFTs, we need to find the set where $Q=1$ has no solution in the Milnor ring. In particular, this means $Q'=1$ or $Q''=1$ also have no solutions in their own Milnor rings.

 \bgroup
 \def\arraystretch{1.3}%
\begin{table} 
 \begin{center}
\begin{tabular} {|c|c|c|c|c |}     \hline 
 $ $ &      $W$ &   $(q_x,q_y)$ &  monomial basis in Milnor ring & $Q_{k,l}$   \\ \hline
 $ \mathsf X$ & $x^a+y^b$   &$( \frac1 a,\frac 1b)$& $x^k y^l  \;(k=0, \cdots a-2, l=0, \cdots, b-2)$ & $\frac k a+\frac l b$  \\ \hline 
  $\mathsf Y$ & $x^a+xy^b$   &$(\frac 1a,\frac{a-1}{ab})$& \makecell{$x^k y^l  \;(k=0, \cdots a-1, l=0, \cdots, b-2$ \\$\text{ or } k=0, l=b-1)$} & $\frac k a+\frac{(a-1)l}{ab}$  \\ \hline 
   $ \mathsf Z$ & $x^a y+x y^b$   &$(\frac{b-1}{ab-1},\frac{a-1}{ab-1})$& $x^k y^l  \;(k=0, \cdots a-1, l=0, \cdots, b-1)$ &
    $\frac{(b-1)k+(a-1)l}{ab-1}  $  \\ \hline 
  \end{tabular} 
\end{center}
\caption{The building blocks of the first six types of singularities. We list the defining polynomials, the weights, the monomial basis in the   Milnor ring, and their corresponding $Q$s.   }
\label{XYZtype}
\end{table}
\egroup

 For type $\mathsf X$, the condition for the theory being isolated is just the condition for $(A_{p-1},A_{q-1})$ SCFTs discussed in \eqref{ApAqisolated} 
\be\label{Xtype}
 \gcd(a,b)=1~, \qquad \text{ or }\qquad a= 2 ~, \qquad  \text{ or } \qquad b= 2~,  \qquad \text{ or }  \qquad (a,b)= (3,3)~.
\ee

 For type $\mathsf Y$,  the equation $Q =1$ is equivalent to $(a - 1)l =b(a-k)$, which can be further rewritten as
 \be
\frac{k'}{a-1}=\frac{l }{b }~, \qquad    k'=a-k = 1, \cdots, a~, \; l  =1, \cdots, b-2~.
\ee
Then the resulting condition for the theory to be isolated can be found to be
\be\label{Ytype}
\gcd(a-1,b )=1~,\qquad \text{ or } \qquad b=2~ .
\ee

 For type $\mathsf Z$,  the equation $Q =1$ is equivalent to $(a - 1) ( b-l) = (b - 1) (k - 1) $ which can be further rewritten as
 \be
\frac{k'}{a-1}=\frac{l'}{b-1}~, \qquad    k'=k-1=-1, \cdots a-2, \; l'=b-l=1~, \cdots b~.
\ee
 It is then easy to find the condition for the theory to be isolated
\be\label{Ztype}
\gcd(a-1,b-1)=1~.
\ee

\bigskip
\noindent{\bf Type I:} The type I singularity is given by 
\be
W(x,y,u,v)=x^a + y^b + u^c + v^d~,
\ee
where $a, b, c, d \ge 2$ so that there is an isolated singularity at the origin. The weights and the Milnor number are given by 
\be
\vec q=\Big( \frac 1a, \frac1b, \frac1c, \frac1d \Big)~, \qquad \mu=(a-1)(b-1)(c-d) (d-1)~.
\ee

The condition $\sum_i q_i>1$ imposes constraints on the possible values of $a,b,c,d$. The solution was found in \cite{Xie:2015rpa}. 
It includes several infinite sequences listed in table~\ref{Iinf0} and a finite number of sporadic cases. Since the sporadic cases are finite in number, one can check the conjecture explicitly case-by-case. Henceforth, we will not list the solution corresponding to the sporadic cases. 

As discussed in the main text, from table~\ref{Iinf0}, we learn that the infinite sequences of Type I singularities are just certain $(\mathfrak{g},\mathfrak{g}') $ theories  and the $(p,k)$ SCFTs considered before. We have already shown how they satisfy our main claim. Therefore, in addition to the checks of the sporadic cases we have performed, the conjecture holds for SCFTs realized via type I singularities. 
 
 \bigskip 
 \noindent
 {\bf Type II:} The type II singularity is given by 
  \be
  W= x^a+   y^b +  u^c + uv^d~,
  \ee
 where  we require $a,b,c\geq 2,~d\geq 1$ in order to have an isolated singularity at the origin. The weights and the Milnor numbers are given by 
 \beqn
 \vec q =\left(\frac{1}{a},\frac{1}{b},\frac{1}{c},\frac{c-1}{c d}\right)~, \qquad \mu=(a-1) (b-1) (c (d-1)+1)~.
 \eeqn
A monomial basis  of the Milnor ring can be obtained from that of $\mathsf X_{a,b}$ and $\mathsf Y_{c,d}$ in Table~\ref{XYZtype}.

 Note that there are actually  some overlaps with Type I singularities. If there exits an
integer $n$ such that ${n (c-1)\over d c}=1$, then $v^n$ is an exactly marginal deformation and the singularity can be put in the equivalent form $\tilde W=x^a+y^b+u^c+v^n$. In this case, the type II singularity reduces to the type I singularity. The reducibility condition is\footnote{In particular, this means that if $c>2$, the   condition \eqref{Ytype} is volated, implying that the SCFTs is not isolated.  }
\begin{equation}
{d\over c-1}\in \mathbb{Z}~.
\label{equi2}
\end{equation}

Imposing the condition  $\sum q_i>1$ as well as $a\leq b$ and $c\leq d$ to remove redundancy, one gets the infinite sequences in table~\ref{IIinf} and a finite number of sporadic cases which will not be listed explicitly. In particular, we have marked all the reducible cases satisfying \eqref{equi2} in gray. Note that exchanging $(c,d)$ leaves $\sum_i q_i$ invariant, but  this is not a symmetry of the SCFTs. Therefore, in table~\ref{IIinf} we also list another sub-table where the entries of $(c,d)$ are exchanged. In this way, we give a complete    list of SCFTs of type II (for the infinite sequences). 

The finite sporadic cases can be checked explicitly one by one and indeed our claim holds. So we next check the infinite sequences in table~\ref{IIinf}. In particular, we only need to consider the irreducible ones there because reducible entries in gray are just the type I we considered already.

\begin{table}[H]
\begin{center}
\begin{tabular}{|C|C|C|C|C| }
\hline
( a,b,c,1 )& (2,2,c,d) & \mk (2,b,2,d) & \mk (3,b,2,2)&\mk (4,b,2,2) \\ \hline
(2,3,3, d ) &(2,3,4, d )& (2,3,5, d )&(2,3,6, d )&(2,4,3, d ) \\ \hline
(2,5,3, d )&(2,6,3, d )&(2, b ,3,3)&\mk (2, b ,3,4)&\mk (3,3,2, d ) \\ \hline
\mk (3,4,2, d  )& \mk (3,5,2, d )&\mk (3,6,2, d )& \mk(3,b,2,3)&(2,4,4, d ) \\ \hline
\mk (4,4,2, d )&(3,3,3, d  )&&&\\  \hline
\end{tabular}  \\[10pt]
\begin{tabular}{|C|C|C|C|C| }
\hline
 &   &   (2,b,c,2) &  & \\ \hline
(2,3,c, 3 ) &(2,3,c, 4 )& (2,3,c, 5 )&(2,3,c,6  )&(2,4,c, 3 ) \\ \hline
(2,5,c, 3 )&(2,6,c, 3 )& &\mk (2, b ,4,3)&  (3,3,c, 2 ) \\ \hline
  (3,4,c, 2  )&   (3,5,c,2  )&  (3,6,c, 2 )&\mk  (3,b,3,2)&(2,4,c, 4 ) \\ \hline
  (4,4,c, 2 )&(3,3,c, 3  )&&&\\ \hline
\end{tabular}
\caption{Infinite sequences of type II singularities. The reducible cases are marked in gray. The lower table is obtained from the upper table by exchanging $c$ and $d$, and the empty entries are either symmetric under exchange or  have $c=1$. This table is adapted from  \cite{Xie:2015rpa}.}
\label{IIinf}
\end{center}
\end{table}

We will first need to find the isolated set and then check the 1-form symmetry. In practice, our procedure is as follows: by considering specific types of monomials in the Milnor ring, we can find some necessary condition for  the theory being isolated; then for an irreducible weight (namely a weight satisfying \eqref{qVU})  we apply our formulas \eqref{Ri} and \eqref{Gi} to compute the 1-form symmetry. As discussed in the main text, to show the absence of    1-form symmetry, we only need to show that $g_i=0$ or $r_i=1$ for each $i=1,2,3,4$.

  \para{(a,b,c,1)}
 The weight vector
  \beqn
 \vec q =\left(\frac{1}{a},\frac{1}{b},\frac{1}{c},\frac{c-1}{c }\right)~, 
 \eeqn
  is irreducible,
 so $r_3=r_4=1,g_1=g_2=0$ and this theory has no 1-form symmetry.
 
   \para{(2,2,c,d)}
 The weight vector can be written  as an irreducible form 
  \beqn
 \vec q =\left( \frac{1}{2},\frac{1}{2},\frac{1}{c},\frac{c-1}{c d} \right) 
 =\left( \frac{1}{2},\frac{1}{2},\frac{1}{c},\frac{\tilde c }{c \tilde d} \right)~,\qquad \gcd(\tilde c, \tilde d)=1~.
 \eeqn
 Then one finds $r_1=r_2=r_3=1,g_4=0$, so this theory has no 1-form symmetry.

 \para{(2,3,3,d)}
  
There is always a marginal term $y u^2$ with $Q=1$.

 \para{(2,3,4,d)}
A necessary condition for being isolated is $d=2$ or $\gcd(3,d)=1$. In both cases, there is no 1-form symmetry.

 \para{(2,3,5,d)} For $d=1,2$ there is no 1-form symmetry as $g_i=0$. Consider the monomial $yu^2 v^l$ in the Milnor ring.  $Q=1$ leads to $d=3l$. It is then easy to figure out a necessary condition for being isolated: $\gcd(3,d)=1$, in addition to the condition   \eqref{Ytype}   $\gcd(4,d)=1 $, which is equivalent to $\gcd(2,d)=1$.  Then one finds $g_i=0$ and trivial 1-form symmetry.

 \para{(2,3,6,d)}

There is always a marginal term $y u^4$ with $Q=1$.

 \para{(2,4,3,d)}  From the necessary condition for being isolated \eqref{Ytype}, we have  $\gcd(2,d)=1$ (the $d=2$  case leads to trivial 1-form symmetry).
Then we find $g_1=g_2=g_4=0,r_3=1$, meaning no 1-form symmetry. 
  
 \para{(2,5,3,d)}  For $d=1,2,3$, we find $g_i=0$, implying trivial 1-form symmetry. For $d>3$, consider $y^3 v^l$. $Q=1$ leads to $3d=5l$.
Then a  necessary condition for being isolated is found to be  $\gcd(5,d)=\gcd(2,d)=1$. The  weight vector is thus irreducible. This leads to  $g_i=0$, implying no 1-form symmetry. 
  
 \para{(2,6,3,d)}  It is easy to see $y^4 u$ has $Q=1$, so this theory is never isolated. 
 
  \para{(2,b,3,3)}  For $b=2,3$, it has no 1-form symmetry. For $b>3$, the $y,u$ part gives a necessary isolated condition   $\gcd(3,b)=1$.  Then $g_1=g_2=g_4=0,r_3=1$. Thus the theory has no 1-form symmetry.
    
 \para{(3,3,3,d)}     The monomial $xyu$ has $Q=1$, so  this theory is never isolated.

 \para{(2,b,c,2)}              
Consider $y^k u^l$. The condition $Q=1$ leads to $b(c-l)=kc$, namely
\be
\frac{l'}{c}=\frac{k}{b}~, \qquad k=0, \cdots, b-2, \; l'=c-l=1,\cdots, c~.
\ee
So a necessary condition for being  isolated   is 
$\gcd(b,c)=1$ or $  b=2 $.

If $c$ is even, $\gcd(c,2)=2$,  and the weight is irreducible.  We then have $r_1=r_3=1,g_2=0,g_4=\gcd(b,c)-\gcd(b,2)=0$ for either $\gcd(b,c)=1$ or $  b=2 $.  So the 1-form symmetry is always trivial.
                
If $c$ is odd, $c=2p+1$,  the weight vector
\be
\vec q=\left(\frac{1}{2},\frac{1}{b},\frac{1}{2 p+1},\frac{p}{2 p+1}\right)~,
\ee
  is irreducible.   Then   $\gcd(b,c)=\gcd(b,2p+1)=1$    leads to  $g_1=g_2=g_4=0, r_3=1$, implying no 1-form symmetry.

      \para{(2,3,c,3)}         For $c=2,3$, one can check that the 1-form symmetry is trivial. For $c>3$, considering the $y,u$ and $u,v$ part respectively, one finds a necessary condition for being isolated  is $\gcd(3,c) =\gcd(c-1,3)=1$. Next, consider $y u^k v$. $Q=1$ leads to $c+1=3k$ which leads to another necessary  condition for being isolated: $\gcd(c+1,3)=1$.
These three conditions cannot be satisfied simultaneously. So this theory cannot be isolated. 

     \para{(2,3,c,4)}              For $c=2,3$, one can check that the 1-form symmetry is trivial.   For $c>3$, considering the $y,u$ and $u,v$ parts respectively, one finds a necessary condition for being isolated  is $\gcd(3,c) =\gcd(c-1,4)=1$, which is equivalent to  $\gcd(3,c)=1, \gcd(2,c)=2$.  Then $g_i=0$ and thus the one-form symmetry is trivial.

     \para{(2,3,c,5)}     For $c=2,3$, one can check that the 1-form symmetry is trivial.    For $c>3$,     one finds that  a necessary condition for being isolated  is          $\gcd(3,c) =\gcd(c-1,5)=1$,     which leads to  $g_i=0$ and thus trivial 1-form symmetry.

     \para{(2,3,c,6)}   For $c=2,3$, one can explicitly check that it has a marginal deformation.       For $c>3$, as above, one finds that  a necessary condition for being isolated  is          $\gcd(3,c) =1$ and $ \gcd(c-1,6)=1$ which is equivalent to $\gcd(c-1,2)=\gcd(c-1,3)=1$. Next consider $y u^k v^2$. $Q=1$ leads to $c+1=3k$ which leads to another necessary  isolated condition: $\gcd(c+1,3)=1$.
These three conditions  $\gcd(c+1,3)=\gcd(c-1,3)=\gcd(c ,3)=1$ cannot be satisfied simultaneously. So this theory cannot be isolated.

     \para{(2,4,c,3)}            For $c=2$, the 1-form symmetry is trivial.         For $c>2$, one finds that  a necessary condition for being isolated  is  $\gcd(4,c) =\gcd(c-1,3)=1 $,  which leads to $g_1=g_2=g_4=0,r_3=1$ and thus  trivial 1-form symmetry.

                  \para{(2,5,c,3)}           For $c=2$, the 1-form symmetry is trivial.   For $c>2$, one  finds that  a necessary condition for being isolated  is   $ \gcd(5,c) =\gcd(c-1,3)=1$.             
 Then $g_1=g_2=g_4=0,r_3=1$, and the 1-form symmetry is trivial.

                  \para{(2,6,c,3)}           For $c=2$, the 1-form symmetry is trivial.                     
   For $c>2$, one  finds that  a necessary condition for being isolated  is   $ \gcd(6,c) =\gcd(c-1,3)=1$, which leads to $\gcd(3,c)=\gcd(2,c)=1$.            
 Then $r_1=r_2=r_3=1,g_4=0$ and 1-form symmetry is trivial.

                  \para{(3,3,c,2)}      For $c=2,3$, the 1-form symmetry is trivial.         
                  For $c>3$,    a necessary condition for being isolated is  $\gcd(3,c) =\gcd(c-1,2) =1$.
     Then the weight vector is irreducible. 
   As a result,  $r_1=r_2=r_3=1,g_4 =0$,  and 1-form symmetry is trivial.

             \para{(3,4,c,2)}    For $c=2,3$, the theory has a marginal deformation.     
           For $c>3$,   the $y,u$ part leads to the condition    $\gcd(4,c)  =1$. So $c$ is odd and we can set     $c=2p+1$ and get the irreducible weight vector
              \be
              \vec q=\left(\frac{1}{3},\frac{1}{4},\frac{1}{2 p+1},\frac{p}{2 p+1}\right)~. 
              \ee  
 Consider $x u^k$. $Q=1$ leads to $2c=3k$. To be isolated, we also need $\gcd(3,c)=1$. Then $g_i= 0$, and the 1-form symmetry is trivial.

                        \para{(3,5,c,2)}            If $c=2$, the 1-form symmetry is trivial. 
                        For $c>2$, a necessary condition for being isolated is $\gcd(5,c) =\gcd(c-1,2) =1$. 
So  $c$ is even and the weight vector is irreducible. Then  $g_1=g_2=g_4=0,r_3=1$, leading to trivial 1-form symmetry.

             \para{(3,  6,c,2)}              For $c=2$, the theory has a marginal deformation. 
For $c>2$, a necessary  condition for being isolated is $\gcd(6,c)=1,\gcd(c-1,2)=1$, which  leads to $\gcd(2,c)=\gcd(2,c-1)=1$. Obviously there is no solution.  So the theory can not be isolated.
                
         \para{(4,4,c,2)}      
            
            This is not isolated as $\gcd(4,4)=4>1$.

         \para{(3,3,c,3)}    For $c=2,3$, the theory is not isolated. For $c>3$, a necessary condition for being isolated is $\gcd(3,c)=1$, $\gcd(c-1,3)=1$.      Then we have
$r_1=r_2=r_3=1,g_4 =0$ with trivial 1-form symmetry.

 \bigskip
 \noindent{\bf Type III:} In this case, 
 \be
W=x^a+y^b+u^cv+uv^d~,
 \ee
 and we require $a,b,c,d\geq 2$ in order to have an isolated singularity at the origin. The weights and Milnor number are 
 
  \be
  \vec q=\Big( \frac{1}{a},\frac{1}{b},\frac{d-1}{c d-1},\frac{c-1}{c d-1}\Big)~, \qquad \mu=(a-1)(b-1)cd~.
  \ee
The reducibility condition is 
\begin{equation}
{c-1\over d-1}~\text{ or}~{d-1\over c-1} \in \mathbb{Z}~.
\end{equation}
Imposing the constraint $\sum_i q_i>1$ as well as   $b\geq a$ and $d\geq c$ to remove redundancy, one gets the infinite sequences in table~\ref{IIIinf} as well as sporadic cases we do not list but have checked are consistent with our claim.

  \begin{table}[H]
\begin{center}
\begin{tabular}{|C|C|C|C|C| }
\hline
(2,2,c,d)&\mk (2,b,2,d)&\mk  (3,b,2,2) &(2,3,3,d)&(2,3,4,d) \\ \hline
(2,3,5,d)&(2,3,6,d)&(2,4,3,d)&(2,5,3,d)&(2,6,3,d)\\ \hline
\mk (2,b,3,3)& \mk (3,3,2,d)&\mk (3,4,2,d)& \mk (3,5,2,d)& \mk (3,6,2,d) \\ \hline
(2,4,4,d)&\mk (4,4,2,d)&(3,3,3,d)&~&~\\ \hline
\end{tabular}
\caption{Infinite sequences of type III singularities. Gray entries are reducible to the previous types of singularities. This table is adapted from  \cite{Xie:2015rpa}.}
\label{IIIinf}
\end{center}
\end{table}

In this case, the monomial basis of the Milnor ring can be obtained from that in $\mathsf X_{a,b}$ and $\mathsf Z_{c,d}$ in Table~\ref{XYZtype}. A necessary condition for being isolated is given by equations of the form \eqref{Xtype} and  \eqref{Ztype}.

  \para{(2,2,c,d)}
 A necessary  condition for the theory being isolated is $\gcd(c-1,d-1)=1$. In this case, the weight vector is irreducible and  $r_i=1$, which implies trivial 1-form symmetry.
   
\para{(2,3,3,d)} A necessary condition for being isolated is $\gcd(2,d-1)=1$.  Then one finds  $g_i=0$, implying no 1-form symmetry. 
   
\para{(2,3,4,d)}  A necessary condition for being isolated is $\gcd(3,d-1)=1$.  Then one finds  $g_i=0$, implying no 1-form symmetry. 
         
                      \para{(2,3,5,d)} A necessary condition for being isolated is $\gcd(4,d-1)=1$, which is equivalent to   $\gcd(2,d-1)=1$. Furthermore, consider $yu^2v^k$. $Q=1$ leads to $ 1 +   d =3 k  $. So another necessary condition is  $\gcd(3,d+1)=1$. Then one finds  $g_i=0$, implying no 1-form symmetry.

                \para{(2,3,6,d)}  The weight is is irreducible and  $g_i=0$, so the 1-form symmetry is trivial. 
               
          \para{(2,4,3,d)}     A necessary condition for being isolated is $\gcd(2,d-1)=1$.  One finds $ g_2=0, r_1=r_3=r_4=1$,  so the 1-form symmetry is trivial.

               \para{(2,5,3,d)}         A necessary condition for being isolated is $\gcd(2,d-1)=1$.     Consider $y^3 v^k$.         $Q=1 $ is equivalent to  $3d-1=5k$. So to be isolated, one should  also have $\gcd(3d-1, 5)=1$.  Then   $g_i=0$, so 1-form symmetry is trivial.

                  \para{(2,6,3,d)}      A necessary condition for being isolated is $\gcd(2,d-1)=1$. Then we have                  $r_1=r_3=r_4=1,  g_2=  0$, implying trivial 1-form symmetry. 
 
   \para{(2,4,4,d)}       We have $g_1=g_2=0, r_3=r_4=1$,  implying trivial 1-form symmetry.

     \para{(3,3,3,d)}                 We have       $r_i=1$ and thus trivial 1-form symmetry.

\bigskip
\noindent
{\bf Type IV:} In this case  
    \be
  W=x^a + x y^b + u^c + u v^d~.
  \ee
  We require $a, c \geq 2$ and $ b,d\geq 1$ in order to get an isolated singularity at the origin. The weights and Milnor number are given by 
  \be
  \vec q=\left( \frac{1}{a},\frac{a-1}{a b},\frac{1}{c},\frac{c-1}{c d}\right)~, \qquad \mu= (a(b-1)+1) (c(d-1)+1)~.
  \ee
  The reducibility condition is
  \begin{equation}
{b \over a -1 }\in \mathbb{Z}~~\text{or}~~{d\over c-1}\in \mathbb{Z}~.
\end{equation}  
 Imposing the constraint  $\sum_i q_i>1$ as well as $d\geq b$, $b\ge a, d\ge c$, one gets the infinite sequences in table~\ref{IVinf} as well as sporadic cases we have checked separately. Note that exchanging $a\leftrightarrow b$ or  {$ c\leftrightarrow d$} gives rise to different SCFTs in general.

\begin{table}[h]
\begin{center}
\begin{tabular}{|C|C|C|C|C| }
\hline
  (a,1,c,d)&   (a,1,c,1)& \mk (2,2,2,d) &\mk (2,2,3,d)&\mk (2,2,4,d) \\ \hline
 \mk (2,b,2,d)& \mk   (2,3,3,d)&(\mk 2,b,3,3)& \mk(2,b,3,4)& \\ \hline
\end{tabular}
\\[10pt]
\begin{tabular}{|C|C|C|C|C|C| }
\hline
 &  & & & \\ \hline
\mk(a,2,2,d)&    \mk (3,2,3,d)&(a,2,3,3)& \mk   (a,2,3,4)& \\ \hline
\end{tabular}
\\[10pt]
\begin{tabular}{|C|C|C|C|C| }
\hline
 &  & \mk (2,2,c,2) & \mk (2,2,c,3)&\mk(2,2,c,4) \\ \hline
\mk(2,b,c,2)& \mk (2,3,c,3)& &\mk  (2,b,4,3)& \\ \hline
\end{tabular}
\caption{Infinite sequences of type IV singularities. The second and third tables are obtained from the first by the exchanges $(a,b)$ and $(c,d)$ respectively. The empty entries correspond to either $a=1$ or $c=1$ or invariance under exchange. Gray entries are reducible to previous types of singularities. This table is adapted from  \cite{Xie:2015rpa}.}
\label{IVinf}
\end{center}
\end{table}

The monomial basis in this case can be obtained from that in $\mathsf Y_{a,b}$ and $\mathsf Y_{c,d}$ in Table~\ref{XYZtype}.
 
    \para{(a,1,c,d)}  We can rewrite the weight vector in an irreducible form 
  \be
  \vec q=\left( \frac{1}{a},\frac{a-1}{a  },\frac{1}{c},\frac{c-1}{c d}\right)
  =\left( \frac{1}{a},\frac{a-1}{a  },\frac{1}{c},\frac{\tilde c}{c \tilde d}\right)~, \qquad
  \gcd(\tilde c, c)=\gcd(\tilde c, d)=1~.
  \ee
Then we can compute $r_1=r_2=r_3=1,g_4=0$. Therefore there is no 1-form symmetry.

   \para{(a,1,c,1)}  In this case, the weight vector is irreducible. Then we find $r_i=1$, so there is no 1-form symmetry. 
  
      \para{(a,2,3,3)} 
      For $a=2$, the weight is irreducible. Then we can compute $g_1=g_2=g_4=0, r_3=1$, giving rise to trivial 1-form symmetry.
                   
 If $a>2$,  consider $x^k u$ with $k=0, \cdots, a-1$ in the Milnor ring. $Q=1$ leads to $2a=3k$. It is then easy to figure out a necessary condition for being isolated is $\gcd(a,3)=1$.         
     
     If $a$ is even, the weight  vector
     \be
  \vec q=\left(   \frac{1}{a},\frac{a-1}{2 a},\frac{1}{3},\frac{2}{9}\right)
     \ee
      is irreducible. Then one finds $r_1=r_3=1$, $g_2=g_4=0$ due to $\gcd(3,a)=1$. So the 1-form symmetry is trivial. 
      
      If $a$ is odd, $a=2p+1$,         the weight vector becomes 
      \be
      \vec q=\left(\frac{1}{2 p+1},\frac{p}{2 p+1},\frac{1}{3},\frac{2}{9}\right)
      \ee      
      and is irreducible. Then we find $r_1=r_2=r_3=1,g_4=0$. Again, the 1-form symmetry is trivial. 
 
\bigskip
\noindent
{\bf Type V:} Here the singularity is given by the polynomial 
  \be
W=x^a y + x y^b + u^c + u v^d~,
  \ee
and we need  $a, b, c\geq 2,~d\geq 1$ to ensure an isolated singularity at the origin.  The weights and Milnor number are given by 
\be
\vec q=\left( {b-1\over ab-1}~, {a-1\over a b-1},{1\over c},{c-1\over cd}\right), \qquad 
\mu=ab(c(d-1)+1)~.
\ee
The reducibility condition is 
  \begin{equation}
{a-1 \over b -1}\in \mathbb{Z}~\text{ or }{b-1 \over a-1}\in \mathbb{Z}~\text{ or }~{d\over c-1}\in \mathbb{Z}~.
\end{equation}
   
Imposing $\sum_i q_i>1$ and $d\ge c, b\ge a$ leads to table~\ref{Vinf} as well as sporadic cases we have checked separately. In general, exchanging $c\leftrightarrow d$ gives different SCFTs.

 \begin{table}[h]
\begin{center}
\begin{tabular}{|C|C|C|C|C| }
\hline
(a,b,c,1)&\mk (2,2,2,d)&\mk (2,2,3,d)&\mk (2,b,2,d)&\mk (3,b,2,2) \\ \hline
\mk (4,b,2,2)&\mk (2,b,3,3)&\mk(2,b,3,4)&\mk(3,3,2,d)&\mk(3,b,2,3)\\ \hline
\end{tabular}
\\[10pt]
\begin{tabular}{|C|C|C|C|C| }
\hline
 &\mk (2,2,c,2)&\mk (2,2,c,3)&\mk (2,b,c,2)& \\ \hline
 & &\mk(2,b,4,3)&\mk(3,3,c,2)&\mk (3,b,3,2)\\ \hline
\end{tabular}
\caption{Infinite sequences of type V singularities. The second table is obtained from the first by exchanging $c,d$. The empty entries are either invariant under exchange or have $c=1$. Gray entries are reducible to previous types of singularities. This table is adapted from  \cite{Xie:2015rpa}.}
\label{Vinf}
\end{center}
\end{table}

A monomial basis of the Milnor ring  can be obtained from that in $\mathsf Z_{a,b}$ and $\mathsf Y_{c,d}$ in Table~\ref{XYZtype}.

    \para{(a,b,c,1)} From \eqref{Ztype}, we learn that a necessary condition to have an isolated theory is $\gcd(a-1,b-1) =1$. The weight vector is then irreducible, and $r_i=1$. Therefore  the theory has no 1-form symmetry.

\bigskip
\noindent
{\bf Type VI:}  
Here
\be
W=x ^a y+xy^b+u^c v+u v^d~,
\ee
and  $a, b, c,d\geq 2$ in order to have an isolated singularity at the origin. The weights and Milnor number are given by 
\be
\vec q=\left( {b-1\over ab-1}, {a-1\over a b-1},{d-1\over c d-1},{c -1\over cd-1}\right)~ , \qquad \mu=abcd~.
\ee

Imposing $\sum_i q_i>1$, $b\geq a$, $d\geq c$, and
$c\geq a$ leads to table~\ref{VIinf} and sporadic solutions (which we have checked separately).  However, all of the entries in the table are reducible to previous types.
 
  \begin{table}[h]
\begin{center}
\begin{tabular}{|C|C|C|C| }
\hline
\mk (2,2,2,d)& \mk  (2,2,3,d)& \mk  (2,b,2,d)&\mk  (2,b,3,3) \\ \hline
\end{tabular}
\caption{Infinite sequences of type VI singularities. Gray entries are reducible to previous types of singularities. This table is adapted from  \cite{Xie:2015rpa}.}
\label{VIinf}
\end{center}
\end{table}

 \bigskip
 \noindent
 {\bf Type VII:} The singularity has the form 
    \be
W=x^a+y^b+y u^c+uv^d~.
\ee
Here $a , b\geq 2, c , d\geq 1$ in order to have an isolated singularity at the origin.

The weights and Milnor number are 
\begin{equation}
\vec q=\left( {1\over a}, {1\over b},{b-1\over b c},{b(c-1)+1\over bcd}\right)~, \qquad 
\mu=(a-1) (bc(d-1)+b-1)~.
\end{equation}
The singularity can be reduced to     type II if
\begin{equation}
{c\over b-1}\in \mathbb{Z}~ .
\end{equation} 

Solving $\sum_i q_i>1$ and $d\ge b$, one gets the infinite sequences in table~\ref{VIIinf} and sporadic solutions (which we have checked separately). In general, exchanging $b\leftrightarrow d$ leads to different SCFTs. 

  \begin{table}[h]
\begin{center}
\begin{tabular}{|C|C|C|C|C| }
\hline
  (a,b,1,d)&(a,b,c,1)&(2,b,2,d)&\mk  (2,2,c,d)&\mk  (3,2,2,d)\\ \hline
\mk  (4,2,2,d)&\mk  (a,2,2,2)&\mk  (a,2,2,3)&\mk  (a,2,c,2)&(2,3,3,d) \\ \hline
(2,4,3,d)&(3,3,2,d)&(2,3,4,d)&(2,3,c,3)&(2,3,c,4) \\ \hline
(2,3,c,5)&(2,3,c,6)&\mk (3,2,3,d)&\mk (3,2,c,3)&\mk (3,2,c,4) \\ \hline
\mk  (3,2,c,5)&\mk  (3,2,c,6)&\mk  (4,2,c,3)&\mk  (5,2,c,3)&\mk  (6,2,c,3) \\ \hline
(2,4,c,4)&(4,2,c,4)&(3,3,c,3)&& \\ \hline
\end{tabular}
\\[10pt]
\begin{tabular}{|C|C|C|C|C| }
\hline
 &\mk(a,1,c,d)& &   (2,b,c,2)&   (3,b,2,2)\\ \hline
   (4,b,2,2)& & \mk  (a,3,2,2)& &(2,b,3,3) \\ \hline
(2,b,3,4)&(3,b,2,3)&(2,b,4,3)& &(2,4,c,3) \\ \hline
(2,5,c,3)&(2,6,c,3)&  (3,b,3,3)&  (3,3,c,2)&  (3,4,c,2) \\ \hline
   (3,5,c,2)&   (3,6,c,2)&   (4,3,c,2)&   (5,3,c,2)&   (6,3,c,2) \\ \hline
 &(4,4,c,2)& && \\ \hline
\end{tabular}
\caption{Infinite sequences of type VII singularities. The second table is obtained from the first by  exchanging $b,d$. Gray entries are reducible to previous types of singularities or $b=1$. This table is adapted from  \cite{Xie:2015rpa}.}
\label{VIIinf}
\end{center}
\end{table}

    The monomial basis of the Milnor ring is $    x^m y^n u^k v^l$ where
    \be
 m=0, \cdots, a-2, \quad
  n=0, \cdots, b-1, \quad k=0, \cdots, c-1, \quad l=0, \cdots d-2~,
    \ee
    or
    \be
     m=0, \cdots, a-2, \quad  n=0, \cdots ,b-2, \quad  k=0,\quad l=d-1~.
    \ee

        \para{(a,b,1,d)}      The weight vector is irreducible and one finds $r_2=r_3=1, g_1=g_4=0$. So the 1-form symmetry is trivial.

       \para{(a,b,c,1)}  The weight vector can be rewritten in the form 
    \be
  \vec q=\left(   \frac{1}{a},\frac{1}{b},\frac{b-1}{b c},\frac{b (c-1)+1}{b c}\right) 
  =\left(   \frac{1}{a},\frac{1}{b},\frac{\tilde b }{b \tilde c},\frac{b \tilde c -\tilde b}{b \tilde c}\right)~, \qquad \gcd(\tilde b,\tilde c)=\gcd(\tilde b, b)=1~,
    \ee
    which is irreducible now. 
    Then one finds $r_2=r_3=r_4=1$, $g_1=0$, implying trivial 1-form symmetry.

         \para{(2,b,2,d)}    
Consider $y^k u$ with $k=0, \cdots b-1$. $Q=1$ leads to $b+1=2k$. So a necessary condition for being isolated is $\gcd(2, b+1)=1$, which is equivalent to $\gcd(2,b)=2$. The weight vector can be written in an irreducible form 
\be
\vec q=\left(\frac{1}{2},\frac{1}{b},\frac{b-1}{2 b},\frac{\tilde b}{2 b \tilde d}\right), \qquad
\gcd(\tilde b,b)=\gcd(\tilde b,d)=\gcd(\tilde b,d)=1~.
\ee
 Then $r_1=r_2=r_3=0, g_4=0$, implying trivial 1-form symmetry.

         \para{(2,3,3,d)}    
Consider $ y v^k$. $Q=1$ leads to $6d=7k$. So a necessary condition for being isolated is $\gcd(7,d)=1$. Then the weight
\be
q=\left(\frac{1}{2},\frac{1}{3},\frac{2}{9},\frac{7}{9 d}\right)
\ee 
is irreducible. As a result,  $r_2=r_3=1, g_1=g_4=0$, implying trivial 1-form symmetry. 

         \para{(2,4,3,d)}   In this case, $y^2 u^2$ has $Q=1$, so the theory is not isolated.  
        
         \para{(3,3,2,d)}   This theory is not isolated because it has an exactly marginal deformation $xyu$ with $Q=1$.

         \para{(2,3,4,d)}    One can show that  $y^2 u^2$ with $Q=1$ is in the Milnor ring. So this theory is not isolated. 

         \para{(2,3,c,3)  , (2,3,c,4) , (2,3,c,5) ,  (2,3,c,6)  } 
        
     Consider $y^2 u^k$.  $Q=1$ leads to $c=2k$. So a necessary condition for being  isolated   is $\gcd(c,2)=1$.

        Then one can show  that   for $d=2,3,4,6$,   $\gcd(3c-2,d)=1$.   For $d=5$, consider $u^k v^3$. $Q=1$ leads to  $3(c+1)=5k$. So a necessary condition for being  isolated   is $\gcd( c+1,5)=1$. Then $\gcd(3c-2,d)=1$ also holds for $d=5$. 
        
        In all cases, the weight 
         \be
\vec q=\left(\frac{1}{2},\frac{1}{3},\frac{2}{3 c},\frac{3 c-2}{3 c d}\right)~,
\ee
is irreducible, and we have $r_2=r_3=1, g_1=g_4=0$, which gives trivial 1-form symmetry.

  \para{(2,4,c,4)} Consider  $y^3 u^k$. $Q=1$ leads to $c=3k$. So we get a necessary condition for being isolated:  $\gcd(3,c)=1$.  Then the weight is irreducible and $r_1=r_2=r_3=1,g_4=0$, giving no 1-form symmetry.

 \para{(4,2,c,4)} In this case  $x^2 y $ has $Q=1$, and we see this theory is not isolated.  

 \para{(3,3,c,3)} In this case $xy^2 $  has $Q=1$, and we again see this theory is not isolated.  
 
 \para{(2,b,c,2)}

From $y^b+ yv^c$, which is of  $\mathsf Y$ type, we learn a necessary condition for being isolated \eqref{Ytype}:
$\gcd(b-1, c)=1$ or $c=2$.

For $c=2$, consider $y^k u$. $Q=1$ leads to $b+1=2k$. So another necessary condition for the theory to be isolated is $\gcd(2,b+1)=1$, which is equivalent to $\gcd(2,b)=2$. The weight is also irreducible. Then $r_1=r_2=r_3=1,g_4=0$, meaning trivial 1-form symmetry.

In the case of  $\gcd(b-1,c)=1$, one can check that $1-b+bc$ is always odd.  Furthermore, we have $\gcd(1-b+bc,2bc)=  1$. Therefore, the weight
\be
\vec q=\left(\frac{1}{2},\frac{1}{b},\frac{b-1}{b c},\frac{b c-b+1}{2 b c}\right)~,
\ee
 is irreducible. 
Then
 $ r_1=r_2=r_3=1$, $2g_4=\gcd(2,bc)-\gcd(2,b)$.  
If $\gcd(b,2)=1$ then $\gcd(c,2)=1$ because $\gcd(b-1,c)=1$ and thus $g_4=0$. 
If $\gcd(b,2)=2$, we also have $ g_4=0$. As a result, the 1-form symmetry is always trivial. 

 \para{(3,b,2,2)} Consider $y^k u$. $Q=1$ leads to $b+1=2k$. So another necessary condition to be isolated is $\gcd(2,b+1)=1$.  The weight is irreducible, and $g_1=g_2=g_4=0, r_3=1$. Therefore, we have trivial 1-form symmetry.

 \para{(4,b,2,2)} As in the previous case, we still have even $b$ which is a necessary condition for being isolated. But then $x^2 y^{b/2}$ has $Q=1$. So this theory is not isolated.
 
  \para{(2,b,3,3) } Considering the $y,u$ part gives a necessary condition for being isolated: $\gcd(b-1,3)=1$. The weight vector  is irreducible, and we have $g_1=g_2=g_4=0, r_3=1$. As a result, the 1-form symmetry is trivial.

 \para{(2,b,3,4) } Again, we have a necessary condition of being isolated: $\gcd(b-1,3)=1$. The weight   is irreducible and  $r_1=r_2= r_3=1,g_4=0$, implying no 1-form symmetry.

 \para{(3,b,2,3) }
 
If $b=2$, then $yuv$ is marginal. If $b=3$, then $xyu$ is marginal.  So in order to have an isolated theory, we need to have $b>3$ and a necessary condition is $\gcd(3,b)=1$ \eqref{Xtype}.

Consider $y^ku$ with $k=0, \cdots, b-1$. $Q=1$ leads to $b+1=2k$, so another necessary condition to be isolated is $\gcd(b+1,2)=1$.
Next, consider $y^kuv$. $Q=1$ leads to $b+1=3k$, so a further necessary condition for an isolated theory is $\gcd(b+1,3)=1$.
Then  the weight is irreducible and  $r_1=r_2=r_3=1, g_4=0$.  The 1-form symmetry is thus trivial.

 \para{(2,b,4,3) } A necessary isolated condition is $\gcd(b-1,4)=1$, which is equivalent to  $\gcd(b,2)=2$. This guarantees that the weight is irreducible.   Then we find $r_1=r_2=r_3=1,2g_4= 0$, and the 1-form symmetry is trivial.

 \para{(2,4,c,3) } A necessary condition for being isolated is $\gcd(3 ,c)=1$.  Then one finds the weight is irreducible. We can compute $r_1=r_2=r_3=1, g_4= 0$, implying the 1-form symmetry is trivial.

 \para{(2,5,c,3) } The first necessary condition for being isolated is  $\gcd(4 ,c)=1$, which is equivalent to $\gcd(2,c)=1$. Consider $y^2 u^kv$. $Q=1$ leads to $c+1=3k$. Therefore, another necessary condition for being   isolated   is $\gcd(c+1,3)=1$.  The weight is irreducible and  $r_2=r_3=1, g_1=g_4=0$,  implying no 1-form symmetry.

 \para{(2,6,c,3) } A necessary condition for being isolated is  $\gcd(5 ,c)=1$. 
The weight is irreducible and  $r_1=r_2=r_3=1, g_4=0$,  implying no 1-form symmetry.

 \para{(3,b,3,3) }   For $b=2$, $xyu$ has $Q=1$, so this theory is not isolated. For $b=3$, one can check this theory is also not isolated.
 
 If $b>3$, two necessary conditions for being isolated are 
  $\gcd(3,b)=1$ and $\gcd(b-1,3)=1$, as one can see from $x,y$ part and $y,u$ part. 
Consider $xy^k u$. $Q=1$ leads to $b+1=3k$. So another necessary  condition for being isolated is $\gcd(b+1,3)=1$. These three conditions cannot hold simultaneously. As a result, this theory is always part of a conformal manifold.

 \para{(3,3,c,2) } A necessary condition for being isolated   is $\gcd(2,c)=1$. Then  the weight is irreducible and   $r_1=r_2=r_3=1, g_4=0$, implying no 1-form symmetry.

 \para{(3,3,c,2),(3,4,c,2),(3,5,c,2),(3,6,c,2)} In these $(3,b,c,2)$ theories, a necessary condition for being isolated is $\gcd(b-1,c)=1$ as can be seen from $y^{b-1} u^k$. Then one can show that $\gcd(bc-b+1, 2bc)= \gcd(bc-b+1,2)$, which can be further shown to be 1 for $b=3,4,5,6$. So in all cases, the weight is irreducible, and $r_2=r_3=1, g_1=0, 2g_4=\gcd(3,bc)-\gcd(3,b)$. For $b=3,4,6$, one can easily show that $g_4=0$. For $b=5$, consider $xy^2 u^k$. $Q=1$ leads to $c=3k$. So another necessary isolated condition is $\gcd(c,3)=1$, which gives $g_4=0$. 

 Therefore, in all cases the 1-form symmetry is trivial. 

 \para{(4,3,c,2) } 
A necessary    condition for being isolated is $\gcd(2,c)=1$. Then  the weight is irreducible and $r_2=r_3=1,g_1=0,g_4=0$. This gives no 1-form symmetry.

 \para{(5,3,c,2) } The first condition for being isolated is $\gcd(2,c)=1$. Then  the weight is irreducible. 
Furthermore, consider $x^2 y u^k$. $Q=1$ leads to $2c=5k$. So another   condition for being isolated is $\gcd(5,c)=1$. 
 This gives $r_2=r_3=1,g_1=g_4=0$ and therefore no 1-form symmetry.

 \para{ (6,3,c,2) ,(4,4,c,2) }
These theories have conformal manifolds because of the $x,y$ part.

  \bigskip
\noindent
{\bf Type X: } The singularity is given by
  \be\label{typeX}
  W=x^a + y^b u + u^c v + y v^d~,
  \ee
and we require $a\geq 2$ and $b,c,d\geq 2$  to get  an isolated singularity at the origin. The weights and Milnor number are given by
    \be
   \vec q=\left( \frac{1}{a},\frac{(c-1) d+1}{b c d+1},\frac{b (d-1)+1}{b c d+1},\frac{(b-1) c+1}{b c d+1} \right)~, \qquad
    \mu=(a-1)bcd~.
    \ee
The singularity is reducible to a previous type if
\begin{equation}
{b(d-1)+1\over d(c-1)+1}\in \mathbb{Z}~,~~\text{or}~~{c(b-1)+1\over b(d-1)+1}\in \mathbb{Z}~~\text{or}~~{d(c-1)+1\over c(b-1)+1}\in \mathbb{Z}~.
\end{equation}

Imposing $\sum_i q_i>1$ and $b\leq c\leq d$ leads to the infinite sequences in table~\ref{Xinf}  and the sporadic solutions we have checked separately. Note that \eqref{typeX} is cyclically symmetric in $b,c,d$. In general, there is another inequivalent SCFTs obtained by exchanging any two in $b,c,d$.

  In this case, the Milnor ring is generated by the following monomials:
    \be
    x^m y^p u^k v^l~, \qquad m=0, \cdots a-2~, \quad p=0, \cdots, b-1~,  \quad  k=0, \cdots, c-1~, \quad l=0, \cdots, d-1~. 
    \ee

 \begin{table}[h]
\begin{center}
\begin{tabular}{|C|C|C|C|C| }
\hline
\mk  (a,1,c,d)&(2,2,c,d)&(3,2,2,d)&  (4,2,2,d) &\mk (a,2,2,2)\\ \hline
(2,3,3,d)&(2,3,4,d)&(2,3,c,4)&(3,2,3,d)&(3,2,c,3) \\ \hline
\end{tabular}
\caption{Infinite sequences of type X singularities. The table is invariant under exchanging   two of $b,c,d$.  Gray entries are reducible to previous types of singularities. This table is adapted from  \cite{Xie:2015rpa}.}
\label{Xinf}
\end{center}
\end{table}
   
           \para{(2,2,c,d)} 
    \be
  \vec  q=\left( \frac{1}{2},\frac{c d-d+1}{2 c d+1},\frac{2 d-1}{2 c d+1},\frac{c+1}{2 c d+1}\right)~. 
       \ee
 Consider $y   u^k v^l$. $Q=1$ leads to   $(1 + c) (d - l) = (2 d - 1) k$. So a necessary condition for being isolated is  $      \gcd(c+1,2d-1)=1 $.  Then one can show that the weight above is irreducible.  This enables us to find $g_i=0$, implying that the 1-form symmetry is trivial.

             \para{(3,2,2,d)}
    
  \be
 \vec  q=\left( \frac{1}{3},\frac{d+1}{4 d+1},\frac{2 d-1}{4 d+1},\frac{3}{4 d+1} \right)
  \ee
Consider $yu v^k$. $Q=1$ leads to $d+1=3k$. So a necessary condition for being isolated is  $      \gcd(3,d+1)=1 $.  Then one can show that the weight above is irreducible.  As a result  $g_1=0,r_2=r_3=r_4=1$, implying that the 1-form symmetry is trivial.

               \para{(4,2,2,d)}
As in the   above $(3,2,2,d)$ case, a necessary condition for being isolated is  $      \gcd(3,d+1)=1 $.  Then one can show that the weight above is irreducible.  As a result  $g_i=0 $, implying that the 1-form symmetry is trivial.

                \para{(2,3,3,d)}
   
  \be
 \vec q =\left(   \frac{1}{2},\frac{2 d+1}{9 d+1},\frac{3 d-2}{9 d+1},\frac{7}{9 d+1} \right)
  \ee   
 Consider $y  u^2 v^k$. $Q=1$ leads to $d+4=7k$. Therefore, a necessary condition for being isolated is $\gcd(7,d-3)=1$. Then the weight vector can be shown to be irreducible, and we find  $r_2=r_3=r_4=1, 2g_1=0$, which implies trivial 1-form symmetry.
   
                  \para{(2,3,4,d)}
   \be
   \vec q=\left(\frac{1}{2},\frac{3 d+1}{12 d+1},\frac{3 d-2}{12 d+1},\frac{9}{12 d+1}\right)~.
   \ee
    The weight vector above is always  irreducible, and we have $r_2=r_3=r_4=1,g_1=0$, giving trivial 1-form symmetry.

                     \para{(2,3,c,4)}

    \be
    \vec q=\left( \frac{1}{2},\frac{4 c-3}{12 c+1},\frac{10}{12 c+1},\frac{2 c+1}{12 c+1}  \right)~.
    \ee
      Consider $yv^2 u^k$. $Q=1$ leads to $2c+1=5k$. So a necessary condition for being isolated is $\gcd(5,2c+1)=1$. As a result, the weight vector above is irreducible.  Then we find $g_i=0$, which implies trivial 1-form symmetry.     
    
                \para{(3,2,3,d )} 
    
    \be
\vec q=\left(    \frac{1}{3},\frac{2 d+1}{6 d+1},\frac{2 d-1}{6 d+1},\frac{4}{6 d+1}\right)~.
    \ee
The weight above is irreducible and leads to $g_i=0$, implying no 1-form symmetry.

                \para{(3,2,c,3  )}   
    \be
    \vec q=\left(\frac{1}{3},\frac{3 c-2}{6 c+1},\frac{5}{6 c+1},\frac{c+1}{6 c+1}\right)
    \ee    
    Consider $yv^2 u^k$. $Q=1$ leads to $c+1=5k$. So a necessary    condition for being isolated is $\gcd(c+1,5)=1$.   The weight above is then  irreducible  and $g_i=0$. Therefore, the 1-form symmetry is trivial.

 \bigskip
 \noindent
{\bf Type XI:} The singularity is given by 
  \be
  W= x^a+ x y^b +y u^c + uv^d~,
 \ee
 and we require $a\geq 2$, $b,c,d\geq 1$ in order to get an isolated singularity at the origin. The weights and Milnor number are given by  
 \beqn
 \vec q =\Big( \frac{1}{a},\frac{a-1}{a b},\frac{a (b-1)+1}{a b c},\frac{a b (c-1)+(a-1)}{a b c d} \Big) ~, \qquad
 \mu=abc(d-1) +a(b-1)+1~. \qquad
 \eeqn
The singularity can be reduced to the previous type VII  if
\begin{equation}
{b\over a-1}\in \mathbb{Z}~.
\end{equation}

Imposing $\sum_i q_i>1$, one gets the infinite sequences in table~\eqref{XIinf} and sporadic solutions we have checked separately.

\begin{table}[h]
\begin{center}
\begin{tabular}{|C|C|C|C|C| }
\hline
(a,1,c,d)&(a,b,1,d)&(a,b,c,1)&\mk  (2,2,2,d)&\mk  (2,2,3,d) \\ \hline
\mk  (2,2,c,2)&\mk  (2,2,c,3)& \mk  (2,2,c,4)&\mk  (2,b,2,d)&\mk (3,2,2,d) \\ \hline
(a,2,c,2)&(a,2,2,2)&(a,2,2,3)&(3,b,2,2)&(4,b,2,2) \\ \hline
(a,3,2,2)&\mk (2,3,c,3)&\mk (3,2,c,3)&\mk (2,b,3,3)&\mk (2,b,3,4) \\ \hline
\mk (2,b,4,3)&(3,3,b,2)&\mk  (3,4,c,2)&\mk (4,3,c,2)&(3,b,2,3) \\ \hline
(3,b,3,2)&&&&\\ \hline
\end{tabular}
\caption{Infinite sequences of type XI singularities. Gray entries are reducible to previous types of singularities. This table is adapted from \cite{Xie:2015rpa}.}
\label{XIinf}
\end{center}
\end{table}

 The monomial basis of the Milnor ring is given by 
\be
x^m y^n u^k v^l, \qquad m=0, \cdots, a-1, \quad n=0, \cdots, b-1, \quad k=0, \cdots, c-1, \quad  l=0, \cdots, d-2~,
\ee
or
\be
x^m y^n   v^{d-1}, \qquad m=0, \cdots, a-1, \quad n=0, \cdots,b-2, \quad k=0,   \quad  l= d-1~,
\ee
or 
\be
 b^{b-1} v^{d-1}=y^n v^l, \qquad m=0, \quad n=  b-1, \quad k=0,  \quad  l=d-1~.
\ee

 \para{(a,1,c,d)}
 
  \be
 \vec q =\left( \frac{1}{a},\frac{a-1}{a},\frac{1}{a c},\frac{a c-1}{a c d} \right)
 \ee
 Consider $u^k v^l$ with $k=0, \cdots ac-1, \; l =0, \cdots,d-2$. The condition $Q=1$ leads to 
   \be
 \frac{k'}{ac-1}=\frac{l}{d}~, \qquad k'=ac-k=1, \cdots ac~.
 \ee
 So a necessary condition for being isolated is $\gcd(ac-1,d)=1$ or $d=2$. 
 
 If $\gcd(ac-1,d)=1$, then the weight above can be shown to be irreducible, and we have $r_1=r_2=r_3=1,g_4=0$,  which implies trivial 1-form symmetry.
 
 If $d=2$ and $ac$ is even,  then the weight above  is  irreducible, and we have $r_1=r_2=r_3=1,g_4=0$,  which implies trivial 1-form symmetry.
 
 If $d=2$ and $ac$ is odd, $ac=2p+1$, then the weight vector becomes 
 \be
 \vec q=\left(\frac{1}{a},\frac{a-1}{a},\frac{1}{2 p+1},\frac{p}{2 p+1}\right)~.
 \ee
 and is irreducible. One can compute $r_i=1$, implying trivial 1-form symmetry.
 
   \para{(a,b,1,d)}
 
  \be
 \vec q=\left( \frac{1}{a},\frac{a-1}{a b},\frac{a (b-1)+1}{a b},\frac{a-1}{a b d}\right)
 \ee
 For $b=d=1$,  the weight vector is irreducible and $r_i=1$, so there is no 1-form symmetry.
  For $b=1,d=2$ or $b=2,d=1$, one can also check that  the 1-form symmetry is trivial. 
  
For $bd>2$, consider $x^k v^l$ with $k=0,\cdots a-1, \; l=0, \cdots bd-2$. $Q=1$ leads to 
 \be
 \frac{l}{bd}=\frac{k'}{a-1}~, \qquad k'=a-k=1, \cdots, a~.
 \ee
 So a necessary condition for being isolated is  $\gcd(bd, a -1)=1$, which is equivalent to  $\gcd(b  , a -1)=\gcd( d, a -1)=1$. Then one can show that the weight  vector above is irreducible, and one can find  $r_1=r_2=r_3=1  ,g_4=0$. As a result, the 1-form symmetry is trivial.  
  
     \para{(a,b,c,1)}
 
  \be
 \vec q=\left(  \frac{1}{a},\frac{a-1}{a b},\frac{a b-(a-1)}{a b c},\frac{abc-a b+(a-1) }{a b c}\right) 
  \ee
 Consider $x^k y^l$. $Q=1$ leads to 
  \be
 \frac{k'}{a-1}=\frac{l}{b}~, \qquad k'=a-k~.
 \ee
Therefore,  a necessary condition for being  isolated   is $\gcd(a-1, b)=1$. 

We can rewrite the weight vector as
 \be
 \vec q=\left(  \frac{1}{a},\frac{a-1}{a b},\frac{\tilde s}{a b \tilde c}, \frac{ab\tilde c-\tilde s}{a b \tilde c} \right)~, 
 \qquad \frac{a b-(a-1)}{      c}=\frac{\tilde s}{  \tilde c}~, \qquad \gcd(\tilde s, \tilde c)=\gcd(\tilde s, ab )=1~.
 \ee
 Then the above weight vector is irreducible. One can show that $r_i=1$, which implies no 1-form symmetry.

    \para{(a,2,c,2)}
 
 \be
 q=\left(\frac{1}{a},\frac{a-1}{2 a},\frac{a+1}{2 a c}, \frac{ 2 a c-a-1}{4 a c} \right)
 \ee
If $c=1$, we can check that the 1-form symmetry is always trivial. This can be seen by observing that $q_y+q_u=1$ and using the lemma in section \ref{LGmass}.
 
 For $a,c>1$, consider $x^l u^k$ with $l=0, \cdots, a-1$, $k=0, \cdots, c-1$. $Q=1$ leads to $(a+1)k=2c(a-l)$ which can be rewritten as
 \be
 \frac{l'}{a+1}=\frac{k}{2c}, \qquad l'=a-l=1, \cdots, a~.
 \ee
Then it is easy to figure out a necessary condition for being isolated   is $\gcd(a+1, 2c)=1$, which  is equivalent to $\gcd(a+1,2)=\gcd(a+1,c)=1$. Then one can show the weight is irreducible.  And we have $r_1=r_2=r_3=1,g_4=0$, which implies no 1-form symmetry.

    \para{(a,2,2,2)}  Consider $x^k y$. $Q=1$ leads to $a+1=2k$.  So a necessary condition for being isolated   is $\gcd(a+1, 2 )=1$.  Then one can show the weight vector is irreducible, and we have $r_1=r_2=r_3=1,g_4=0$, which implies no 1-form symmetry.

     \para{(a,2,2,3)}  The discussion is identical to the previous $(a,2,2,2)$ case.
 
       \para{(3,b,2,2)} 
 \be
 \vec q=\left(\frac{1}{3},\frac{2}{3 b},\frac{3 b-2}{6 b},\frac{3 b+2}{12 b}\right)~.
 \ee
 The $x,y$ part in the form of \eqref{Ytype} gives the necessary condition for being isolated $\gcd(2,b)=1$.  Then the weight vector is irreducible, and  $r_1=r_2=r_3=1,g_4=0$. Therefore, the 1-form symmetry is trivial. 
  
       \para{(4,b,2,2)} 
 The $x,y$ part in the form of \eqref{Ytype} gives the necessary condition for being isolated $\gcd(3,b)=1$.  
The weight vector is  then irreducible, and  $r_1=r_2=r_3=1,g_4=0$, implying trivial 1-form symmetry.
  
         \para{(a,3,2,2)} 
    
     The $x,y$ part in the form of \eqref{Ytype} gives the necessary condition for being isolated $\gcd(a-1,3)=1$.  
 Then one can show that the   weight  vector is irreducible, and  $r_1=r_2=r_3=1,g_4=0$, implying trivial 1-form symmetry.

          \para{(3,3,c,2)} 
         \be
         \vec q=\left( \frac{1}{3},\frac{2}{9},\frac{7}{9 c},\frac{9 c-7}{18 c} \right)~.
         \ee
      Consider $xy u^k $. $Q=1$ leads to $4c =7k $. So a necessary  condition for being isolated   is $\gcd(4c, 7)=\gcd(c,7)=1$.
      
      If $c$ is even, then the weight vector above is irreducible. One finds    $r_1=r_2=r_3=1,g_4=0$, giving no 1-form symmetry. 
      
      If $c$ is odd,  $c=2p+1$, the weight vector becomes 
      \be
      \vec q=\left(\frac{1}{3},\frac{2}{9},\frac{7}{18 p+9},\frac{9 p+1}{18 p+9}\right)~,
      \ee
      and is irreducible.  Then one finds $r_i=1$, meaning no 1-form symmetry.  
               
             \para{(3,b,2,3)} 
         \be
         \vec q=\left(  \frac{1}{3},\frac{2}{3 b},\frac{3 b-2}{6 b},\frac{3 b+2}{18 b}\right)~.
         \ee
        The $x,y$ part in the form of \eqref{Ytype} gives a necessary condition for being isolated: $\gcd(2,b)=1$.  
Then the weight vector is irreducible and  we have   $r_1=r_2=r_3=1,g_4=0$, which implies trivial 1-form symmetry.
  
              \para{(3,b,3,2)}  
             \be
          \vec   q=\left(\frac{1}{3},\frac{2}{3 b},\frac{3 b-2}{9 b},\frac{3 b+1}{9 b}\right)~.
             \ee      
 The $x,y$ part in the form of \eqref{Ytype} gives a necessary condition for being isolated: $\gcd(2,b)=1$.  
Then the weight vector is irreducible, and we have $r_i=1$, meaning trivial 1-form symmetry.

 \bigskip
 \noindent
 {\bf Type XIX:} The singularity is defined by 
\be
W= x^a  u + x y^b + v u^c + y v^d~.
\ee
 The weight vector is
 \be
 \vec q=\left(\frac{b ( d(c-1)  +1)-1}{a b c d-1},\frac{d    ( c(a-1)  +1)-1}{a b c d-1},\frac{a (b (d-1)+1)-1}{a b c d-1},\frac{c (a (b-1)+1)-1}{a b c d-1}\right)~,
 \ee
 and the Milnor number is $\mu=abcd$. There are some sporadic cases that we have checked separately, and we include the infinite sequences in table \ref{XIXinf}.
 
 \begin{table}[h]
\begin{center}
\begin{tabular}{|C|C|C|}
\hline
(1,b,c,d)&(2,b,c,2)&(2,2,c,3) \\ \hline
\end{tabular}
\caption{Infinite sequences of type XIX singularities. This table is adapted from \cite{Xie:2015rpa}.}
\label{XIXinf}
\end{center}
\end{table}

 A monomial basis in the Milnor ring is 
\be
x^m y^n u^k v^l, \qquad m=0, \cdots, a-1, \quad n=0, \cdots, b-1, \quad k=0, \cdots, c-1, \quad  l=0, \cdots, d-1~.
\ee

 \para{(1,b,c,d)}
The weight vector vector becomes 
 \be\label{wt1bcd}
 \vec q=\left( \frac{b ((c-1) d+1)-1}{b c d-1},\frac{d-1}{b c d-1},\frac{b (d-1)}{b c d-1},\frac{b c-1}{b c d-1}\right)~, \qquad b c,\, d\ge 2~.
 \ee
 The Milnor ring is
  \be
 \mathcal R=\mathbb C[v,y]/\langle b (-1)^c c \, v y^{b c-1}+v^d,  (-1)^c \, y^{b c}+d\, y v^{d-1}\rangle~.
 \ee
 The monomial basis can be chosen as
  \be
 y^k v^l~, \qquad k=0, \cdots, bc-1~, \qquad l=0, \cdots, d-1~. 
 \ee
The equation  $Q=1$ is equivalent to 
 \be
 \frac{k'}{bc-1} =\frac{l'}{d-1}~, \qquad k'=bc-k=1,\cdots bc~, \qquad l'=l-1=-1, \cdots d-2~.
 \ee

Then it is not difficult to show that  the condition for an isolated theory is 
\be
\gcd(bc-1, d-1)=1~.
  \ee
 This guarantees that the weight vector \eqref{wt1bcd} is irreducible, and one can check $r_i=1$, so the 1-form symmetry is trivial. 
 
 \para{(2,b,c,2)}
 \be
 \vec q=\left( \frac{-2 b c+b+1}{1-4 b c},\frac{2 c+1}{4 b c-1},\frac{2 b+1}{4 b c-1},\frac{-2 b c+c+1}{1-4 b c}\right)~, 
 \qquad bc\ge 2~.
 \ee

The Milnor ring has monomial basis
 \be
 y^k u^l x^m v^n~, \qquad k=0, \cdots, b-1~, \quad  l=0, \cdots, c-1~, \quad m,n=0,1~.
 \ee
The equation $Q=1$ leads to
 \be
 B (C (m+n-2)+2 l-2 m-n+2)+C (2 k-m-2 n+2)=0~,\qquad B=2b+1~, \quad C=2c+1~.
 \ee
Then one can show that the condition to be isolated is $\gcd(B,C)=\gcd(2b+1,2c+1)=1$.  This result also implies $\gcd(2c+1,c-b)=\gcd(2c+1,b+c+1 )=1$. These conditions are enough to show that the weight is irreducible.  Moreover, we can compute $r_i=1$, which implies vanishing 1-form symmetry.

 \para{(2,2,c,3)}
 
 \be
 \vec q=\Big(\frac{ 6 c-5}{ 12 c-1},\frac{3 c+2}{12 c-1},\frac{9}{12 c-1},\frac{ 3 c-1}{ 12 c-1} \Big)
 \ee
 In this case, the weight is always irreducible. Moreover, we can compute $r_i=1$, showing the absence of 1-form symmetry. 

\subsection{Singularities with more than four monomials}\label{moreM}
In this section, we consider theories with more than four monomials in $W$, following the classification of \cite{yau2005classification}. We will extend our discussion to theories arising in a refinement of this classification in \ref{beyondClass} (although, as we will see, our goal does not require us to make this classification explicit).

By analyzing the weights of each type of singularity, we find that if any of $a,b,c,d$ are allowed to equal one, then there are two weights whose sum equals one. Then using the lemma in section \ref{LGmass}, we learn that there is no 1-form symmetry. In particular, the case $\sum_i q_i>3/2$, which corresponds to isolated SCFTs  according to Claim 3 in section~\ref{conjectures}, only arises when a specific $a,b,c,d$ equals to 1.  As a consequence, Corollary 1 in section~\ref{conjectures} holds in these situations.

As a result, we only need to consider cases where all $a,b,c,d\ge 2$.  For these cases, we have numerically checked that, for all $2 \le a,b,c,d\le 13$ giving an isolated singularity,\footnote{More explicitly,  we impose $ (i)$ the Milnor number is an integer, $ (ii)$ the solution  $p,q$ (and $r,s$) to the weight 1 condition exists, and $(iii)$  $dW=0 \Leftrightarrow x_i=0$.}  the corresponding Milnor ring always has a monomial with weight 1. This numerical result   motivates us  to make the following conjecture:

 \noindent{\bf Conjecture:} For the isolated singularities in table~\ref{Full19type} defined with 5 or more monomials in $W$ and   $a,b,c,d\ge2$,   the  corresponding  Milnor ring always has a marginal  monomial with weight one.
 
In the rest of this appendix, we will prove a slightly weaker statement.  Namely, we will show that the conjecture holds in most situations, except for some special values of $a,b,c,d$. For those exceptional case we show that the singularities reduce to previous types or that the SCFTs have no 1-form symmetry. This statement is enough for us to establish a complete proof of our main claim  for SCFTs related to all these singularities. We complete our proof for the most general theories in \ref{beyondClass}.

We begin with a simple observation:

\bigskip
 \noindent
 {\bf Claim:} \emph{Consider those singularities involving 5 or more monomials in $W$. The weights are fixed by $a,b,c,d$ in table~\ref{Full19type} and are independent of the remaining parameters. If no entry in the weight vector can be written as a non-negative linear combination of the rest of the entries, namely, $\forall i, q_i\neq\sum_{j\neq i } M_{ij}q_j,\; M_{ij}\in\mathbb N $, then the theory has a conformal manifold.}
\bigskip

\noindent \emph{Proof:}  We need to consider elements of the Milnor ring (i.e., monomials which are non-trivial even when $dW=0$ is imposed). We are particularly interested in weight-one constraints since these will affect the exactly marginal elements of the Milnor ring. Such constraints take the following form
 \be
0= X_i \p_{X_i} W~, \qquad X_i\in\left\{x,y,u,v\right\}~.
 \ee
 Since we assume that  $\sum_{j\neq i} n_j q_j/q_i\neq 1$, these are the only constraints.\footnote{If, say, $q_i/q_j=k\in\mathbb N$, then we can have a further constraint, $q_j^k\p_{x_i} W$, which is trivial in the Milnor ring and has weight 1.}
 
 Therefore, at weight one there are at most four constraints.  However, for these types of singularities, by construction, we have at least five monomials with weight one appearing in $W$. As a result, there is at least one  non-trivial element with weight one in the Milnor ring. This is an exactly marginal deformation, so the theory is not isolated. \hfill $\square$
 
 \bigskip
 The more difficult situation to deal with then is one in which there are integral relations among weights:
 \be\label{qirealtion}
 q_i=\sum_{j\neq i } M_{ij}q_j~,\qquad M_{ij}\in\mathbb N~.
 \ee
We denote  the number of such distinct relations for each $i$ as $N_i\ge0$. Each such relation enables us to generate  an additional weight-one constraint  on the Milnor ring:\footnote{Some of these constraints may be redundant.} 
 \be
 0=\Big( \prod_{j\neq i } X_j^{M_{ij}}  \Big) \p_{X_i} W~.
 \ee
 As a result, we generate (at most) $N_x+N_y+N_u+N_v$ additional constraints at weight one on top of the four constraints, $X_i \p_{X_i} W$.
 
Therefore, to show that a theory of the type we are considering has a marginal deformation, it is enough to show that we can generate one more distinct weight one monomial for each relation in \eqref{qirealtion}. 
 In such a case, we are always left with at least one non-trivial element with weight one in the Milnor ring, which thus gives rise to a marginal deformation.

 \bigskip
 \noindent
 {\bf Type VIII}: Let us start with type VIII
 \be\label{WVIII}
W=x^a+y^b +y u^c+yv^d+u^p v^q~.
\ee
We want to show that for $a,b,c,d\ge 2$, the theory has a marginal deformation ($c,d\ge2$ or else the lemma of section \ref{LGmass} guarantees the 1-form symmetry is trivial; the singularity cannot be isolated unless $a,b\ge2$). Instead of directly analyzing the Milnor ring,  $\mathcal R$, of $W$, we consider the subring  $\mathcal R_{sub}=\mathcal R/\langle x\rangle$. Our strategy will be to show  that there is a weight one monomial in this subring. Therefore, the Milnor ring  $\mathcal R$ itself will have at least one weight one monomial. Practically speaking, going to the subring can be done by setting $x\to 0$. As a result, \eqref{WVIII} reduces to 
\be\label{VIIIyuv}
y^b +y u^c+yv^d+u^p v^q~.
\ee

To simplify the discussion, we introduce the notion of a \emph{monomial vector}, which  represents each  monomial in terms of a vector of entries corresponding to the power of a given variable in the monomial. For example, we can represent the monomials in \eqref{VIIIyuv} as
\beqn
V_b &=&(b, \; 0,\; 0)~, \\
V_c &=&(1, \; c,\; 0)~, \\
V_d &=&(1, \; 0,\; d)~, \\
V_{pq} &=&(0, \; p,\; q)~.  
\eeqn

There are also \emph{replacement vectors}, which represent each weight relation \eqref{qirealtion} in terms of a vector 
\beqn
S_b^j &=&(-1, \; m_j,\; n_j)~, \qquad j=1~, \cdots, N_b~,\\
S_c^j &=&(e_j, \; -1,\; f_j)~,  \qquad j=1~, \cdots, N_c~,\\
S_d^j &=&(k_j, \; l_j,\; -1)~, \qquad j=1~, \cdots, N_d~,  
\eeqn
where $m,n,e,f,k,l$ are non-negative integers subject to $\vec q \cdot S=0$. The meaning of $(-1,m,n)$ is that we can replace $y \to u^mv^n$.

Therefore, we can use the $V+S$ combinations to generate  new monomials. Each replacement vector enables us to generate one more monomial. To show that there are marginal deformations, it is enough to show that  the new monomial vectors generated in this way are all different from each other (and from the five terms appearing in $W$). 

It turns out we can consider the following special combinations:
\beqn
\tilde V_b^j &= V_b+b S_b^j \qquad\;\; &=(0, \;\star , \; \star)~, \qquad \star\neq 1~,\\
\tilde V_c^j &= V_c+(c-1) S_c^j & =(\#, \; 1 , \; *)~, \qquad   \# \ge 1~,\\
\tilde V_d^j &= V_d+(d-1) S_d^j &=(\#, \; * , \; 1)~, \qquad   \# \ge 1~,
\eeqn
where $*,\star,\#\in \mathbb N$, and $\#>0, \star \neq 1$.  Comparing all the $V,\tilde V$ above, it is not difficult to see that they are distinct except for the possibilities that $V_{pq}=\tilde V_b$ and $\tilde V_c=\tilde V_d$ (note $a,b,c,d\ge 2$).

 First consider  the case $V_{pq}=\tilde V_b$. This scenario arises if and only if 
 \be
 (0,p,q) =(b,0,0)+b(-1,m,n) \qquad\Rightarrow\qquad  p=mb~, \qquad q=nb~.
 \ee 
 We assume $p,q>0$ (if either $p=0$ or $q=0$, this theory reduces to type II, which we have dealt with in section \ref{fourM}). So $m,n\ge 1 $ and we can replace 
 \be
 V_{pq}= (0,p,q) \qquad \to  \qquad   (b,0,0)+(b-1)(-1,m,n) =(1,  \; (b-1)m,\; (b-1)n)~.
 \ee
 Since $(b-1)m,(b-1)n>0$,  the new vector may only overlap with $\tilde V_c,\tilde V_d$.  If $b\ge 3$, $(b-1)m,(b-1)n\ge 2$, and it would be different from all the rest of monomial vectors (including $\tilde V_c,\tilde V_d$).
So we only need to worry about $b=2$. 

For $b=2$, we have $y^2+yu^c+y v^d+u^p v^q$. 
In this case, the weight vector is given by 
\be\label{w81}
\vec q=\left(\frac 1 a, \frac1 2, \frac{1}{2c},\frac{1}{2d}\right)~.
\ee
This reduces to a type I theory, and it also has an exactly marginal deformation since the $u,v$ part has $\gcd(2c,2d)>1$.

For the case $\tilde V_c=\tilde V_d$, we have
\be
(1,c,0)+(c-1) (e,-1,f) =(1,0,d)+(d-1) (k,l,-1) =( *, 1,1)~.
\ee
It is easy to find that this equation implies 
\be
c=d=2~, \qquad f=l=1~.
\ee
Solving $p (b-1)/(2b) +q(b-1)/(2b)=1$, one finds that $b=2$ or $b=3$. We have already considered $b=2$. For $b=3$, we have 
\be\label{3eqnVIII}
y^3+yu^2+yv^2+uv^2~.
\ee
Since $q_y=q_u=q_v$, claim 2 (proven in appendix \ref{twoq}) implies that the one-form symmetry is trivial (it is also easy to check there is an exactly marginal deformation).
 
\bigskip
\noindent
{\bf  Type IX:}
In this case, we can again consider the subring gotten by setting $x\to 0$. For convenience, we reshuffle the variables by exchanging $y\leftrightarrow v, b\leftrightarrow d, p\leftrightarrow q$. Then the problem boils down to analyzing the subring corresponding to
\be\label{IXyuv}
 v y^b+y u^c+y v^d+u^p v^q~.
\ee
This is almost identical to \eqref{VIIIyuv} in the type VIII case except for  the first monomial, $vy^b$. The monomial vectors are
\beqn
V_b &=&(b, \; 0,\; 1)~, \\
V_c &=&(1, \; c,\; 0)~, \\
V_d &=&(1, \; 0,\; d)~, \\
V_{pq} &=&(0, \; p,\; q)~,  
\eeqn
and the replacement vectors are 
\beqn
\tilde V_b^j &= V_b+b S_b^j \qquad\;\; &=(0, \;\star , \; \#)~, \qquad \star\neq 1~, \quad \#\ge 1~,\\
\tilde V_c^j &= V_c+(c-1) S_c^j & =(\#, \; 1 , \; *)~, \qquad   \# \ge 1~,\\
\tilde V_d^j &= V_d+(d-1) S_d^j &=(\#, \; * , \; 1)~, \qquad   \# \ge 1~,
\eeqn
where $*,\star,\#\in \mathbb N$, and $\#>0, \star \neq 1$.
All the vectors above are distinct except possibly when $\tilde V_b=V_{pq}$, $\tilde V_c=\tilde V_d$, or $V_b=\tilde V_d $.

In the case that $\tilde V_b=V_{pq}$, we have 
\be
(b,0,1)+b(-1,m,n)=(0,p,q) \qquad \Rightarrow \qquad p=mb~, \quad q=nb+1~.
\ee
If $b=2$, we have $(d-1)/(b-1)\in \mathbb Z$, and this reduces to the previous type VII.
Thus we just need to consider $b>2$. 
For $b>2$, we can replace 
\be
\tilde V_b \to  \hat V_b= (b,0,1)+(b-2)(-1,m,n)=(2,m(b-2),n(b-2)+1)~.
\ee
It may happen that this still coincides with $\tilde V_c, \tilde V_d$.

If $  \hat V_b=\tilde V_c$, one can show that $b=3, c=2$.  Solving the equation for $p,q$, one further finds that $d=2s+1$ must be odd. This case is reducible to the previous type VII singularity as $(d-1)/(b-1)=s\in \mathbb Z$ \cite{Xie:2015rpa}. 

 If $  \hat V_b=\tilde V_d$, namely
 \be
(2,*,1)= (2,m(b-2),n(b-2)+1)=(1,0,d)+(d-1)(k,l,-1)  ~,
 \ee
then we should have  $d=2, k=1$, and we can replace 
 \be
 \tilde V_d \to  (1,0,2)+2 (1,l,-1)  =(3,2l,0)~.
 \ee
 This vector is different from all the rest.

In the case that $\tilde V_c=\tilde V_d$,  we are led to $c=d=2$. Solving the equation for $p,q$ we find the only solution $b=2,p+q=3$. So we   have
$
  v y^2+y u^2+y v^2+u v^2
$.
 One can explicitly check that this theory has a weight 1 marginal monomial in the Milnor ring. 
 
Finally if $V_b=\tilde V_d $, then
\be
 (b,0,1) =(1,0,d)+(d-1)(k,0,-1) \qquad \longrightarrow\qquad  \frac{b-1}{d-1}=k\in\mathbb Z~.
\ee
Therefore this case reduces to the type VII considered before \cite{Xie:2015rpa}.
 
\bigskip
\noindent
{\bf  Type XII:} We can consider the subring $ \mathcal R/\langle v\rangle$ and therefore set $v\to 0$.  The resulting $W$  is identical to  \eqref{VIIIyuv} in the type VIII case after reshuffling the variables. 
  
\bigskip
\noindent
{\bf Type XIII:}
The monomial vectors are given by
\beqn
V_a &=&(a, \; 0,\; 0, \;0   )~, \\
V_b &=&(1, \; b,\; 0, \;0  )~, \\
V_c &=&(0, \; 1,\; c, \;0  )~, \\
V_d &=&(0, \; 1,\; 0,\; d)~, \\
V_{pq} &=&(0, \; 0,\; p, \;q  )~.
\eeqn
Using replacement vectors, we can generate the new   weight 1  monomial  vectors in the following form
\beqn
\tilde V_a  &=&(0, \;\star,\; \star, \;\star   )~, \qquad \star\neq 1~, \label{Va}\\
\tilde V_b &=&(\#, \; 1,\; *, \;*  ) ~, \qquad   \#\ge 1~, \label{Vb}\\
\tilde V_c &=&(*, \; \#,\; 1, \;*  )~, \qquad   \#\ge 1~, \label{Vc}\\
\tilde V_d &=&(*, \; \#,\; *,\; 1) ~, \qquad   \#\ge 1~. \label{Vd} 
\eeqn
where $\star,\#,*\in \mathbb N$ and $\star \neq 1, \#\ge 1,*\ge 0$. All the vectors $V, \tilde V$ above are distinct except possibly $V_{pq} =\tilde V_a$, $\tilde V_b=\tilde V_c$, $\tilde V_b=\tilde V_d$, or $\tilde V_c=\tilde V_d$.

If  $V_{pq} =\tilde V_a$, then 
\be
(0,0,p,q)=(a,0,0,0)+a(-1,k,m,n)~,
\ee
and we have $k=0,p=ma, q=na$. These equalities enable us to replace
\be\label{Vanew}
\tilde V_a  \to (a,0,0,0)+ (a-1) \; (-1,k,m,n)=(1, 0,(a-1)m, (a-1)n )~.
\ee
Obviously this new vector is distinct from all the rest of the vectors above. 

In the case of $\tilde V_b=\tilde V_c$ or $\tilde V_b=\tilde V_d$, we must have $b=2$.  If $\tilde V_c=\tilde V_d$, we should have $c=d=2$. 

Let us discuss the case of $ c=d=2$ first. In this case, the weight vector can be rewritten as
\be
\vec q=\left(\frac{1}{a},\frac{a-1}{a b},\frac{a (b-1)+1}{2 a b},\frac{a (b-1)+1}{2 a b}\right)
=\left(\frac{1}{a},\frac{\tilde a}{a \tilde b},  \frac{a \tilde b-\tilde a}{2a \tilde b}, \frac{a \tilde b-\tilde a}{2a \tilde b} \right)~, 
\ee
where $\gcd(\tilde a,a)=\gcd(\tilde a, \tilde b)=1$. Thus $\gcd(\tilde a, a\tilde b)=1$ and 
$\gcd(a\tilde b, a\tilde b-\tilde a)=1$. If $\gcd(2 , a\tilde b-\tilde a)=1$, then the above weight vector is irreducible.  One can check $r_i=1$ and so the 1-form symmetry is trivial.  If $\gcd(2 , a\tilde b-\tilde a)=2$, we can set 
$a\tilde b-\tilde a=2k$ and rewrite the weight vector as 
\be
\vec q=\left( \frac{1}{a}  , \frac{\tilde a}{\tilde a+2k}  ,  \frac{k}{\tilde a+2k} ,  \frac{k}{\tilde a+2k}   \right)~,
\ee
which is irreducible as $\gcd(\tilde a, k)=1$. One then finds $r_i=1$, implying trivial 1-form symmetry. 

If $\tilde V_b=\tilde V_c$ and $\tilde V_b\neq \tilde V_d$, then $b=2$ and
\be
(*,1,1,*)=(1,2,0,0)+(n,-1,1,m)=(0,1,c,0)+(c-1)(k,0,-1,l)~.
\ee
Therefore, we can replace 
\be
\tilde V_b\to (1,2,0,0)+ 2(n,-1,1,m)=(\#,0,2,*)~, \qquad \#\ge 1~.
\ee
This new vector is different from all the rest except, potentially, for \eqref{Vanew} which happens when $a=b=2$.

Similarly if $\tilde V_b=\tilde V_d$ and $\tilde V_b\neq \tilde V_c$, then $b=2$, and
  we can replace 
\be
\tilde V_b\to  (\#,0, *,2)~, \qquad \#\ge 1~.
\ee
This vector is also  different from all the rest except, potentially, for \eqref{Vanew} which happens when $a=b=2$.

If $\tilde V_b=\tilde V_d$ and $\tilde V_b= \tilde V_c$,  we then have $\tilde V_c=\tilde V_d$  with $c=d=2$ which we have already considered. 

Therefore, we only need to study the case $a=b=2$, which has weight vector 
\be
\vec q=\left(\frac{1}{2},\frac{1}{4},\frac{3}{4 c},\frac{3}{4 d}\right)~.
\ee 
The weights coincide with those of the type VIII theory.

\bigskip
\noindent{\bf Type XIV:} We can consider the subring $ \mathcal R/\langle v\rangle$ and thus set $v\to 0$.  The resulting $W$  is identical to  \eqref{VIIIyuv} for the type VIII case after reshuffling the variables. 
  
\bigskip
\noindent{\bf Type XV, XVI, XVII:} For these three types of singularities, we can consider the subring $ \mathcal R/\langle v\rangle$ and thus set $v\to 0$. The resulting $W$  is identical to  \eqref{IXyuv} for the type IX case after reshuffling the variables.

\bigskip
\noindent
{\bf Type XVIII:}
The monomial vectors are given by
\beqn
V_a &=&(a, \; 0,\; 1, \;0   )~,\\
V_b &=&(1, \; b,\; 0, \;0  )~, \\
V_c &=&(0, \; 1,\; c, \;0  )~, \\
V_d &=&(0, \; 1,\; 0,\; d)~, \\
V_{pq} &=&(0, \; 0,\; p, \;q  )~. 
\eeqn
Using the replacement vectors, we can generate the new   weight one monomial  vectors of the following form
\beqn
\tilde V_a  &=&(1, \;*,\; \#, \;*   )~,  \\
\tilde V_b &=&(\#, \; 1,\; *, \;*  )~, \\
\tilde V_c &=&(*, \; \#,\; 1, \;*  )~,   \\
\tilde V_d &=&(*, \; \#,\; *,\; 1)~,  
\eeqn
where  $\#,*\in \mathbb N$ and $\#\ge 1, *\ge 0$. 

One can easily check that for $a,b,c,d\ge 2$, all the monomial vectors above are distinct unless $a,b,c,d$ equals to 2.  More precisely, the two vectors above coincide if: $\tilde V_a=\tilde V_b :  a=2$;
$\tilde V_a=\tilde V_c: c=2$;
$\tilde V_a=\tilde V_d : d=2$;
 $\tilde V_b=\tilde V_c  \text{ or }\tilde V_b=\tilde V_d  :  b=2$;
$\tilde V_c=\tilde V_d :  c=d=2$.

Therefore for all $a,b,c,d\ge 3$, the corresponding SCFTs always have exactly marginal deformations, and our claim on isolated SCFTs and 1-form symmetry trivially holds.  We need only check the case when any of $a,b,c,d$ equals to 2. 

In the case of $a=b=c=d=2$, we have 
\be
W= x y^2+u x^2+u^2 y +v^2 y+u^2 v~, \qquad \vec q=\left(\frac13,\frac13, \frac13, \frac13\right)~.
\ee
The singularity is isolated and one can explicitly check that all the monomials in $W$ are in the Milnor ring. 

On the other hand, we expect the Milnor ring has the property that it increases with increasing $a,b,c,d$ (where we define \lq\lq increasing" to be in the sense of \eqref{Rincreasing} below).\footnote{One simple explicit example illustrating this property is when $a=b=c=2$. The singularity is isolated and  defined by $ W=u v^d+y v^d+u^2 y+u x^2+x y^2 $. One can check that the Milnor ring has monomial basis $x^k y^l u^m v^n$ 
with $k, l=0,1, \; m=0, 1, 2,  \;  n=0, \cdots, d-2$ or  $k, l=0,1, \; m=0,   \;  n=d-1$. The Milnor ring and the Milnor number $\mu=12 d-8$ increase with $d$.  }
This is because whenever one increase the value of $a,b,c,d$, the constraints $dW=0$ become weaker due to the higher power of the variables. The only concern is that $p,q$ may not increase with $a,b,c,d$. However, the Milnor ring should be independent of the specific value of $p,q$ as long as the   singularity is isolated. This statement is consistent with the fact that the Milnor number $\mu$ is a monotonically increasing function of $a,b,c,d$. As a consequence, we have 
\be\label{Rincreasing}
\{ x y^2, \; u x^2  , \;  u^2 y , \;  v^2 y \}\subset \mathcal R_{(2,2,2,2)}\subset   \mathcal R_{(a,b,c,d)}~, \qquad a,b,c,d\ge 2~.
\ee
We can then immediately show that as long as at least one of $a,b,c,d$ equals to two, there is a weight one monomial in the Milnor ring.  For example, when $a=2$, we have $ux^2 \in \mathcal R_{(2,2,2,2)}\subset   \mathcal R_{(2,b,c,d)}$, and $ux^2$ has weight one. 
Therefore, we justify that even when one of the $a,b,c,d$ equal two, there is still at least a marginal monomial with weight one in the Milnor ring. The corresponding SCFT  thus has a conformal manifold. 

Alternatively, we can also justify our claim without relying on the property \eqref{Rincreasing}. Indeed, if three of $a,b,c,d$ equal to two, we have explicitly checked the conjecture holds.  Then, if two of $a,b,c,d$ equal to 2, by analysing the cases where $\tilde V$'s coincide, one can show that the possibilities are $(a,b,2,2)$,$(a,2,c,2)$, and $(a,2,2,d)$. For $(a,b,2,2)$, one can show the the 1-form symmetry of the corresponding SCFT  is always trivial.\footnote{This is because the denominators of all the weights are the same.}
The $(a,2,c,2)$ and $(a,2,2,d)$ cases are more complicated to analyze  directly, but can be discussed by imposing the further requirement that $\sum q_i>1 $ for SCFTs (which we did not impose explicitly before).\footnote{The requirement $\sum_i q_i>1$ then leads to two infinite sequences, $a=2$ or $c=2$ (or $d=2$), as well as a finite number of cases with $a,c\ge 3$ (or $a,d\ge 3$).  If either  of $a,c,d$ equals two, we have at least  three twos in   $(a,b,c,d)$  which  was already considered.  For the finite set, we also need to have a solution for $p,q$ in order to get an isolated singularity (within the scheme of \cite{Xie:2015rpa}). For $(a,2,c,2)$ with $a,c>3$, we get the unique case $(4, 2, 3, 2)$ which indeed has weight one monomial in the Milnor ring. While for $(a,2,2,d)$ wtih $a,d>3$, there is no solution. 
 We can also use the extra  condition, $\sum_i q_i>1$ for SCFTs,  to discuss  the case of $(a,b,2,2)$ as well as other types of singularities. 
 }

Finally, we can discuss the case when only one of $a,b,c,d$ equals two. If $a=2$, we have $ \tilde V_a=\tilde V_b=(1,1,*,*)=(2,0,0,0)+(-1, 1,*,*)$. We can then replace $\tilde V_a \to (2,0,0,0)+2(-1, 1,*,*)=(0,2,*,*)$. This new vector never coincides with any other vectors. In particular, it is different from $\tilde V_c, \tilde V_d$ due to our assumption that $c,d>2$. If $b=2$,  and $ \tilde V_b=\tilde V_d$, we can replace $\tilde V_b\to (*,0,*,2)$.
 If $b=2$ and $ \tilde V_b=\tilde V_c$, we can replace $\tilde V_b\to (*,0,  2,*)$.
 If $c=2$,  namely $ \tilde V_a=\tilde V_c$, we can replace $\tilde V_c\to (2,*,0,  *)$.
  If $d=2$,  namely $ \tilde V_a=\tilde V_d$, we can replace $\tilde V_d\to (2,*,*, 0)$.
 
To conclude, we have shown that for $a,b,c,d\ge 2$, the theory has marginal deformations and our claim holds. This completes the proof of our main claim for the theories discussed in \cite{Xie:2015rpa}.

 \subsection{1-form symmetry and more general singularities with five or more monomials}\label{beyondClass}
 In this appendix, we wish to study the theories covered in \ref{moreM} having five or more monomials in $W$ while allowing for more general possibilities regarding the extra monomials rendering the singularities isolated (e.g., see the recent discussion in \cite{Davenport:2016ggc,Closset:2020afy}).

Therefore, we study these singularities directly from their weight vectors without relying on particular choices for the extra monomials.  Since the weights are determined by $a,b,c,d$, they are independent of the extra monomials.  {\it As a result, we neither need nor attempt to perform a classification of the extra terms.}

To have a well-defined SCFT, the condition $\sum_i q_i>1$ should be imposed as well, and the resulting solutions were classified in \cite{Xie:2015rpa} (here we only allow for more general additional monomials rendering the singularity isolated than those considered in \cite{Xie:2015rpa}). We will use the results there and discuss each set of solutions one-by-one. Note that not all the solutions in \cite{Xie:2015rpa} give well-defined SCFTs.  In particular, we will impose the condition that the corresponding Milnor number is integral and that the corresponding Poincare polynomial indeed truncates to a polynomial (instead of an infinite series). These are all necessary conditions for a well-defined SCFT.

\bigskip
\noindent 
  {\bf Type VIII:}
 The singularity is defined by the polynomial\footnote{Note that we  can impose $c\ge d$, because exchanging $c\leftrightarrow d$ is a symmetry.}
 \be\label{typeVIIIB3}
 W=x^a+y^b+y u^c+y v^d + \cdots, \qquad a,b \ge 2~,  \qquad c, d\ge 1 ~,
 \ee
 where the dots represent extra weight 1 monomials to make the singularity isolated. It is easy to show that, as discussed in \cite{Davenport:2016ggc,Closset:2020afy}, these extra terms must include monomials of the form $u^p v^q$ or $x u^p v^q$ (we have explicitly carried out an analysis akin to that conducted in \ref{moreM} for $u^pv^q\to xu^pv^q$).\footnote{To understand this statement, let us denote the ellipses in \eqref{typeVIIIB3} as $f(x,y,u,v)$. Suppose that $u^pv^q\not\subset f(x,y,u,v)$ and that there is no $x$-dependent and $y$-independent term in $f(x,y,u,v)$ either, i.e., $x^{
 \alpha}u^pv^q\not\subset f(x,y,u,v)$. Then, setting $x=y=0$, we solve $\partial_uW=\partial_vW=\partial_xW=0$. This means that the final constraint yields $0=\partial_yW|_{x=y=0}=g(u,v)$, and the singularity is not isolated. Similar comments apply if all $x^{\alpha}u^pv^q$ terms appear with $\alpha>1$. \label{isolated}} Besides similar cursory comments in the type IX case, we will not attempt to systematically classify the possible additional terms.
 
The weight vector and Milnor number are 
\be
\vec q=\left(\frac{1}{a},\frac{1}{b},\frac{b-1}{b c},\frac{b-1}{b d}\right)~, \qquad \mu=\frac{(a-1) (b (c-1)+1) (b (d-1)+1)}{b-1}~.
\ee
Note that $ \sum_i q_i>3/2$ only holds when $c=1 $ or  $d=1$ (therefore, the corresponding SCFTs are isolated by our claim 3).\footnote{Note that this  is a necessary but not sufficient condition. There may still be no $p,q$ leading to an isolated singularity. On the other hand,  $ \sum_i q_i>3/2$ does not lead to an empty set of well-defined SCFTs. One example is $W=u v^3+u y+v^6 y+x^2+y^3$ which has isolated singularity and satisfies  $ \sum_i q_i>3/2$. }
  If $c=1$, we have $q_y+q_u=1$.  If $d=1$, we have $q_y+q_v=1$.  This gives trivial 1-form symmetry.  So we only need to focus on $a,b,c,d\ge 2$. Finally, note that the singularity is reducible if 
\be
\frac{c}{b-1}\in \mathbb Z~, \qquad \text{or} \qquad  \frac{d}{b-1}\in \mathbb Z~.
\ee

In addition to the  finite number of sporadic cases which we have checked separately, the infinite sequences solving the condition $\sum_i q_i>1$ are given in Table~\ref{VIIIinf}.
 \begin{table}[H]
\begin{center}
  \begin{tabular}{|C|C|C|C |C| }
    \hline
\mk (a,b,1,d)&\mk (a,b,c,1)&\mk(2,2,c,d)&\mk(3,2,2,d)&\mk(4,2,2,d)\\ \hline 
(2,b,2,d)&\mk(a,2,2,2)&(a,b,2,2)&(2,3,3,d)&\mk(2,3,4,d)\\ \hline
\mk(3,2,3,d)&\mk(2,4,3,d)&(2,b,3,3)&(2,b,3,4)&(2,b,3,5) \\ \hline
(2,b,3,6)&\mk(3,3,2,d)&(3,b,2,3)&(3,b,2,4)&(3,b,2,5) \\ \hline
(3,b,2,6)&(4,b,2,3)&(5,b,2,3)&(6,b,2,3)&(2,b,4,4) \\ \hline
(4,b,2,4)&(3,b,3,3)&&& \\ \hline
  \end{tabular}
  \caption{Infinite sequences of type VIII singularities. The gray entries are either reducible to previous types or have at least one of $c,d=1$, which gives trivial 1-form symmetry. This table is adapted from \cite{Xie:2015rpa}.}
  \label{VIIIinf}
\end{center}
\end{table}

\para{(2,b,2,d)} The Milnor number is 
\be
\mu=2d-1+b(d-1)+\frac{2d}{b-1}~.
\ee
If $d/(b-1)\in \mathbb Z$ then it reduces to previous types. So we  just need to consider $d/(b-1)\not\in \mathbb Z$ but $2d/(b-1)\in \mathbb Z$ in order to have an integral Milnor number. Thus
\be
\frac{ d}{b-1}=\frac{2m+1}{2}\qquad \Rightarrow \qquad b=2k+1~, \qquad d=(2m+1)k~.
\ee
Then 
\be
\vec q=\left(\frac{1}{2},\frac{1}{2 k+1},\frac{k}{2 k+1},\frac{2}{(2 k+1) (2 m+1)}\right)~.
\ee
In order to make the singularity isolated, we need to have a weight-one term of the form $u^p v^q$.\footnote{As discussed in footnote \ref{isolated}, the ellipses in \eqref{typeVIIIB3} must include either a term of the form $u^pv^q$ or else a term of the form $xu^pv^q$. Since $q_x=1/2$ in this case, we see that if we have a term of the form $xu^p v^q$, then we have $u^{2p}v^{2q}$ as well. \label{isolated2}} So
\be
kp +2\frac{q}{2m+1} =1+2k~.
\ee
We must have 
\be
\frac{q}{2m+1} =s\in \mathbb N,\qquad kp  =1+2(k-s)
\ee
Clearly $k=2\tilde k+1$ must be odd. Then, we find 
\be
  s=\tilde k+1~, \qquad p=1~,\qquad q=(\tilde k+1)(2m+1)~,
\ee
and 
\be
W=u v^{( \tilde k+1) (2 m+1)}+y v^{(2 \tilde k+1) (2 m+1)}+y^{2 (2 \tilde k+1)+1}+u^2 y+x^2~.
\ee
One can compute the Milnor ring and find that there is always an exactly marginal deformation.~\footnote{There is a simpler way to show this by noticing that $u^2y \in \mathcal R_{(2,3,2,3)} \subset \mathcal R_{(2,b,2,d)}$. Since $u^2y$ always has weight 1 for $(2,b,2,d)$, it is thus an exactly marginal deformation. }

\para{(a,b,2,2)}
If $b=2,3$, the theory reduces to type I.  Since $q_u=(b-1)/(2b )=q_v$, we have $\frac13<q_{u,v}<\frac12$ for $b> 3$. According to lemma 2.8 of \cite{Davenport:2016ggc}, we should have another $q_i=1-2q_u=1/b, q_j=1-2q_v=1/b$.
Therefore $a=b$, and there is no 1-form symmetry by the discussion in appendix \ref{twoq}.

\para{(2,3,3,d)} The Milnor number is 
\be
\mu=\frac{21 d}{2}-7~,
\ee
and so we must have $d=2n$ in order to have an integer.  Then $d/(b-1)=n\in \mathbb Z$, and the singularity reduces to a previous type. 

\para{(2,b,3,3)}  By requiring the Milnor number to be integral, we find $b=2,4,10$. By computing the Poincare polynomial, one finds that for $b=2,4$ there is a weight 1 term. For $b=10$, the Poincare polynomial does not truncate to a polynomial, so the singularity is not well defined.

\para{(2,b,3,4)}   One finds the solutions $b=2,3,4,5, 7,13$. By computing the Poincare polynomial, one finds there is always a weight one term in the Milnor ring.

\para{(2,b,3,5)}   The integral Milnor number requirement leads to $b=2,4,6,16$, and there is always a weight-one term in the Milnor ring.

\para{(2,b,3,6)}  One finds $b=2, 3, 4, 7, 10, 19$.  In the first four cases, there is a weight-one term in the Milnor ring. For the last two, the singularity is not well defined.

\para{(3,b,2,3)}    One finds $b=2, 3, 4, 5, 7, 13$.  The first five cases have a weight one term, while the last one is not well defined. 

\para{(3,b,2,4)}   One finds $b=2, 3, 5, 9, 17$.  The first four cases have a weight one term, while the last one is not well defined.

\para{(3,b,2,5)}   One finds  $b=2, 3, \cancel  { 5}, 6, 11, 21$.\footnote{From now on, we will use the notation $\cancel x$ to indicate that the corresponding singularity is not well-defined due to a lack of truncation in the Poincare polynomial.  For the rest of the cases, there is always a weight-one term, but we will not repeat this statement any more. }

\para{(3,b,2,6)}  $b=2, 3, 4, 5, 7, \cancel 9,\cancel{ 13},\cancel{ 25}$.

\para{(4,b,2,3)}  $b=2, 3, 4, 7, 10, \cancel{19}$.

\para{(5,b,2,3)}  $b=2, 3, 4, 5, 7,\cancel  9, \cancel{13}, 25$.

\para{(6,b,2,3)}  $b=2, 3, 4, 6, 7, \cancel{11}, \cancel{16},\cancel{ 31}$.

\para{(2,b,4,4)}  $b=2, 3, 5, \cancel 9, \cancel{17}$.

\para{(4,b,2,4)}  $b=2, 3, 4, 5,\cancel 7,\cancel 9,\cancel{ 13},\cancel{ 25}$.

\para{(3,b,3,3)}  $b=2, 3, 4, \cancel 7, \cancel{10},\cancel{ 19}$.

\bigskip
 \noindent
{\bf Type IX:} The singularity is defined by the polynomial
 \be
 W=x^a+v y^b+v u^c+y v^d +\cdots, \qquad a,b,d\ge 2~, \quad c\ge 1~.
 \ee
The weight vector and Milnor number are 
 \be
\vec q=\left(\frac{1}{a},\frac{d-1}{b d-1},\frac{b (d-1)}{c (b d-1)},\frac{b-1}{b d-1}\right)~, \qquad
 \mu= \frac{(a-1) d (b (c-1) d+b-c)}{d-1}~.
 \ee
We have that  $ \sum_i q_i>3/2$ holds only when $c=1$. This theory has trivial 1-form symmetry since $q_u+q_v=1$.

Our singularity reduces to previous types if 
\be
\frac {b-1}{d-1}\in \mathbb Z~, \qquad \frac {d-1}{b-1}\in \mathbb Z~, \qquad \frac {c(b-1)}{b(d-1)}\in \mathbb Z~.
\ee

\begin{table}[H]
\begin{center}
  \begin{tabular}{|C|C|C|C|C| }
    \hline
\mk (a,b,1,d)& \mk(2,2,c,d)&(2,b,2,d)&\mk(a,2,2,d)\mk&\mk(3,2,c,2)\\ \hline 
 \mk(3,b,2,2)&\mk(4,b,2,2)&(2,3,3,d)&(2,3, 4,d)&(2,3,5,d) \\ \hline
 (2,3,6,d)&\mk (2,3,c,3)&\mk (3,2,3,d)&\mk(3,2,4,d)&\mk(3,2,5,d) \\ \hline
\mk (3,2,6,d)&(2,4,3,d)&(2,5,3,d)&(2,6,3,d)&\mk (4,2,3,d) \\ \hline
\mk (5,2,3,d)&\mk(6,2,3,d)&(2,b,3,3)&(2,b,3,4)&(2,b,4,3) \\ \hline
 (3,3,2,d)&(3,4,2,d)&(3,5,2,d)&(3,6,2,d)&(4,3,2,d) \\ \hline
 (5,3,2,d)&(6,3,2,d)&(3,b,2,3)&\mk(3,b,3,2)&(2,4,4,d) \\ \hline
\mk (4,2,4,d)&(4,4,2,d)&(3,3,3,d)&&\\ \hline
  \end{tabular}
  \caption{Infinite sequences of type IX singularities. The gray entries are either reducible to previous types or have $c=1$, which gives trivial 1-form symmetry. This table is adapted from \cite{Xie:2015rpa}.}
  \label{IXinf}
\end{center}
\end{table}

\para{(2,b,2,d)} The Milnor number is 
\be
\mu=\frac{2 (b-1)}{d-1}+b(d+2)-2~.
\ee
If $(b-1)/(d-1)\in \mathbb Z$ it reduces to previous types. So we must have 
\be
\frac{   b-1 }{d-1} =\frac{2m+1}{2}\qquad\Rightarrow \qquad d=2k+1~, \qquad b=(2m+1)k+1~.
\ee
In this case, the weight vector is 
\be
\vec q=\left(\frac{1}{2},\frac{2}{k (4 m+2)+2 m+3},\frac{2 k m+k+1}{4 k m+2 k+2 m+3},\frac{2 m+1}{4 k m+2 k+2 m+3}\right)~,
\ee
We need to add weight one term $y^p u^q$ to make the singularity isolated: 
$
k (2 m+1) (q-2)-2 m+2 p+q-3=0
$.
When $k$ is odd one can show that it has no solution. So $k=2\tilde k$ and one solution is 
\be
q=1, \qquad p=1 + \tilde k + m + 2 \tilde k m~.
\ee
One can compute the Milnor ring and find that there is always an exactly marginal deformation.

\para{(2,3,3,d)} $d=2, 3, \cancel 4,\cancel 7$.

\para{(2,3,4,d)}  $d=2, 3, 5, 9$.

\para{(2,3,5,d)}  $d=2, 3, 6, 11$.

\para{(2,3,6,d)} $d=2, 3, \cancel 4, 5, \cancel 7, \cancel{13}$.

\para{(2,4,3,d)} $d=2, 4, 10$.

\para{(2,5,3,d)} $d=2, 3, 4, 5, 7, 13$.

\para{(2,6,3,d)} $d=2, 4, 6, 16$.

  \para{(2 ,b,3,3)}  The Milnor number is $\mu=\frac{3}{2} (7 b-3)$. To be an integer, we should require $b=2k+1$  to be odd. Then $(b-1)/(d-1)=k\in\mathbb Z$, so this case reduces to previous types. 
  
  \para{(2 ,b,3,4)}  To be isolated, we must have an extra $y^p u^q$ term (this statement follows from an argument analogous to the one in footnotes \ref{isolated} and \ref{isolated2}). Using an argument similar to one in the previous subsection after eq.~\ref{IXyuv}, we conclude this theory has an exactly marginal deformation.
   
  \para{(2 ,b,4,3)}  The same as above.
 
  \para{(3,3,2,d)}  $d=2, 3, 5,\cancel 9$.
 
  \para{(3,4 ,2,d)}  $d=2, \cancel3, 4, \cancel5, 7,\cancel 13$.
 
   \para{(3,5,2,d)} $d=2, 3, 5, 9, 17$.
   
     \para{(3,6,2,d)} $d=2,\cancel  3, \cancel 5, 6,\cancel{ 11}, \cancel{21}$.
     
       \para{(4,3,2,d)}  $d=2, 3, \cancel 4, 5, 7, \cancel{13}$.

        \para{(5,3,2,d)} $d=2, 3, 5, \cancel 9, 17$.
       
            \para{(6,3,2,d)}    $d=2, 3, 5, \cancel 6,\cancel{ 11},\cancel{ 21}$.
            
         \para{(3,b,2,3)}         If $b$ is odd, the singularity reduces to previous types. Therefore, we can consider $b=2k$. One finds $r_2=r_3=r_4=1,2 g_1=1$, which is not meaningful. This result implies that the theory cannot have 1-form symmetry. One can check that the Poincare polynomial is not finite, so the singularity is actually not well-defined.

          \para{(2,4,4,d)}        $d=2,\cancel  3,   4,\cancel 5,\cancel 7,\cancel{ 13}$
          
       \para{(4,4,2,d)}        $d=2, \cancel 3, 4, \cancel 7, \cancel{10},\cancel{ 19}$.
                   
             \para{(3,3,3,d)}      $d=2, 3, \cancel 4,\cancel 5,\cancel 7,\cancel{ 13}$.

\bigskip 
\noindent
 {\bf Type  XII:}  The singularity is defined by the polynomial
 \be
 W=x^a+x y^b+x u^c+y v^d +\cdots, \qquad a\ge 2~, \quad  b,c,d \ge 1~.
 \ee
The weight vector and Milnor number are
\be
\vec q=\left(\frac{1}{a},\frac{a-1}{a b},\frac{a-1}{a c},\frac{a (b-1)+1}{a b d}\right)~,\qquad 
\mu=\frac{(a (c-1)+1) (a b (d-1)+a-1)}{a-1}~.
\ee
We find that $\sum_i q_i>3/2$ holds only when $c=1 $ or  $d=1$.  If $c=1$, we have $q_x+q_u=1$.  If $d=1$, we have $q_y+q_v=1$. When $b=1$, we also have $q_x+q_y=1$, but $ \sum_i q_i>3/2$  does not hold. Therefore, by our lemma, 1-form symmetry is trivial in all three cases.

If $a=2$ the singularity reduces to type II. Moreover, it reduces to type IV if 
\be
\frac{b}{a-1}\in\mathbb Z~.
\ee

\begin{table}[H]
\begin{center}
  \begin{tabular}{|C|C|C|C|C | }
    \hline
\mk (a,1,c,d)&\mk (a,b,1,d)&\mk(a,b,c,1)&\mk(2,2,2,d)&\mk(2,2,c,2) \\ \hline
\mk (2,2,c,2)&\mk(2,2,c,3)&\mk(2,b,c,2)&(a,2,2,d)&\mk(3,2,c,2) \\ \hline
\mk (2,b,2,2)&\mk (2,b,2,3)&\mk(2,b,2,4)&(a,2,3,2)&(a,2,4,2)\\ \hline
 (a,b,2,2)&\mk(2,b,3,3)&(a,2,3,3)&(3,b,2,3)&(3,b,3,2) \\ \hline
 (3,b,4,2)&(a,3,2,3)&(a,3,2,4)&(a,3,3,2)&(4,b,3,2) \\ \hline
 (a,3,3,2)&(4,b,3,2)&(a,4,2,3)&& \\ \hline
  \end{tabular}
  \caption{Infinite sequences of type XII singularities. The gray entries are either reducible to previous types or have at least one of $b,c,d=1$, which gives trivial 1-form symmetry. This table is adapted from \cite{Xie:2015rpa}}
  \label{XIIinf}
\end{center}
\end{table}

\para{(a,2,2,d)} If $a=2,3$, the singularity reduces to type II. If $a>3$, we have $1/3<q_y=q_u=(a-1)/(2a)<1/2$, so we must have $q_x=q_v=1-2 q_y=1/a$. According to the lemma, this theory has no 1-form symmetry. 

\para{(a,2,3,2)} $a=2, 3, 4, 7$.

\para{(a,2,4,2)} $a=2, 3, 5, \cancel 9$.

\para{(a,b,2,2)}
\be
\mu=1 + a + 2 b + a b+2\frac{b}{a-1}
\ee
To have an irreducible singularity and integral Milnor number, we should satisfy 
\be
\frac{b}{a-1}=\frac{2m+1}{2} \qquad \Rightarrow \qquad b=(2m+1)k, \quad a=2k+1~.
\ee
One finds that if $k$ is even, the singularity is not well-defined.  So $k=2\tilde k+1$ and $a=4\tilde k+3,b=(2m+1)(2\tilde k+1)$.  Then, one has an extra weight-one monomial $y^{(2m+1) (\tilde k+1)} u$. Adding this to $W$ makes the singularity isolated.  Furthermore, one finds that there is a weight 1 monomial in Milnor ring, so the SCFT has a conformal manifold. 

\para{(a,2,3,3) } $a=2, 3, 4, 5, 7, \cancel {13}$.

\para{(3,b,2,3)  }   We can assume $b=2k+1$ is odd, otherwise the singularity is reducible. In this case, one finds $r_1=r_2=r_3=1, g_4=0$, implying trivial 1-form symmetry. 

\para{(3,b,3,2) } Since  $\mu=7 + (21 b)/2$, we must have even $b$. Then this case is reducible since $b/(3-1)\in \mathbb Z$. 

\para{ (3,b,4,2) }   We can assume $b=2k+1$ is odd, otherwise the singularity is reducible. In this case, one finds $r_i=1$, implying trivial 1-form symmetry

\para{(a,3,2,3)  } $a=2, 3, 4, \cancel 5, 7, 13$.

\para{(a,3,2,4)  } $a=2, 3, 4, 7,\cancel{ 10}, \cancel{19}$.

\para{(a,3,3,2)  } $a=2, 4, \cancel{10}$.

\para{(4,b,3,2) }  If $b=3k$, this singularity is reducible. For  $b=3k+1,3k+2$,  one always finds $r_1=r_2=r_3=1, g_4=0$, implying trivial 1-form symmetry.

\para{ (a,4,2,3)} $a=2, 3, 5, \cancel 9, \cancel{17}$.

\bigskip 
\noindent
   {\bf Type  XIII:} The singularity is defined by the polynomial
 \be\label{defXIII}
 W=x^a+x y^b+y u^c+y v^d +\cdots, \qquad a\ge 2, \quad  b,c,d \ge 1 
 \ee
Before proceeding as in the other cases, let us give a quick proof that the corresponding SCFTs have trivial 1-form symmetry or an exactly marginal deformation if $a,b,c,d\ge2$. We will do this by directly building on our results in \ref{moreM}:

\medskip
\noindent
\emph{Proof:} Let us denote the ellipses in \eqref{defXIII} as $f(x,y,u,v)$. By the discussion in \ref{moreM}, we either have trivial 1-form symmetry or can compensate all possible relations amongst variables with new marginal monomials in \eqref{Va}, \eqref{Vb}, \eqref{Vc}, and  \eqref{Vd} that have the following property: if the monomial depends on $x$, then it will also depend on $y$. Now, if we have a marginal term of the form $u^pv^q\subset f(x,y,u,v)$, we are back to the case analyzed in \ref{moreM}. Therefore, let us assume there is no such term. In order for the singularity to be isolated, we still need $f(x,y,u,v)$ to contain some $y$-independent term that is, at the same time, $x$ dependent (otherwise, setting $x=y=0$ solves $\partial_uW=\partial_vW=\partial_xW=0$). We have seen there is no such candidate amongst \eqref{Va}, \eqref{Vb}, \eqref{Vc}, and \eqref{Vd}. $\square$
 
\medskip
\noindent
 Alternatively, we can proceed as in other cases treated here and first note that we can impose $c\ge d$ (this exchange is a symmetry). The weight vector and Milnor number are
\bea
\vec q&=&\left(\frac{1}{a},\frac{a-1}{a b},\frac{a (b-1)+1}{a b c},\frac{a (b-1)+1}{a b d}\right)~, \cr
\mu&=&\frac{(a b (c-1)+a-1) (a b (d-1)+a-1)}{a (b-1)+1}~.
\eea
We find that $\sum_i q_i>3/2$ only holds when $c=1 $ or  $d=1$.   If $c=1$, we have $q_y+q_u=1$.  If $d=1$, we have $q_y+q_v=1$. When $b=1$, $q_x+q_y=1$ and $ \sum_i q_i\le3/2$. In all three cases, the 1-form symmetry is trivial by our lemma.
 
 The singularity reduces to type VIII if 
 \be
 \frac{b}{a-1}\in \mathbb Z~.
 \ee
 
 \begin{table}[h]
\begin{center}
  \begin{tabular}{|C|C|C|C|C| }
    \hline
 \mk (a,1,c,d)&\mk(a,b,1,d)&\mk(a,b,c,1)&\mk(2,2,2,d)&\mk(2,2,3,d) \\ \hline
 \mk(2,b,2,c)&\mk(3,2,2,d)&(a,2,2,2)&(a,b,2,2)&\mk(2,b,3,3) \\ \hline
\mk (2,b,3,4)&\mk(2,b,3,5)&\mk(2,b,3,6)&(3,b,2,3)&(3,b,2,4) \\ \hline
 (3,b,2,5)&(3,b,2,6)&(4,b,2,3)&(5,b,2,3)&(6,b,2,3) \\ \hline
\mk (2,b,4,4)&(4,b,2,4)&(3,b,3,3)&& \\ \hline
  \end{tabular}
  \caption{Infinite sequences of type XIII singularities. The gray entries are either reducible to previous types or have at least one of $b,c,d=1$, which gives trivial 1-form symmetry. This table is adapted from \cite{Xie:2015rpa}.}
  \label{XIIIinf}
\end{center}
\end{table}

\para{(a,2,2,2)} $a=3, 7, \cancel{ 15}$.

\para{(a,b,2,2)}     For $b\ge 3$,  we have  $1/3<q_u=q_v=(1 + a (-1 + b))/(2 a b)<1/2$. Therefore, by lemma 2.8 in \cite{Davenport:2016ggc}, we  must have $q_x=q_y=1-2 q_u$. Therefore $b=a-1$, and the 1-form symmetry is trivial according to our lemma.

\para{(3,b,2,3)}  $b=2$.

\para{(3,b,2,4)} $b=2, 6$.

\para{(3,b,2,5)}  $b=2, 4, 14$.

\para{(3,b,2,6)}  $b=2, \cancel 6$. 

\para{(4,b,2,3)}  $b= 3$.

\para{(5,b,2,3)}  $b= 2, 4, 20$.

\para{(6,b,2,3)}  $b= 5$.

\para{(4,b,2,4)} $b= 3$.

\para{(3,b,3,3)} $b= 2$.

 \bigskip
 \noindent 
   {\bf Type  XIV:}
 The singularity is defined by the polynomial
 \be
 W=x^a+x y^b+x u^c+x v^d+ \cdots, \qquad a\ge 2, \quad  b,c,d \ge 1 
 \ee
 We can impose $b\ge c\ge d$, as this exchange is a symmetry. The weight vector and Milnor number are 
\be
\vec q=\left( \frac{1}{a},\frac{a-1}{a b},\frac{a-1}{a c},\frac{a-1}{a d} \right)~, \qquad
\mu=\frac{(a (b-1)+1) (a (c-1)+1) (a (d-1)+1)}{(a-1)^2}~.
\ee
 We find that $ \sum_i q_i>3/2$ holds only when $b=1$, $c=1 $, or  $d=1$. In these cases, $q_x+q_y=1$, $q_x+q_u=1$, or $q_x+q_v=1$, respectively. Therefore, the 1-form symmetry is trivial.
 \begin{table}[H]
\begin{center}
  \begin{tabular}{|C|C|C|C|C| }
    \hline
 \mk(a,1,c,d)&\mk(a,b,1,d)&\mk(a,b,c,1)&\mk(2,2,2,d)&(a,2,2,d) \\ \hline
 (a,2,3,3)&(a,2,3,4)&(a,2,3,5)&& \\ \hline
  \end{tabular}
  \caption{Infinite sequences of type XIV singularities. The gray entries are either reducible to previous types or have at least one of $b,c,d=1$, which gives trivial 1-form symmetry. This table is adapted from \cite{Xie:2015rpa}.}
  \label{XIVinf}
\end{center}
\end{table}

\para{(2,2,2,d)} This case reduces to type I, and it has a weight-one term in the Milnor ring. 

\para{(a,2,2,d)}  In this case, we have
\be
\vec q=\left(\frac{1}{a},\frac{a-1}{2 a},\frac{a-1}{2 a},\frac{a-1}{a d}\right)~.
\ee
If $a=3$, it reduces to type II. If $a>3$, then $1/3<q_y=q_u<1/2$, so we should have $q_x=q_v=1-2 q_u=1/a$ according to lemma 2.8 of \cite{Davenport:2016ggc}. As result, we have $q_u=q_y, q_x=q_v$, which gives no 1-form symmetry

\para{(a,2,3,3)}  Imposing integral Milnor number gives $a=2, 3, 4, 7$.

\para{(a,2,3,4)}  Imposing integral Milnor number gives $a=2, 3, 4, 5, 13$.

\para{(a,2,3,5)}  Imposing integral Milnor number gives $a=2, 3, 4, 6, 31$.

\bigskip 
\noindent
  {\bf Type  XV:}
 The singularity is defined by the polynomial
\be
W=y x^a+x y^b+x u^c+u v^d +\cdots~, \qquad a,b\ge 2~, \qquad c,d\ge 1~.
\ee
The weight vector and Milnor number are
\bea
\vec q&=&\left(\frac{b-1}{a b-1},\frac{a-1}{a b-1},\frac{(a-1) b}{c (a b-1)},\frac{c (a b-1)-(a-1) b}{c d (a b-1)}\right)~, \cr
\mu&=&\frac{a (b (a c (d-1)+a-1)+c (-d)+c)}{a-1}
\eea
We have that $\sum_i q_i>3/2$ holds only when $c=1$ or $d=1$. We then have $q_x+q_u=1$ and   $q_u+q_v=1$ respectively.
 
 If $a=2$, the singularity reduces to type  XI$(2b-1,2,c,d)$. If $b=2$, it reduces to XII$(2a-1,c,2,d)$.
\begin{table}[H]
\begin{center}
  \begin{tabular}{|C|C|C|C|C | }
    \hline
\mk (a,b,1,d)&\mk(a,b,c,1)&\mk(2,2,2,d)&\mk(2,2,c,2)&\mk(2,2,c,3) \\ \hline
\mk(2,b,c,2)&\mk(a,2,2,d)&\mk(2,b,2,2)&\mk(2,b,2,3)&(3,b,2,2) \\ \hline
(a,3,2,2)&(a,3,2,2)&(a,4,2,2)&\mk(a,2,3,3)&\mk(a,2,3,4) \\ \hline
\mk(a,2,4,3)&(3,3,c,2)&(a,3,2,3)&(a,3,3,2)& \\ \hline
  \end{tabular}
  \caption{Infinite sequences of type XV singularities. The gray entries are either reducible to previous types or have at least one of $c,d=1$, which gives trivial 1-form symmetry. This table is adapted from \cite{Xie:2015rpa}.}
  \label{XVinf}
\end{center}
\end{table}

\para{(3,b,2,2)} For $b=2k+1$, one finds $r_1=r_2=r_3=1, g_4=0$, giving trivial 1-form symmetry. For $b=2k$, one finds $r_1=r_2=r_3=1, 2g_4=1$ which is not meaningful. In fact, one can compute the Poincare polynomial in this case, and it turns out that it does not truncate (thereby implying that the singularity is not well-defined). 

\para{(a,3,2,2)} $a=2,3,5$.

\para{(a,4,2,2)} $a=2, \cancel 3, 4, \cancel7$.

\para{(3,3,c,2)} This case is reducible to VII$(4,4,c,2)$.

\para{(a,3,2,3)} $a=2, 3, 5, 9$.

\para{(a,3,3,2)} $a=2, 3, \cancel 4, \cancel 7$.

\bigskip 
\noindent
   {\bf Type  XVI: }
   The singularity is defined by the polynomial
 \be
 W=y x^a+x y^b+x u^c+x v^d +\cdots~, \qquad a,b \ge 2~, \quad   c,d \ge 1~. 
 \ee
 We can impose $ c\ge d$, as this is a symmetry.
 
 The weight vector  and Milnor number are 
\bea
\vec q&=&\left(\frac{b-1}{a b-1},\frac{a-1}{a b-1},\frac{(a-1) b}{c (a b-1)},\frac{(a-1) b}{d (a b-1)}\right)~,\cr
 \mu&=&\frac{a (a b (c-1)+b-c) (a b (d-1)+b-d)}{(a-1)^2 b}~.
\eea
We find that $ \sum_i q_i>3/2$ holds only when    $c=1 $ or  $d=1$. Then, $q_x+q_u=1$ or $q_x+q_v=1 $ respectively.
 
 If $a=2$ the singularity reduces to type XIII. On the other hand, if $b=2$ it reduces to type VIV.
 
\begin{table}[H]
\begin{center}
  \begin{tabular}{|C|C|C|C|C| }
    \hline
\mk(a,b,1,d)&\mk(a,b,c,1)&\mk(2,2,2,d)&\mk(a,2,2,d)&\mk(2,b,2,2) \\ \hline
(a,b,2,2)&\mk(a,2,3,3)&\mk(a,2,3,4)&\mk(a,2,3,5)&(a,3,2,3)\\ \hline
(a,3,2,4)&(a,3,2,5)&(a,4,2,3)&(a,5,2,3)& \\ \hline
  \end{tabular}
  \caption{Infinite sequences of type XVI singularities. The gray entries are either reducible to previous types or have at least one of $c,d=1$, which gives trivial 1-form symmetry. This table is adapted from \cite{Xie:2015rpa}.}
  \label{XVIinf}
\end{center}
\end{table}
 
 \para{(a,b,2,2)} In this case, we have
 \be
 \vec q=\left(\frac{b-1}{a b-1},\frac{a-1}{a b-1},\frac{(a-1) b}{2 (a b-1)},\frac{(a-1) b}{2 (a b-1)}\right)~.
 \ee
 Since $a=2$ is reducible, let us consider $a\ge 3$.  Then $1/3<q_u=q_v<1/2$, and we must have $q_x=q_y=1-2 q_u$. Therefore $a=b$, and the 1-form symmetry is trivial. 
 
 \para{(a,3,2,3)} $a=2, 3, 5$.
 
  \para{(a,3,2,4)} $a=3, 9$.
  
   \para{(a,3,2,5)} $a=3, 21$.
   
   \para{(a,4,2,3)} $a=2, 4, 10$.
   
    \para{(a,5,2,3)}  $a=5, 25$.

\bigskip
\noindent
  {\bf Type  XVII:}
  The singularity is defined by the polynomial
 \be
 W=y x^a+x y^b+y u^c+x v^d+\cdots~, a,b \ge 2~, \quad   c,d \ge 1~.
  \ee
  The weight vector and Milnor number are 
 \bea
 \vec q&=&\left( \frac{b-1}{a b-1},\frac{a-1}{a b-1},\frac{a (b-1)}{c (a b-1)},\frac{(a-1) b}{d (a b-1)}\right)~,\cr
 \mu&=&\frac{(a b (c-1)+a-c) (a b (d-1)+b-d)}{(a-1) (b-1)}~.
  \eea
We find that $ \sum_i q_i>3/2$ holds only when      $c=1$ or  $d=1$. In these cases, $q_y+q_u=1$ or $q_x+q_v=1$ respectively. Therefore, the 1-form symmetry is trivial.
  
 The reducibility condition to type XII is 
 \be
 \frac{b-1}{a-1} \in \mathbb Z~, \qquad \text{or} \qquad  \frac{a-1}{b-1} \in \mathbb Z~.
 \ee
 
 \begin{table}[H]
\begin{center}
  \begin{tabular}{|C|C|C|C|C| }
    \hline
\mk ( a,b,1,d)&\mk (a,b,c,1)&\mk(2,2,2,d)&\mk(2,2,c,2)&\mk(2,b,2,d)\\ \hline
\mk(a,2,c,2)&\mk(2,b,3,2)&\mk(2,b,4,2)&\mk(a,2,2,2)&\mk(a,2,2,3) \\ \hline
\mk(a,2,2,4)&(a,b,2,2)&\mk(2,b,3,3)&\mk(a,2,3,3)&(3,b,2,3) \\ \hline
(3,b,2,4)&(3,b,3,2)&(a,3,2,3)&(a,3,3,2)&(a,3,4,2) \\ \hline
(4,b,2,3)&(a,4,3,2)&&& \\ \hline
  \end{tabular}
  \caption{Infinite sequences of type XVII singularities. The gray entries are either reducible to previous types or have at least one of $c,d=1$, which gives trivial 1-form symmetry. This table is adapted from \cite{Xie:2015rpa}.}
  \label{XVIIinf}
\end{center}
\end{table}

\para{(a,b,2,2)} We find that the Milnor number satisfies
\be
\mu=2 b + a (2 + b)+2\frac{a-1}{b-1}+2\frac{b-1}{a-1}~.
\ee
In order for this to be an integer, one can show that either $\frac{a-1}{b-1}$ is an integer or $\frac{b-1}{a-1}$ is an integer. In both cases, the singularity is reducible. 

\para{(3,b,2,3)}  $b=3, 5, 9$. 

\para{(3,b,2,4)} $b=2, 3, \cancel 4, 5,\cancel  7, \cancel{13}$.

\para{(3,b,3,2)} $b=2, 3, \cancel 4,\cancel 7$.

\para{(a,3,2,3)} $a=2, 3,\cancel 4,\cancel 7$.

\para{(a,3,3,2)} $a=3, 5, 9$.

\para{(a,3,4,2)} $a=2, 3,\cancel 4, 5,\cancel 7,\cancel{ 13}$.

\para{(4,b,2,3)}  $b=2,\cancel  3, 4, \cancel 5, \cancel 7, \cancel{13}$.

\para{(4,b,3,2)}$b=4, 10$.

 \para{(a,4,3,2)}$a=2, \cancel 3, 4, \cancel 5, \cancel 7, \cancel {13}$.
 
\bigskip 
\noindent
  {\bf Type  XVIII:}
 The singularity is defined by the polynomial
\be
W=u x^a+x y^b+y u^c+y v^d +\cdots, \qquad a,b,c,d \ge 1~.
\ee
The weight vector and Milnor number are 
 \bea
 \vec q&=&\left(\frac{b (c-1)+1}{a b c+1},\frac{(a-1) c+1}{a b c+1},\frac{a (b-1)+1}{a b c+1},\frac{c (a (b-1)+1)}{d (a b c+1)}\right)~,\cr
 \mu&=&\frac{a b (c (a b (d-1)+a-1)+d)}{a (b-1)+1}~.
 \eea
 It is easy to check that  $ \sum_i q_i>3/2$ holds   for $a=1$ or $d=1$.  When $a=1$, $q_x+q_u=1$, and when $d=1$, $q_y+q_v=1$.  If $b=1$,  $q_x+q_y=1$, and if  $c=1$,  $q_y+q_u=1$.  In these two cases,  we have $ \sum_i q_i\le3/2$. On the other hand, all four cases have trivial 1-form symmetry by our lemma.

 \begin{table}[H]
\begin{center}
  \begin{tabular}{|C|C|C|C|C| }
    \hline
\mk (1,a,c,d)&\mk (a,1,c,d)&\mk(a,b,1,c)&\mk(a,b,c,1)&(2,2,2,d) \\ \hline
(2,2,c,2)&(2,2,c,3)&(2,b,c,2)&(3,2,c,2)&(2,b,2,2) \\ \hline
(2,b,2,3)&(2,b,2,4)&(a,2,2,2)&(a,b,2,2)&(2,b,3,3) \\ \hline
(3,b,2,3)&(3,b,3,2)&(3,b,4,2)&(4,b,3,2)& \\ \hline
  \end{tabular}
  \caption{Infinite sequences of type XVIII singularities. The gray entries have at least one of $a,b,c,d=1$, which gives trivial 1-form symmetry. This table is adapted from \cite{Xie:2015rpa}.}
  \label{XVIIIinf}
\end{center}
\end{table}

\para{(2,2,2,d)} This case is reducible to type II.

\para{(2,2,c,2)} In order to have integral Milnor number, we should have $c=3k+2$. Then there is a weight 1 monomial $u^{k+1} v^2$ which can make the singularity isolated. Then, one finds that there is a weight 1 term in the Milnor ring. Actually, this case reduces to type XI$(3+4k,2,2,2)$.

 \para{(2,2,c,3)} For $c=3k,3k+1$ one can show that there is no 1-form symmetry. For $c=3k+2$, one can add an extra monomial $u^{4k+3} $ to $W$ to make the singularity isolated. By computing the Milnor ring, one finds that there is always a marginal term with weight one.  Actually, this case reduces to type XI$(3+4k,2,2,3)$.
 
  \para{(2,b,c,2)}   The Milnor number is given by 
  \be
  \mu=\frac{2 (c+1)}{2 b-1}+2 (b c+c+1)~.
  \ee
To be an integer, we must have $c=(2b-1)k-1$. Then it is easy to show that this reduces to type XI$(2bk-1,2,b,2)$.

 \para{(3,2,c,2)}  The weight vector is irreducible and one finds $r_i=1$, implying trivial 1-form symmetry.

\para{(2,b,2,2)}  $b=2 $.

\para{(2,b,2,3)} $b=2, 5$.

\para{(2,b,2,4)} $b=2$.

 \para{(a,2,2,2)} Considering $a=3k,3k+1,3k+2$ separately, we always find   $r_i=1$, which implies the 1-form symmetry is trivial. 
 
 \para{(a,b,2,2)} If $b=2$, one finds that the 1-form symmetry is always trivial, as $r_i=1$. If $b\ge 3$, we have $q_u=q_v=(1 + a (-1 + b))/(1 + 2 a b)$. Since $1/3<q_u=q_v<1/2$, we must have $q_x=q_y=1-2 q_u$. As a result, we have $b=2a-2$, and the 1-form symmetry is trivial. 
 
 \para{ (2,b,3,3)} $b=\cancel 2$.
 
 \para{(3,b,2,3) }  $b=4$.

 \para{(3,b,3,2)}  $b=2, 3, 10$.

  \para{(3,b,4,2) } $b =\cancel 2$.

 \para{(4,b,3,2)} $b=2$.
 
 \bigskip
\noindent
 This concludes our proof of the main claim in this paper. $\square$

\end{appendix}
 
 \newpage
\bibliography{chetdocbib}

\end{document}